\documentclass[a4paper,11pt]{article}
\pdfoutput=1

\usepackage{jcappub}

\usepackage[T1]{fontenc}

\newcommand{\be}{\begin{equation}}
\newcommand{\ee}{\end{equation}}
\newcommand{\bea}{\begin{eqnarray}}
\newcommand{\eea}{\end{eqnarray}}

\title{\boldmath Parametric Resonance in the Early Universe -- A Fitting Analysis}

\author[a]{Daniel G. Figueroa,}
\author[b]{Francisco Torrent\'i}

\affiliation[a]{Theoretical Physics Department, CERN, Geneva, Switzerland}

\affiliation[b]{
    Instituto de F\'isica Te\'orica IFT-UAM/CSIC, Universidad Aut\'onoma de Madrid, Cantoblanco 28049
    Madrid, Spain.}

\emailAdd{daniel.figueroa@cern.ch}
\emailAdd{f.torrenti@csic.es}

\abstract{Particle production via parametric resonance in the early Universe, is a non-perturbative, non-linear and out-of-equilibrium phenomenon. Although it is a well studied topic, whenever a new scenario exhibits parametric resonance, a full re-analysis is normally required. To avoid this tedious task, many works present often only a simplified linear treatment of the problem. In order to surpass this circumstance in the future, we provide a fitting analysis of parametric resonance through all its relevant stages: initial linear growth, non-linear evolution, and relaxation towards equilibrium. Using lattice simulations in an expanding grid in $3+1$ dimensions, we parametrize the dynamics' outcome scanning over the relevant ingredients: role of the oscillatory field, particle coupling strength, initial conditions, and background expansion rate. We emphasize the inaccuracy of the linear calculation of the decay time of the oscillatory field, and propose a more appropriate definition of this scale based on the subsequent non-linear dynamics. We provide simple fits to the relevant time scales and particle energy fractions at each stage. Our fits can be applied to post-inflationary preheating scenarios, where the oscillatory field is the inflaton, or to spectator-field scenarios, where the oscillatory field can be e.g.~a curvaton, or the Standard Model Higgs.}
\begin{document}
\maketitle
\flushbottom

\section{Introduction}

Compelling evidence supports the idea of an inflationary phase in the early Universe~\cite{Planck2015}. The specific particle physics realization of inflation is however uncertain, so the inflationary period is typically parametrized in terms of a scalar field, the inflaton, with a vacuum-like potential. After inflation, the reheating stage follows, converting all inflationary energy into different particle species, which represent all the matter and radiation in the Universe. Eventually, the created particles dominate the total energy budget and 'thermalize', signaling the onset of the 'hot Big Bang' thermal era.

In this paper we consider inflaton potentials with simple monomial shapes, as this gives rise to one of the most important particle creation phenomena in the early universe: parametric resonance. This is the case of chaotic inflation models, where the inflaton rolls down a monomial potential during the whole inflationary period. Although these scenarios are under tension with cosmological data~\cite{Planck2015}, the simple addition of a small non-minimal gravitational coupling reconcile them with the observations~\cite{Tsujikawa:2013ila}. Some scenarios which fit perfectly well the observational data, e.g.~Higgs-Inflation~\cite{Bezrukov:2007ep,Bezrukov:2010jz} and Starobinsky inflation~\cite{Starobinsky:1980te}, also exhibit a monomial potential with a single minimum, but only during the stages following inflation.

In all the scenarios we consider, soon after the end of inflation, the inflaton is in the form of a homogeneous condensate, and starts oscillating around the minimum of its potential. Each time the inflaton crosses zero, all particle species sufficiently strongly coupled to the inflaton, are created in energetic bursts. In the case of bosonic species, the production of particles is resonant, and the energy transferred grows exponentially within few oscillations of the inflaton~\cite{Traschen:1990sw,Kofman:1994rk,Shtanov:1994ce,Kaiser:1995fb,Kofman:1997yn,Greene:1997fu,Kaiser:1997mp,Kaiser:1997hg}. In the case of fermionic species, there is also a significant transfer of energy~\cite{Greene:1998nh,Greene:2000ew,Peloso:2000hy,Berges:2010zv}, but Pauli blocking prevents resonance from developing. The production of particles in this way, either of fermions or bosons, represents the archetypical example of what is meant by an initial 'preheating' stage of reheating.

Inflationary preheating is however not the only case where parametric resonance takes place in the early Universe. If a light spectator field is present during inflation, this field forms a homogeneous condensate during the inflationary period, and oscillates around the minimum of its potential afterwards. This is the case e.g.~of the curvaton scenario~\cite{Enqvist:Curvaton,Lyth:Curvaton,Takahashi:Curvaton,Mazumdar:2010sa}. The curvaton may decay after inflation via parametric resonance, transferring abruptly all its energy to the particle species coupled to it~\cite{Enqvist:2008be, Enqvist:2012tc, Enqvist:2013qba, Enqvist:2013gwf}. Another example of a spectator field, naturally decaying through parametric resonance after inflation, is the Standard Model Higgs field. If the Higgs is weakly coupled to the inflationary sector, the Higgs is always excited either during inflation~\cite{Starobinsky:1994bd,Enqvist:2013kaa,DeSimone:2012qr}, or towards the end of it~\cite{Herranen:2015ima,Figueroa:2016dsc}. The Higgs is then 'forced' to decay into the rest of the SM species after inflation\footnote{Note that in the case of Higgs-Inflation~\cite{Bezrukov:2007ep,Bezrukov:2010jz}, the Higgs also decays after inflation via parametric resonance, into the rest of the SM fields~\cite{Bezrukov:2008ut,GarciaBellido:2008ab,Figueroa:2009jw}. In this scenario, the Higgs plays however the role of the inflaton. Therefore, the Higgs decay in the case of Higgs-inflation scenarios~\cite{Bezrukov:2008ut,GarciaBellido:2008ab,Figueroa:2009jw,Bezrukov:2014ipa}, should rather be categorized within the context of preheating scenarios.}, via parametric resonance~\cite{Enqvist:2013kaa,Enqvist:2014tta,Figueroa:2014aya,Kusenko:2014lra,Figueroa:2015hda,Enqvist:2015sua,Figueroa:2016dsc}. 

In this paper, independently of the context, we will often refer to the oscillatory field as the 'mother' field, and to the created species as the 'daughter' fields. Particle production of daughter fields via parametric resonance, corresponds to a non-perturbative effect, which cannot be captured by perturbative coupling expansions, not even if the couplings involved are small~\cite{Kofman:1997yn}. During the initial stage of parametric resonance, the system is linear, and analytical methods can be applied. As the particle production is exponential for bosonic species, the daughter field(s) eventually 'backreact' onto the mother field, making the system non-linear. In order to fully capture the non-linearities of the system, we need to study this phenomenon in the lattice. The approach of classical field theory real-time lattice simulations can be considered valid as long as the occupation number of the different species is much larger than one, and hence their quantum nature can be ignored~\cite{Khlebnikov:1996mc,Prokopec:1996rr}. Lattice simulations have been, in fact, successfully carried out for different preheating scenarios during the last years, see e.g.~\cite{Allahverdi:2010xz,Amin:2014eta} and references therein. However, each time a new scenario exhibits parametric resonance, a new re-analysis is often required. 

As lattice simulations are computationally expensive and time consuming, and not everybody has the expertise on the appropriate numerical packages~\cite{Latticeeasy-paper,Defrost-paper,Easther:2010qz,Huang:2011gf,GABElink}, many studies often resort to over-simplified analytical analysis, which capture only the initial linear stage. A systematic study of parametric resonance, fitting the dynamics through all the relevant stages, from the initial linear growth till the relaxation towards equilibrium, passing through an intermediate non-linear stage, is missing in the literature. In this work, we fill in this gap. We have used massively parallelized lattice simulations to charaterize the dynamics of parametric resonance through all its stages. We have parametrized the dynamics by scanning over the relevant circumstances and parameters: role of the oscillating field, particle coupling, initial conditions, and background rate of expansion. We have obtained in this way simple fits to the most significant quantities, like the characteristic time scales and energy fractions of the different particle species. Our fitted formulas can be applied to the study of parametric resonance in scenarios where the mother field dominates the energy budget of the universe (i.e.~preheating), or in scenarios where the mother field represents only a sub-dominant component (e.g.~inflationary spectator fields).

As parametric resonance in the context of the early Universe has been well studied in the past, let us emphasize here that with our present work, we simply aim to aid in the analysis of future scenarios exhibiting parametric resonance. The advantage of using our fitted formulas will be twofold: on the one hand skipping the tedious task of running new simulations, and on the other hand preventing the use of over-simplified linear analysis of the problem.

The structure of this work is as follows. In Section \ref{sec:AnalyticalParamRes} we describe general aspects of parametric resonance, while we derive an analytical estimate of the decay time of the mother field, based on a linear calculation. In Section~\ref{sec:Lattice} we preset the numerical results from our lattice simulations. We describe our results for preheating with a quartic potential in Section~\ref{sec:lphi4}, and for preheating with a quadratic potential in Section~\ref{sec:m2phi2}. We compare these results against the analytical estimations from Section~\ref{sec:AnalyticalParamRes}. In Section \ref{sec:specfields} we present the analogous numerical study for scenarios where the mother field represents only a sub-dominant energy component of the Universe. In Section \ref{sec:Summary} we list all fitted formulas together from all scenarios considered. In Section~\ref{sec:discussion} we discuss the context where our results can be useful. In the appendices we present details on the lattice formulation we have used, and discuss briefly the evolution of the field spectra in some of the scenarios considered. 

From now on we consider $\hbar = c = 1$ units, and represent the reduced Planck mass by {\small$m_p^2 = {1/{8\pi G}} \simeq 2.44\cdot10^{18}$ GeV}. We take a flat background with Friedman-Robertson-Walker (FRW) metric {\small$ds^2 =  dt^2 - a^2 (t) dx^i dx^i$}, where {\small$a(t)$} is the scale factor, and {\small$t$} the cosmic time. 

\section{Parametric Resonance: Analytical Calculation}\label{sec:AnalyticalParamRes}

Before we move into the specific scenarios of Sections \ref{sec:lphi4}, \ref{sec:m2phi2}, and \ref{sec:specfields}, let us discuss some general aspects of parametric resonance, while we derive an analytical estimation of the decay time of the mother field. In the following Sections we will compare this analytical estimation with the results obtained from lattice simulations.

Let us begin by considering a scalar field $\phi$ with a quartic potential $V(\phi) = {\lambda\over 4}\phi^4$, coupled to another scalar field $X$ through an interaction $g^2 \phi^2 X^2$, with $g^2$ a dimensionless coupling constant. The equations of motion (EOM) of the system read
\be
\ddot \phi - \frac{1}{a^2} \nabla^2\phi + 3 H \dot \phi + g^2  X^2 \phi  + \lambda\phi^3 = 0 \ , \hspace{0.5cm} \ddot X -  \frac{1}{a^2} \nabla^2 X + 3 H \dot X + g^2 \phi^2 X = 0  \label{eq:generic-eom} \ ,\ee
where $H \equiv \dot{a} / a$ is the Hubble rate. We will consider the field $\phi$ to be initially homogeneous with some initial amplitude $\phi_* \neq 0$, and null initial velocity $\dot\phi_* = 0$, whilst the field $X$ is not excited initially, $X_* = \dot X_* = 0$. If we neglect for the time being the interaction term, the equation for the homogeneous part of the $\phi$ field, corresponds to an anharmonic oscillator in the presence of a friction term. As $V(\phi)$ has a single minimum at $\phi = 0$, the field $\phi$ will start rolling down towards the minimum. If the friction term dominates over the potential term, the system is overdamped and the field rolls down very slowly, in the so called slow-roll regime. Eventually, as the Hubble rate diminishes due to the expansion of the Universe, there will be a time when the system becomes underdamped. This time signals the onset of the mother field oscillations around the minimum of its potential. More specifically, we will define an initial time $t_*$   as the moment when the Hubble rate just becomes smaller than the effective frequency of oscillation. In light of Eq.~(\ref{eq:generic-eom}), the period of oscillation is $T \propto {1/\sqrt{\lambda}\phi_*}$, so we can determine $t_*$ from the condition $H_* \equiv \sqrt{\lambda}\phi_*$, with $H_* \equiv H(t_*)$ and $\phi_* \equiv \phi(t_*)$.

After a convenient conformal transformation of the time and field variables
\begin{eqnarray}
\vec{x} \rightarrow \vec{y} \equiv \sqrt{\lambda}\phi_*\vec{x}\,,~~~~~~ t \rightarrow z \equiv \sqrt{\lambda}\phi_*\tau\,,~~~~\tau \equiv \int {dt\over a(t)}\,,\\
\phi \rightarrow \varphi \equiv a(t){\phi\over\phi_*}\,,~~~~~~ X \rightarrow \chi \equiv a(t){X\over\phi_*} \,,\hspace*{1.5cm}
\end{eqnarray}
the EOM read
\be
\varphi'' + \varphi^3 - \nabla^2\varphi + q \chi^2 \varphi = {a''\over a}\varphi \ , \hspace{0.5cm} \chi'' -  \nabla^2 \chi + q \varphi^2 \chi = {a''\over a}\chi  \label{eq:conformal-eom} \ ,\ee
where $' \equiv d/dz$, $\nabla_i \equiv \partial/\partial y_i$, and $q$ is the so called {\it resonance parameter}, 
\be
q \equiv {g^2\over\lambda} \ .
\ee
Neglecting the interaction term, the EOM of the homogeneous part of $\varphi$ reduces to
\be
\varphi'' + \varphi^3 = {a''\over a}\varphi \label{eq:InflatonConformal-eom} \ . \ee
In the case when the mother field dominates the energy budget of the universe (e.g.~in preheating), the energy density scales as radiation dominated (RD)~\cite{Turner:1983he}, so the scale factor behaves as $a \propto \sqrt{t} \propto z$. In this case, the term on the $rhs$ of Eq.~(\ref{eq:InflatonConformal-eom}) simply vanishes, ${a''/a} = 0$. If the field $\phi$ does not dominate the energy budget of the universe, the behavior of the scale factor depends on the equation of state $w$ of the dominant energy component of the Universe. For fixed $w$, one can find ${a''/a} = {1\over 2}(1-3w)/(1+0.5(1+3w)z)^2$, which either dies away as $a''/a \propto 1/z^2$ if $w \neq 1/3$, or vanishes directly $a''/a = 0$ for $w = 1/3$. We will therefore set ${a''/a} = 0$, for the simplicity of the discussion. The solution of Eq.~(\ref{eq:InflatonConformal-eom}) $\varphi'' + \varphi^3 = 0$ with initial conditions $\varphi_* = 1$, $\varphi_*'= 0$, is the Elliptic function\footnote{In reality, the initial conditions should be $\varphi_* = 1$, $\varphi_*' \neq 0$, with $\varphi_*'$ some value propagated from the past when the field was deep in the slow-roll condition $3H\dot{\phi} + \lambda\phi^3 = 0$. Taking into account this does not change the essence of the oscillatory regime once the field enters into the underdamped regime. Hence, for the easiness of the discussion, we will simply stick here to the solution $\varphi(x) = cn(z;1/2)$.} 
\begin{equation}\label{eq:EllipticSol}
\varphi(z) = {\rm cn}(z;1/2) \ .
\end{equation}

\noindent The equation for the Fourier modes of the field $\chi$ (assuming RD) can be written as
\begin{eqnarray}\label{eq:modeEQ}
    \chi_k''+ \left(\kappa^2+q \varphi(z)^2\right) \chi_k = 0~\,,~~~~~\kappa \equiv {k\over\sqrt{\lambda}\varphi_*} \ .
\end{eqnarray}
In this form the equation for the fluctuations of the $\chi$ field does not depend on the expansion of the universe, and it is completely reduced to a problem in Minkowski space-time\footnote{This is of course only a special feature of the conformally invariant theory ${\lambda\over4}\phi^4 + {1\over2}g^2\phi^2 X^2$.}. Given the behavior of $\varphi(z)$ in Eq.~(\ref{eq:EllipticSol}), Eq.~(\ref{eq:modeEQ}) corresponds to the class of the Lam\'e equations, which has a well-understood structure of resonances. Whenever $q \in {1\over2}[n(n+1),(n+1)(n+2)]$, with $n = 1, 3, 5, ...$ (i.e. $q \in$ [1, 3], [6, 10], ...), there is an infrared band of modes $k \lesssim k_L \sim q^{1/4}H_*$, for which the modes can be exponentially amplified as $\chi_k \propto e^{\mu_kz}$, with $\mu_k$ a parameter known as the Floquet index~\cite{Greene:1997fu}. Considering the mode frequency $\omega_k^2 \equiv \kappa^2 + q\varphi^2$, we can speak of adiabatic modes if the condition $\omega'(k) < \omega_k^2$ is fulfilled. The set of unstable modes $k \lesssim k_L$ correspond to the modes that violate the adiabaticity condition each time $\varphi$ crosses around zero, verifying the opposite condition, $\omega'(k) > \omega_k^2$. The instability $\chi_k \propto e^{\mu_kz}$ of the resonant modes is naturally interpreted as a strong particle creation of the $\chi$ field, as the occupation number grows as $n_k \sim |\chi_k|^2 \propto e^{2\mu_kz}$. 

        \begin{figure}
            \begin{center}
\includegraphics[width=6.7cm]{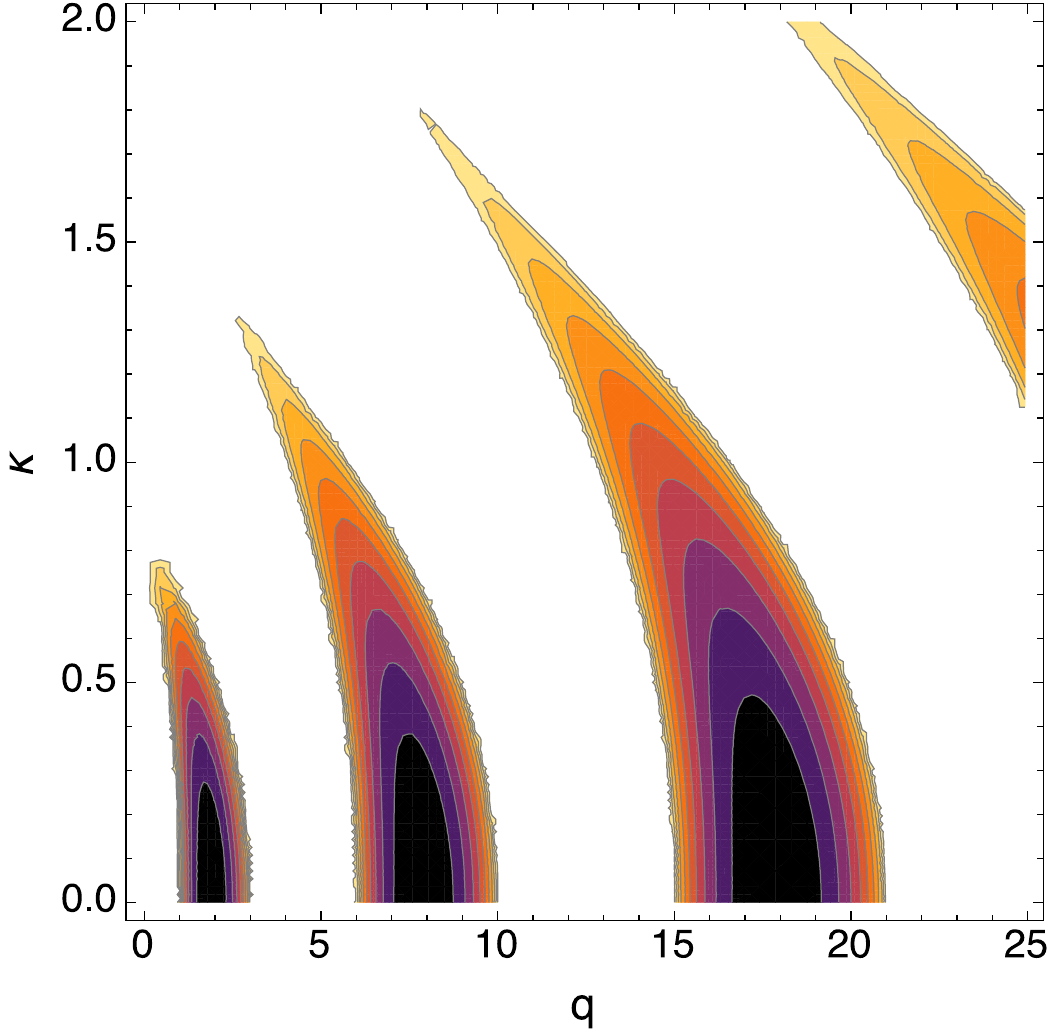} \hspace{0.1cm}                \includegraphics[width=7.9cm]{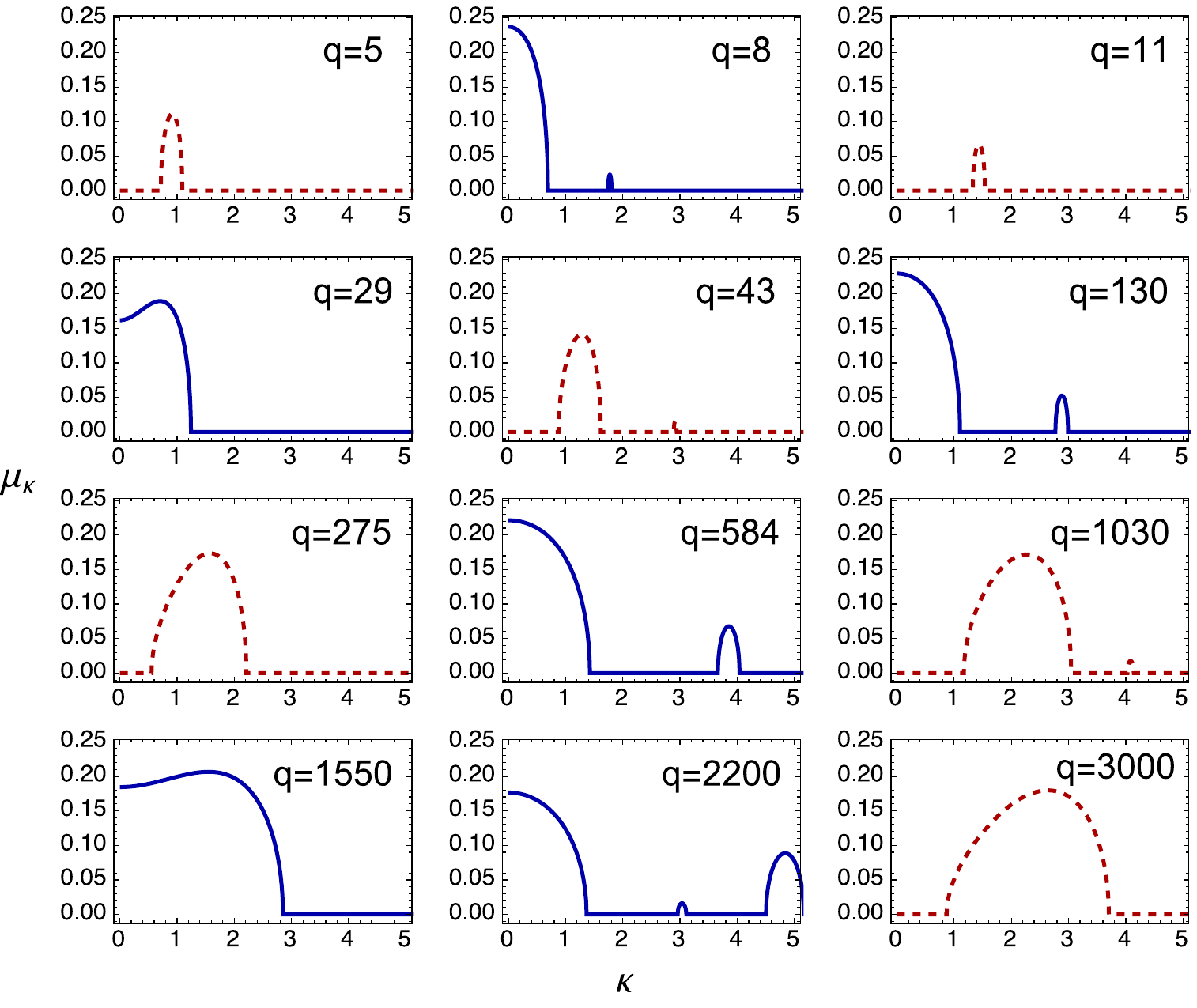}

            \end{center}
                \caption{Left: We show the stability/instability chart of the Lam\'e equation (\ref{eq:modeEQ}). Coloured bands indicate the regions of the ($q$,$\kappa$) parameter space in which the real part of the Floquet index is a positive number $\mathfrak{Re} [\mu_{\kappa} ]>0$ and hence the solution of the Lam\'e equation is exponential. The darker the colour, the greater the index, up to a maximum of $\mu_{\kappa} \approx 0.237$ for black areas. White areas are the regions in which $\mathfrak{Re}[\mu_{\kappa}] = 0$. Right: Some examples of the Floquet index derived numerically from the Lam\'e equation for resonance parameters ranging between $q=5$ and $q=3000$. In each panel, we plot the corresponding Floquet index $\mu_{\kappa}$ as a function of the momentum $\kappa$. We have divided the different $q$'s in two groups: those inside one of the resonance bands $q \in [1, 3]$, $[6, 10]$, $[15, 21], ...$, which excite modes down to $\kappa=0$ (blue solid lines), and those which are in between resonance bands (red dashed lines), which only excite modes down to some minimum momentum $\kappa_{\rm min} > 0$.} 
            \label{fig:lame-bands}
        \end{figure}

If the resonance parameter $q > 1$ is not within one of the resonant bands, but lies in between two adjacent bands, then there is still a resonance of the type $\chi_k \propto e^{\mu_kz}$, but within a shorter range of momenta $k_{\rm min} \leq k \lesssim k_L$, and hence with a smaller Floquet index $\mu_k$. There is a theoretical maximum value for the Floquet index given by $\mu_{k,{\rm max}} \equiv 0.2377...$~\cite{Greene:1997fu}, so that any $\mu_k$ is always constrained as $\mu_k \leq \mu_{k,{\rm max}}$ for $q > 1$. For resonant parameters $q \gg 1$, $\mu_k$ is typically of order $\mathcal{O}(0.1)$, see Fig.~\ref{fig:lame-bands}.

For simplicity, in the remaining of this Section we will consider   the resonance parameter to be within one of the resonant bands, $q \in [1, 3]$, $[6, 10]$, $[15, 21], ...$. The growth of the fluctuations in the initial stages of resonance is described by the linear Eq.~(\ref{eq:modeEQ}). Even if the amplitude of the fluctuations grows exponentially, Eq.~(\ref{eq:modeEQ}) is expected to represent a good description of the field excitation during the initial stages. Of course, one is ignoring in this way the backreaction of the $\chi$ bosons into $\varphi$. This is a good approximation for as long as the energy tranferred into the $\chi$ field is only a marginal fraction of the energy available in the mother field $\varphi$. In our numerical analysis of Section \ref{sec:lphi4}, we will quantify exactly when the linear approximation breaks down. For the time being, to continue with our analytical approach, we will just consider valid the linear regime all the time through.

The energy density of the created particles due to the resonance, is given by
\be
  \rho_{\chi} = {1\over 2\pi^2a^3}\int dk k^2 n_k \Omega_k~,~~~\Omega_k^2 \equiv {k^2\over a^2} + g^2\overline{\phi^2}\,,
\ee
where we have introduced an oscillation-averaged effective mass for the $\chi$ field,
\begin{eqnarray}\label{eq:massX}
m_\chi^2 = g^2\overline{\phi^2} = g^2{\phi_*^2\over a^2} \overline{\varphi^2}\,, ~~~~~~~ \overline{\varphi^2} \equiv {1\over Z_T}\int_{z}^{z+Z_T} dz' \varphi^2(z') \simeq 0.46\,,
\end{eqnarray}
with $Z_T \simeq 7.416$ the oscillation period of $\varphi$~\cite{Greene:1997fu}. From the violation of the adiabaticity condition for $q \gg 1$, i.e.~$\omega_k' > \omega_k^2$, we can determine an estimation of the maximum (comoving) momentum possibly excited in broad resonance, 
\begin{eqnarray}\label{eq:momentumX}
k \lesssim k_L^2 \equiv {q^{1/2}\over \sqrt{2\pi^2}}\lambda\phi_{*}^2 = {q^{1/2}\over \sqrt{2\pi^2}}H_{*}^2 \ ,
\end{eqnarray}
where we have identified $\sqrt{\lambda}\phi_* \equiv H_*$. From Eqs.~(\ref{eq:massX}),(\ref{eq:momentumX}), we conclude that
\begin{eqnarray}
    {m_\chi^2\over (k_L/a)^2} \sim \mathcal{O}(1)q^{1/2} > 1\,.
\end{eqnarray}
In other words, in broad resonance $q \gg 1$, the decay products are always non-relativistic. Correspondingly we can approximate the effective mode frequency as $\Omega_k \simeq m_\chi \sim g{\phi_*\over a}\varphi_{\rm rms}$, where $\varphi_{\rm rms} \equiv \sqrt{\overline{\varphi^2}} \simeq {2\over3}$. If $q$ is within a resonant band, then all modes with momenta $0 \leq k \lesssim k_L$ are excited with some Floquet index varying within $[0,\mu_{k,{\rm max}}(q)]$. This corresponds to the cases with blue solid lines in Fig.~\ref{fig:lame-bands}. We can therefore model the occupation number of the excited modes simply as a step function $n_k = e^{2\overline{\mu}z}\Theta(1-k/k_L)$, with $\overline{\mu} \simeq 0.2$ a mean Floquet index. It follows that\footnote{Notice that the scaling $\rho_{\chi}(z) \propto 1/a^4$ is characteristic of relativistic species, despite the fact that we stated that the decay products are non-relativistic. This is because the energy density of the daughter fields is given by $\rho_\chi \simeq n_\chi\cdot m_\chi$, with $n_\chi$ the number density and $m_\chi$ their mass, as it corresponds to any non-relativistic species. However, while $n_\chi \propto 1/a^3$, the effective mass is also time dependent, $m_\chi \propto 1/a$, and hence the total energy density scales as radiation $\rho_\chi \propto 1/a^4$.}
\begin{eqnarray}\label{eq:GaugeEnergyAtEff}
   \rho_{\chi}(z) \simeq {\varphi_{\rm rms}\over 6\pi^2 a^4} e^{2\overline{\mu} z}\,g\phi_* k_L^3 
    \simeq {q^{5/4}\over 2^{3/4}\cdot 3^2 \cdot \pi^{7/2}} {e^{2\overline{\mu} z}\over a^4}H_*^4 \ .
\end{eqnarray}

This is how the energy density of the daughter fields (those fully within a resonant band) will grow, at least as long as their backreaction into the mother field remains negligible. Using this linear approximation we can estimate the moment $z_{\rm eff}$ at which an efficient transfer of energy has taken place from $\phi$ into the $\chi$ bosons, characterized by $\rho_\chi(z_{\rm eff}) = \rho_\phi(z_{\rm eff})$. This will be just a crude estimate of the time scale of the mother field decay, since by then backreaction and rescattering effects will have become important, invalidating the linear approach. However, the nonlinear effects due to backreaction of the decay products, simply tend to shut off the resonance. Hence, the calculation in the linear regime should provide at least, in principle, a reasonable estimate of the time scale for when the energy has been efficiently transferred into the daughter fields. Whether $z_{\rm eff}$ is also a good estimate of the decay time of the mother field, will be contrasted against our lattice simulations in the next Section. 

The energy of the oscillating field, since the onset of the oscillations, decays as~\cite{Figueroa:2015hda}
\begin{eqnarray}\label{eq:HiggsEnergyAtEff}
    \rho_{\varphi}(z) = {3\over4}{\lambda\phi_*^4\over a^4}\overline{\varphi^4} = {H_*^4\over 4\lambda a^4}\,,
\end{eqnarray}
where in the second equality we have used $\overline{\varphi^4} \simeq 1/3$. We can now find $z_{\rm eff}$ by simply equating Eqs.~(\ref{eq:GaugeEnergyAtEff}) and (\ref{eq:HiggsEnergyAtEff}),
\begin{eqnarray}
    q^{1/4}e^{2\overline{\mu}z} = {2^{-1/4} \cdot 3^2 \cdot \pi^{7/2} \over g^2}\,,
\end{eqnarray}
so that
\begin{eqnarray}\label{eq:EffEnergyTransferTimeScale} 
    z_{\rm eff} \simeq  + {1\over 2\overline{\mu}}\left[6 - \ln \lambda - {5\over 4}\ln q\right] \ . 
\end{eqnarray}
For instance, for chaotic inflation with quartic potential, $\lambda \simeq 10^{-13}$, and hence $\log\lambda \simeq - 30$. Looking at Fig.~\ref{fig:lame-bands}, we see that the Floquet index of the modes $0 \leq k \lesssim k_L$ for which $q$ is within a resonant band (blue solid lines in the figure), can be approximated, as said, by a simple step function $\mu_k \simeq \overline{\mu}\Theta(1-k/k_L)$, with a mean Floquet index $\overline{\mu} \simeq 0.2$. Taking this into account, for $V \propto \phi^4$ chaotic inflation we find 
\begin{eqnarray}\label{eq:EffEnergyTransferTimeScaleApproxPhi4}
z_{\rm eff} \sim  2.5\left(36 - 2.9\log_{10} q\right)~~~~~~\Rightarrow ~~~~~ 83 \gtrsim z_{\rm eff} \gtrsim 18\,,~~{\rm for}~q \in [10,10^{10}] \ .
\end{eqnarray}
It is clear that the larger the $q$, the shorter it takes for the mother field to transfer energy efficiently into the daughter fields. This is expected, as the stronger the interaction is, the faster the decay should be. We see that the decay time, however, according to the above calculation, is always some value of the order $z_{\rm eff}\sim \mathcal{O}(10)$. Therefore, contrary to 'popular wisdom' about parametric resonance, the time scale $z_{\rm eff}$, identified with the decay of the oscillatory field in the linear approximation, is in practice mostly independent of $q$. Though it is certainly true that the larger the $q$ the shorter the decay, the dependence is only logarithmic, see Eq.~(\ref{eq:EffEnergyTransferTimeScale}), so the time scale does not change appreciably. For instance, increasing $q$ in 10 orders of magnitude, only speeds up the decay time in a factor $\sim 1/4$. In the following Section we will check the validity of these estimation by comparing it with the numerical outcome obtained directly from lattice simulations. 

Before we move into the numerical results, let us note that a similar computation can be carried out for a mother field with a quadratic potential $V(\phi) = \frac{1}{2} m^2 \phi^2$. The details are more cumbersome in this case, because contrary to the quartic case previously described, in the quadratic case (when the expansion of the Universe cannot be ignored), the Floquet index is not fixed for a given mode. This is because there is now a new mass scale, $V ''(\phi) = m^2$, which breaks the conformal invariance, making impossible to reduce the problem into a Minkowski analogue~\footnote{Of course if there was no expansion of the Universe, the problem is directly formulated in Minkowski, so the structure of the resonance bands is fixed. In such a case, there is a well defined Floquet index for each mode. However, whenever the expansion of the universe cannot be ignored, as it is the case in $m^2\phi^2$ preheating, each mode scans several resonance bands, and therefore one cannot ascribe a given Floquet index to a given mode.}, as it happened in the quartic case. In the quadratic case the resonance of a given mode is such that each mode scans several resonance bands, and the evolution of a resonant mode function $\chi_k$ is in fact stochastic, see~\cite{Kofman:1994rk} for a detailed explanation on this. Without entering into further details, as the linear computation in the quadratic case was carried out in~\cite{Kofman:1997yn}, we do not repeat it here. We just quote their result, adapting it to our notation. They find that the maximum momentum excited during parametric resonance in a quadratic potential is approximately
\be k \lesssim k_M \equiv \sqrt{\frac{2}{\pi}} q_*^{1/4} m \ . \label{eq:mathieu-kcut}\ee
Taking $\bar\mu \simeq 0.15$ as a reasonable averaged value of the stochastic Floquet index $\mu_\kappa$, for chaotic inflation with $V (\phi) \propto \phi^2$, Eq.~(112) of~\cite{Kofman:1997yn} is equivalent to
\begin{eqnarray}\label{eq:EffEnergyTransferTimeScaleApproxPhi2}
z_{\rm eff} \simeq 8.3(15.1-1.1\log_{10}q_*)~~~~~~\Rightarrow ~~~~~ 89 \gtrsim z_{\rm eff} \gtrsim 34\,,~~{\rm for}~q_* \in [10^4,10^{10}] 
\end{eqnarray}
with $q_* \equiv g^2 \phi_*^2 /(4 m^2)$. As in the quartic case, we see that one expects this scale to be always of the order of $z_{\rm eff} \sim \mathcal{O}(10)$, changing only logarithmically with resonance parameter.

\section{Parametric Resonance: Lattice Simulations}
\label{sec:Lattice}

As mentioned before, parametric resonance in the early Universe can be realized in two main different circumstances: $i)$ when the mother field dominates the energy budget of the Universe, and $ii)$ when the mother field is only a sub-dominant energy component of the Universe. In this Section we will perform lattice simulations of both situations:
\begin{list}{}{}

\item $i)$ {\it Inflaton Preheating}. In this case we identify the field $\phi$ with the field responsible for inflation, the inflaton. We consider single-field slow-roll scenarios where the inflaton has a monomial potential $V_{\rm inf} (\phi)$. Short after inflation ends, when the slow-roll parameters become approximately of order unity, the Hubble rate just becomes smaller than the inflaton mass. As the inflaton has a very large vacuum expectation value (VEV), the inflaton amplitude starts then oscillating around the minimum of its potential. This induces a strong creation of all particles coupled to it, if the coupling strength is sufficiently large. The creation of these particles represents possibly the most important particle creation stage in the history of the Universe: as the inflaton and its decay products are the dominant energy component of the Universe, this stage represent the creation of (most of) the matter in the universe. This adds an extra difficulty, as the time-evolution of the scale factor must be obtained by solving self-consistently the fields EOM together with the Friedmann equations. We consider the two paradigmatic particular models of chaotic inflation, where the inflaton has either a quartic potential (Section \ref{sec:lphi4}) or a quadratic potential (Section \ref{sec:m2phi2}):
    \bea
     V_{\rm inf} (\phi) = \left\{ \begin{array}{ll}
        \frac{1}{4} \lambda \phi^4  , \hspace{1cm} & \lambda \approx 9 \times 10^{-14} , \vspace*{2mm}\\ 
         \frac{1}{2} m^2 \phi^2 , \hspace{1cm} & m \approx 6 \times 10^{-6} m_p \ .
        \end{array} \right. \label{eq:inflation-potentials}
    \eea
The strength of the parameters $\lambda$ and $m$ is fixed by the amplitude of the observed CMB anisotropies. In the quartic model, the energy density of the inflaton scales (after averaging over oscillations) as in a RD background, with $\rho_\phi \propto 1/a^4$, with the scale factor evolving correspondingly as $a(t) \propto t^{1/2}$. In the quadratic model the energy density of the inflaton (again after oscillations-averaging) evolves as in a MD background, with $\rho_\phi \propto 1/a^3$, and the scale factor evolving correspondingly as $a(t) \sim t^{2/3}$. Both scenarios of inflation are in fact challenged by recent CMB measurements~\cite{Planck2015} (the quartic case more severely), but in reality, the simple addition of an non-minimal gravitational coupling to the inflaton can easily reconcile these scenarios with the observations~\cite{Tsujikawa:2013ila}.

\item $ii)$ {\it Inflationary Spectator Fields}. In this second type of scenarios, we consider the field $\phi$ to be just a spectator field during inflation, hence representing a very subdominant component of the energy budget. This does not prevent however the amplitude of these fields to be rather large at the end of inflation (though not as large, in principle, as in single field chaotic inflation scenarios). When inflation ends and the Hubble rate becomes smaller than the effective mass of the spectator field, the amplitude of the field starts oscillating around the minimum of its potential. The expansion rate of the universe after inflation is determined by the inflationary sector, which we will not model explicitly. It is in fact only the evolution of the scale factor that we really need to introduce in the simulations. For instance, for matter-dominated (MD), radiation-dominated (RD) and kination-dominated (KD) universes, the scale factor behaves as $a(t) \propto t^{2/3}$, $a(t) \propto t^{1/2}$, and $a(t) \propto t^{1/3}$, respectively. The most obvious case of a spectator-field is a curvaton, which is normally described with a quadratic potential\footnote{Other polynomial potentials have been considered, but the realization of the curvaton mechanism seems much more contrived in those cases~\cite{Huang:2008zj}.} of the type $V (\phi) = \frac{1}{2} m^2 \phi^2$ in the context of a RD background~\cite{Enqvist:Curvaton,Lyth:Curvaton,Takahashi:Curvaton}. We will restrict our numerical analysis to this case (Section \ref{sec:specfields}), taking $m$ as a free parameter varied over a certain range. A relevant case of a spectator-field with a quartic $\propto \phi^4$ potential, although not a curvaton, is the Standard Model (SM) Higgs in the weak coupling limit~\cite{Starobinsky:1994bd,Enqvist:2013kaa,DeSimone:2012qr,Herranen:2015ima,Figueroa:2016dsc}. The study of the Higgs dynamics after inflation has triggered recently an intense activity~\cite{Enqvist:2013kaa,Enqvist:2014tta,Figueroa:2014aya,Figueroa:2015hda,Enqvist:2015sua,Figueroa:2016dsc}. In particular, in~\cite{Figueroa:2015hda}, the outcome of the dynamics was parametrized in a similar fashion to what we will do here in Section \ref{sec:m2phi2} for the RD quadratic curvaton case. Therefore, we will not repeat the details of the quartic case here, though we will include a summary of those results in Section \ref{sec:Summary}, where we collect the fits from all the cases studied (inflaton or spectator field cases, with quadratic or quartic potential).

\end{list}

In all scenarios, we will always consider a symmetric interaction $g^2\phi^2 X^2$ between the mother field $\phi$ and the daughter field $X$. This interaction is scale free, with $g^2$ a dimensionless coupling constant. This is particularly convenient from the point of view of the lattice, since any other form of interaction would require the introduction of a new mass scale. Besides, this interaction has been often assumed in the context of preheating, and it is the leading interaction term in the context of gauged spectator fields, as demonstrated in~\cite{Figueroa:2015hda} for the case of the SM Higgs. It is also interesting to note that this interaction does not lead to a tree level decay of the mother field into the daughter species, so all the transfer of energy from $\phi$ into $X$ will be due only to the non-perturbative effects characteristic of parametric resonance.

\subsection{Lattice Simulations of preheating with quartic potential}\label{sec:lphi4}

We consider in this section preheating in the case of a massless self-interacting inflaton with potential
\be V_{\rm inf} (\phi) = \frac{1}{4} \lambda \phi^4 \ . \ee
The time $t_*$ for the onset of the oscillatory regime is defined through the condition $H (t_*) = \sqrt{\lambda} \phi (t_*)$. This constitutes the initial time of our lattice simulations. We will write all quantities evaluated at time $t_*$ with a sub-index $*$, so this condition can be simply written as $H_* = \sqrt{\lambda} \phi_*$. From a simple numerical calculation of the homogeneous Klein-Gordon and Friedmann equations, $\ddot\phi + 3(\dot a/a)\dot\phi + {dV\over d\phi} = 0$, $3m_p^2(\dot a/a)^2 = \lbrace V_{\rm inf} (\phi)+(\dot\phi)^2/2 \rbrace$, we obtain $\phi_* \simeq 3.05 m_p$ and $\dot{\phi}_* \simeq - 3.54 m_p^2$. The equations of motion (EOM) of the inflaton and the daughter field can be easily derived, but for convenience, let us first define new field and space-time variables, similarly as in Sec.~\ref{sec:AnalyticalParamRes},
\be \varphi \equiv \frac{a}{\phi_*} \phi \ , \hspace{0.5cm} \chi \equiv \frac{a}{\phi_*} X \ , \hspace{0.5cm} z \equiv H_*\int\frac{dt}{a(t)} \ , \hspace{0.5cm}  \vec{z} \equiv H_* \vec{x}\,, \label{eq:lphi4-variables}\ee
where $x^{\mu} = (t, \vec{x})$ are the old cosmic time and comoving coordinate variables. We denote this set of field and spacetime variables as the 'natural' variables of the problem. We indicate differentiation with respect cosmic/natural time with a dot/prima respectively, so $\dot{} \equiv d / dt$ and $' \equiv d / dz$. Spatial derivatives should equally be understood, from now on, as taken with respect natural variables. In these variables, the EOM are
\bea \varphi'' -\frac{a''}{a} \varphi -  \partial_i \partial_i \varphi + \left( \varphi^{2} + q \chi^2 \right) \varphi = 0  \ , \hspace{0.5cm} \chi'' -  \frac{a''}{a}\chi - \partial_i \partial_i \chi + q \varphi^2 \chi = 0 \ , \label{eq:eom-lame}\eea
where
\be q \equiv \frac{g^2}{\lambda} \label{eq:lphi4-qres}\ee 
is the resonance parameter. These equations are of course the same as Eqs.~(\ref{eq:conformal-eom}) from Sec.~\ref{sec:AnalyticalParamRes}. However, whereas before, in order to gain some insight on the dynamics of parametric resonance, we used the homogeneous part of the equation for $\varphi$ and the Fourier transformed equation of $\chi$, now we will be rather solving the (lattice version) of the full Eqs.~(\ref{eq:eom-lame}) in real space. 

We take $\lambda = 9 \times 10^{-14}$ in this Section, as this is fixed by the observed amplitude of the CMB anisotropies~\cite{Tsujikawa:2013ila}. The strength of the coupling $g^2$ is in principle arbitrary. However, in order not to spoil inflation, radiative corrections in the effective inflaton potential must be under control. This sets a constraint $g \lesssim 10^{-3}$~\cite{Lyth:1998xn}. Unfortunately, in practice we are not capable of simulating resonance parameters outside of the range $0.4 \lesssim q \lesssim 10^4$. Since $q \sim g^2 10^{13}$, this means that we can only simulate couplings $6\cdot 10^{-7} \lesssim g \lesssim 3\cdot 10^{-5}$. The lower limit is due to the natural limitations of the lattice to simulate fields with narrow resonance bands, as we cannot resolve well the relevant dynamical range of momenta with an appropriate number of modes. The upper limit emerges because the required simulation time and number of lattice points grow with $q$. This is discussed in more detail in Appendix \ref{appen:Lattice}, so we refer there to the interested reader. Fortunately, as we shall see, the results for the $q$'s simulated are well described by simple power-law fits, allowing in principle to extrapolate the outcome to larger $q$'s.

\subsubsection{Onset of non-linearities, energy evolution and decay time}
        
Let us briefly recall first the properties of the system from our discussion in Sect.~\ref{sec:AnalyticalParamRes}. As the mode function of the daughter field follows the Lam\'e equation [Eq.~(\ref{eq:modeEQ})], there are unstable solutions of the type $\chi_{\kappa} \sim e^{\mu_{\kappa} z}$, with $\mu_{\kappa}$ the $q$-dependent Floquet index. For certain values of $(q, \kappa)$, $\mathfrak{Re} [\mu_{\kappa}] > 0$, causing an exponential growth of the given field mode, and hence of the occupation number. When $q \in (1,3), (6, 10) , \dots$, the growth of $\chi_{\kappa}$ is much stronger than for other values, as can be seen in the pattern of resonance depicted in Fig.~\ref{fig:lame-bands}. 

\begin{figure}
      \begin{center} 
      \includegraphics[width=7cm]{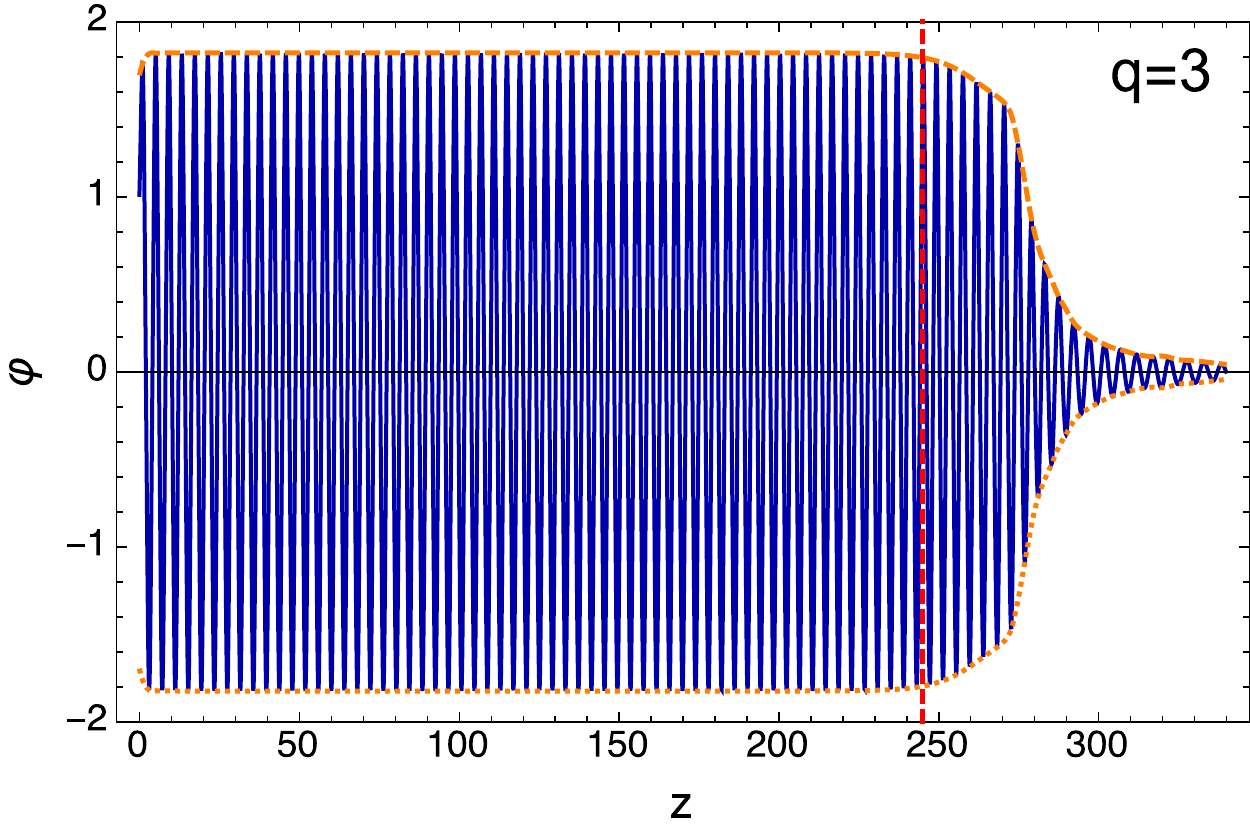} \hspace{0.5cm}
       \includegraphics[width=7cm]{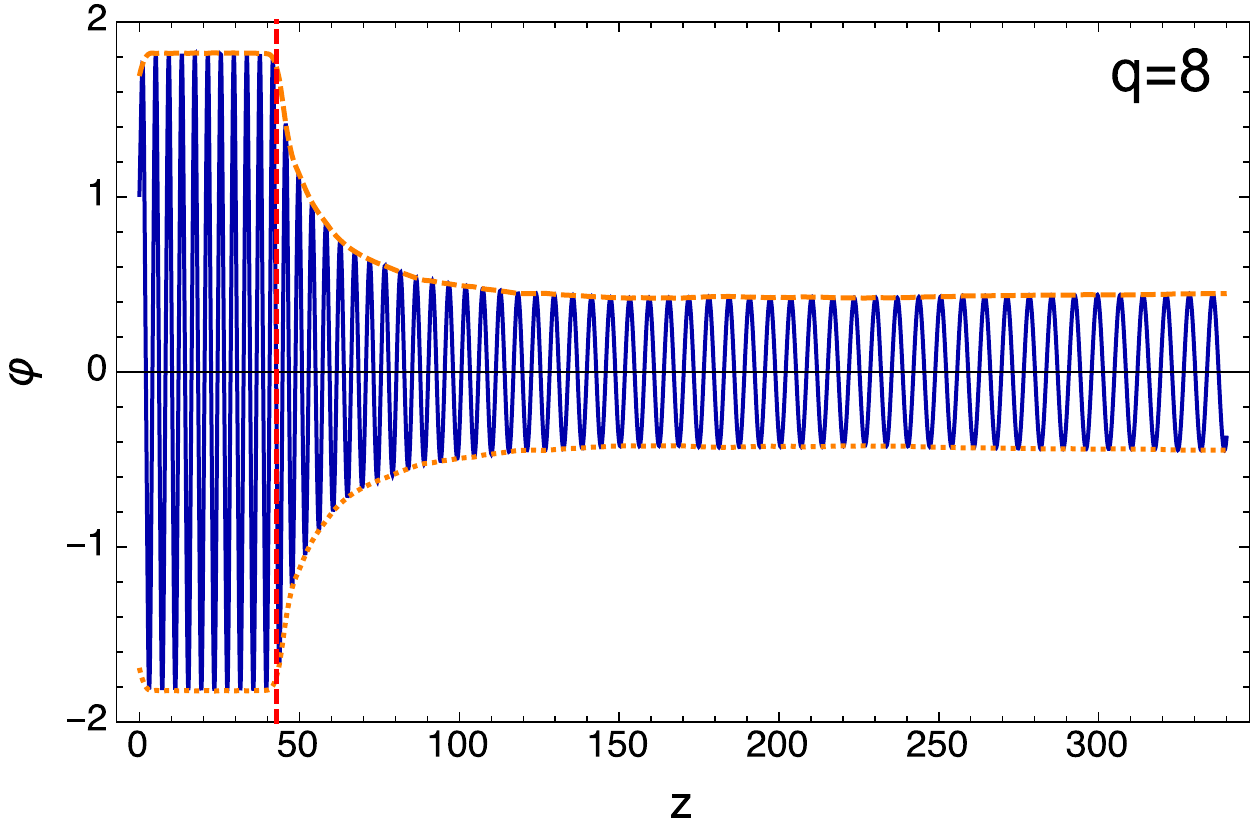}
       \includegraphics[width=7cm]{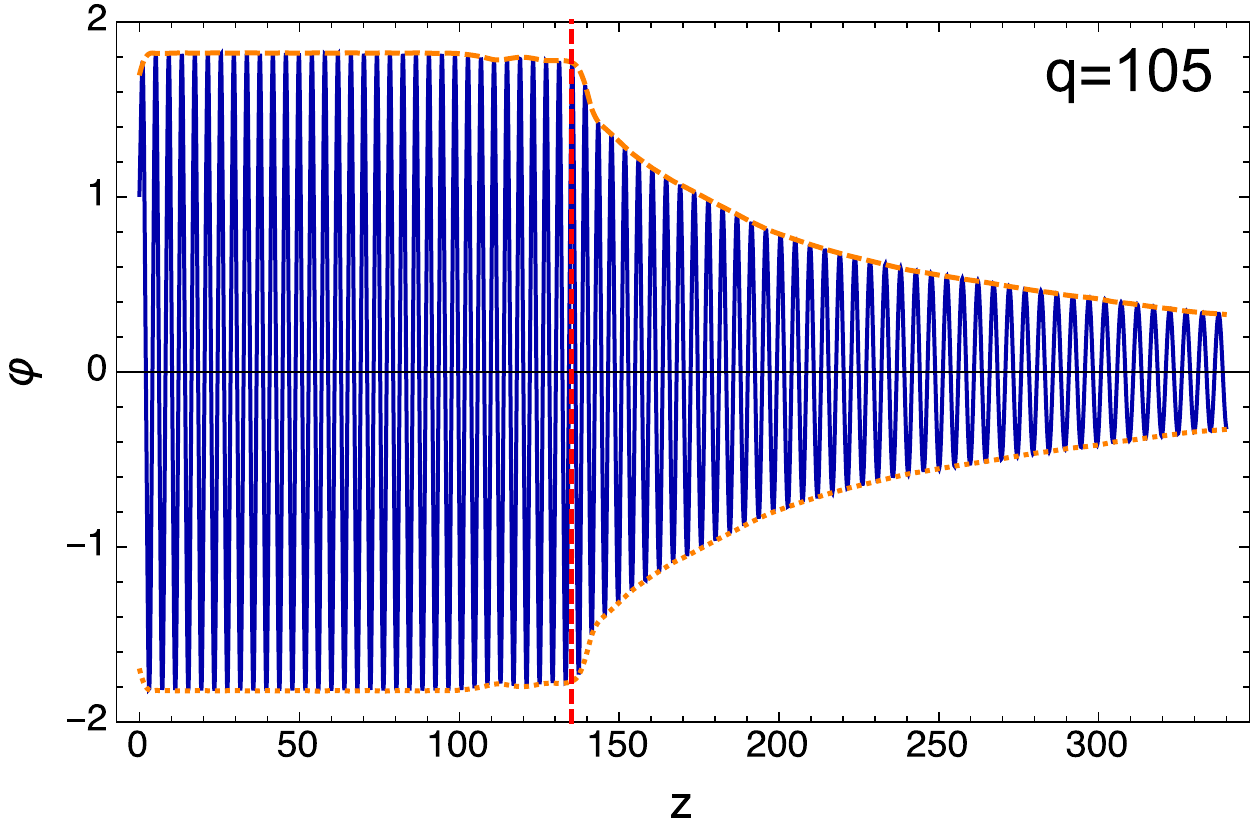} \hspace{0.5cm}
        \includegraphics[width=7cm]{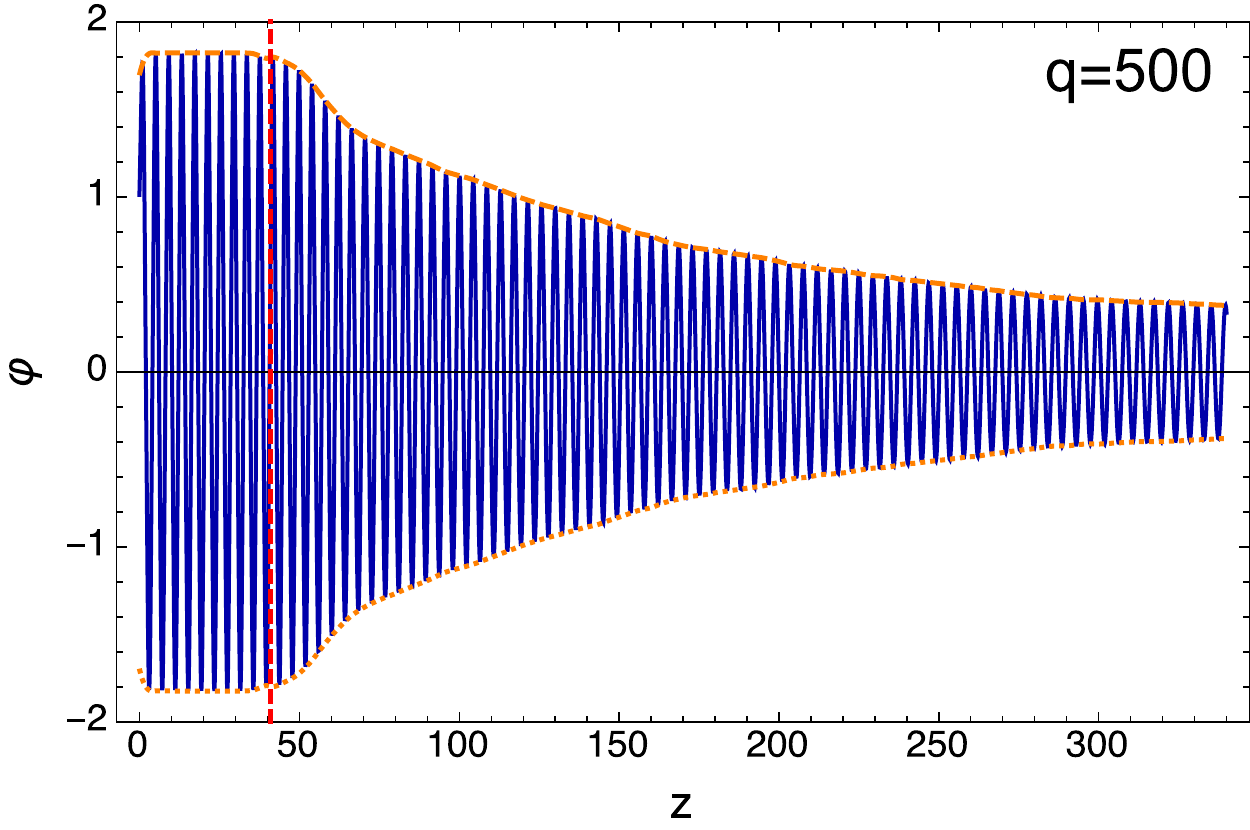}
      
      \end{center} \caption{We show the initial oscillations of the volume-averaged conformal amplitude of the inflaton field $\varphi$. We show the cases $q=3$, $q=8$, $q = 105$, and $q = 500$ for the preheating scenario with quartic potential. We use notation of Eq.~(\ref{eq:lphi4-variables}). The dashed vertical red line indicates the time $z_{\rm br}$, when backreaction of the daughter fields become relevant, triggering the decay of the inflaton amplitude and energy density (see also Fig.~\ref{fig:lame-energy}).} 
            \label{fig:lphi4-init}
 \end{figure}
        
Let us move now into the results from the lattice simulations. In Fig.~\ref{fig:lphi4-init} we plot the conformal amplitude of the inflaton field for the resonance parameters $q = 3, 8, 105$ and $500$. It is clearly appreciated that during a certain number of oscillations, the conformal amplitude of the inflaton $\varphi$ remains just constant, like if it was not coupled to the daughter field(s). However, there is a time (which differs for the different $q$'s) when the amplitude of the conformal inflaton starts decreasing significantly. This is the initial moment when the inflaton starts decaying due to the backreaction from the daughter fields. We shall refer to that time as $z_{\rm br}$ (the {\rm br} subindex meaning \emph{backreaction})\footnote{Let us note that our definition of backreaction differs from the standard condition labeled as 'backreaction' in the seminal paper \cite{Kofman:1997yn}, which corresponds to the moment when $g^2\left\langle\chi^2\right\rangle$ becomes equal to the effective inflaton mass. The latter is a condition that determines the onset of the modulation of the inflaton's frequency of oscillation. However, we prefer to define the moment of backreaction as the onset of the decay of the (conformal) amplitude of the inflaton, because it is then when the presence of the excited field $\chi$ becomes truly noticeable, and hence the inflaton energy start decreasing significantly.}. During the time $0 \leq z \lesssim z_{\rm br}$, the daugther fields have been experiencing parametric resonance, so their energy density has been growing exponentially from initially small quantum fluctuations\footnote{See Appendix~\ref{appen:Lattice} for a discussion about the introduction of initial field fluctuations in the lattice.}. As the energy flows from the mother field into the daughter fields, at $z \simeq z_{\rm br}$ the amount of energy transferred onto the $\chi$ bosons is not anymore a negligible fraction of energy stored in the mother field. Therefore, from then onwards, the (conformal) inflaton amplitude starts to decrease noticeable, see Fig.~\ref{fig:lphi4-init}. The time $z_{\rm br}$ corresponds, in order words, to the onset of the inflaton decay, when the backreaction effects from excited daughter fields become non-negligible. In practice, we have determined $z_{\rm br}$ as the moment when the (conformal) energy of the mother field drops $\sim 5\%$ with respect its initial amplitude.

                        \begin{figure}
            \begin{center}
                \includegraphics[width=11cm]{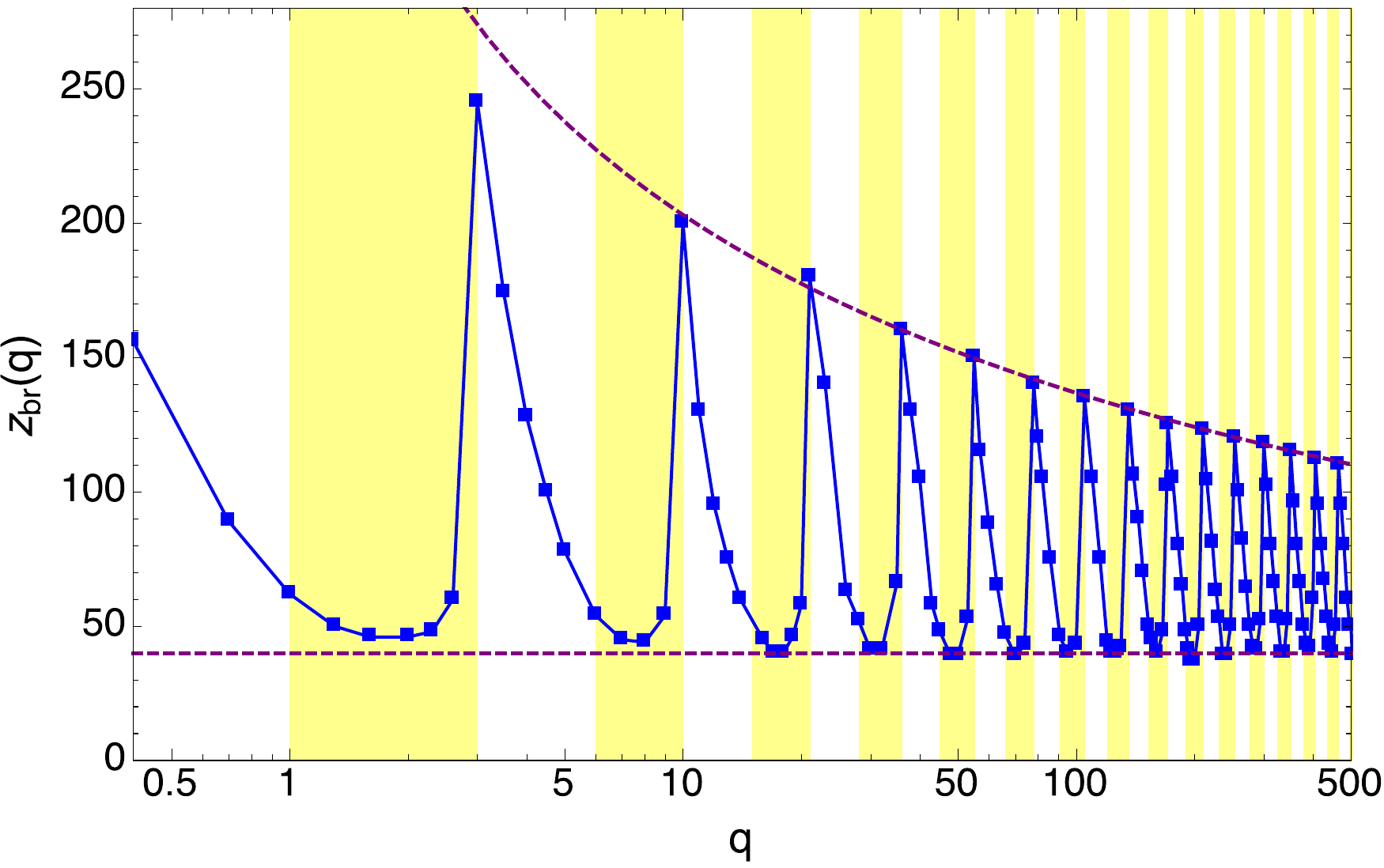}
            \end{center}
                \caption{We depict $z_{\rm br}$ as a function of $q$ for the range $0.4<q<500$. Each point corresponds to the value obtained directly from a lattice simulation, and we have joined the different points with straight lines. Yellow vertical bands indicate the position of the resonance bands of the Lam\'e equation $q \in (1,3), (6,10) \dots$. The dashed, purple, lower line indicate the estimate $z_{\rm br} (q) \approx 40$ [Eq.~(\ref{eq:lphi4-zbtime})] for $q$ values within resonance bands, while the upper one indicates the fit Eq.~(\ref{eq:lphi4-zbtime2}) for the relative maxima.} 
            \label{fig:lame-zbtime}
        \end{figure}

In Fig.~\ref{fig:lame-zbtime} we have plotted the different $z_{\rm br}$'s obtained from our simulations, for several resonance parameters in the range $0.4 < q < 500$. We observe that $z_{\rm br}(q)$ follows a clear oscillatory pattern, in clear correspondence with the particular structure of resonance bands shown in Fig.~\ref{fig:lame-bands}. In general, the wider the resonance band in the Lam\'e equation for a given $q$, the shorter $z_{\rm br}$ is. For those values of $q$ emplaced within resonance bands, we find in fact an almost constant value
\be z_{\rm br} (q) \sim 40 \ , \hspace{0.3cm} q \in  (1,3), (6,10) \dots \label{eq:lphi4-zbtime}\ee
On the other hand, the behavior of $z_{\rm br}$ for $q$ values outside the resonance bands, i.e.~for $q \in [3,6], [10,15], \dots$, is quite different. For $q$ values that are in the left extreme of these intervals, i.e. $q \simeq 3, 10, ...$, $z_{\rm br}$ takes its maximum value, as this corresponds to the right end of a resonance band at $\kappa = 0$, see Fig.~\ref{fig:lame-bands}. We provide the following phenomenological fit to these relative maxima (excluding the particular case $q=3$), which we also plot in the Figure,
\be z_{\rm br} (q) \approx 552 e^{-| \log_{10} q |^{0.48}} \ , \hspace{0.3cm} q = 10, 21, 36 \dots \label{eq:lphi4-zbtime2}\ee
As $q$ increases inside one of the intervals outside the resonance bands, $z_{\rm br}$ decreases until hitting $z_{\rm br} (q) \sim 40$ at the center (more or less) of the nearest resonance band, see Fig.~\ref{fig:lame-zbtime}. In conclusion, we observe a direct translation of the resonance structure of Fig.~\ref{fig:lame-bands} into the lattice simulations. This happens because for $z \lesssim z_{\rm br}$, the backreaction effects of $\chi$ onto the $\varphi$ is negligible, and hence the Lam\'e equation~(\ref{eq:modeEQ}) is really at work. 

Let us compare now this result with the analytical calculation from Sect.~\ref{sec:AnalyticalParamRes}. There, using the linear regime, we derived the time scale $z_{\rm eff}$ in Eq.~(\ref{eq:EffEnergyTransferTimeScale}), and identified it with the decay time of the mother field. However, we see now that this identification is misleading, as $z_{\rm eff}$ rather corresponds to a rough indication of the time scale when the transfer of energy from the mother field to its decay products becomes significant. In other words, it corresponds to the onset of backreaction, which as explained, it only determines the initial moment when the inflaton starts decaying, see Fig.~\ref{fig:lphi4-init}. For the range of $q$ values shown in Fig.~\ref{fig:lame-zbtime}, $z_{\rm eff} \sim 78$, so the analytical prediction only overestimates in a factor $\sim 2$ the actual number $z_{\rm br} \sim 40$, found in the simulations at the onset of backreaction. Failing in a factor $\sim 2$ is not surprising, as the estimation of $z_{\rm eff}$ in Eq.~(\ref{eq:EffEnergyTransferTimeScale}) involved in fact many approximations. However, the relevant observation to make here is not that $z_{\rm eff}$ can be considered as an order of magnitude estimation of $z_{\rm br}$. Rather, the relevant point, is than $z_{\rm eff}$ should not be identified with a decay time, as it rather signals the moment $z_{\rm br}$ of backreaction, when the linear approximation breaks down. The time scale for determining the end of the transfer of energy from the mother field into the decay products, which we shall identified as the truly 'decay time' scale of the inflaton, will be referred to as $z_{\rm dec}$. As we will explain shortly, it corresponds in fact to a much longer time scale, $z_{\rm dec} \gg z_{\rm eff}, z_{\rm br}$, which cannot be estimated analytically, as the dynamics at {\small$z \gtrsim z_{\rm br}$} become non-linear.

To follow the post-inflationary dynamics in the non-linear regime, it is useful to see how the different contributions to the total energy of the system evolve as a function of time. The total energy can be written as a sum of its different contributions as
\be E \equiv \frac{\lambda \phi_*^4 }{a^4} E_t \equiv \frac{\lambda \phi_*^4 }{a^4} \left( E_{K,\varphi} + E_{K,\chi} +  E_{G,\varphi} + E_{G,\chi} + E_{\rm int} + E_{\rm V} \right) \ , \label{eq:lphi4-energy}\ee
with
\be E_{K,f} = \frac{1}{2} \left(  f' - f \frac{a'}{a} \right)^2 \ , \hspace{0.5cm} E_{G,f} = \frac{1}{2} | \nabla f |^2 \ , \hspace{0.5cm} E_{\rm int} = \frac{1}{2} q \varphi^2 \chi^2  \ , \hspace{0.5cm} E_V = \frac{1}{4} \varphi^4 \ , \ee
where $E_{K,f}$ and $E_{G,f}$ are the kinetic and gradient energy of the fields $\phi, \chi$, and $E_{\rm int}$ and $E_{V}$ are the interaction and potential energies, all written in terms of the natural variables of Eq.~(\ref{eq:lphi4-variables}) (i.e.~in terms of the field variables $f = \varphi, \chi$ and derivatives of these with respect $z^\mu$).

 \begin{figure}
      \begin{center}
                  \includegraphics[width=7.7cm]{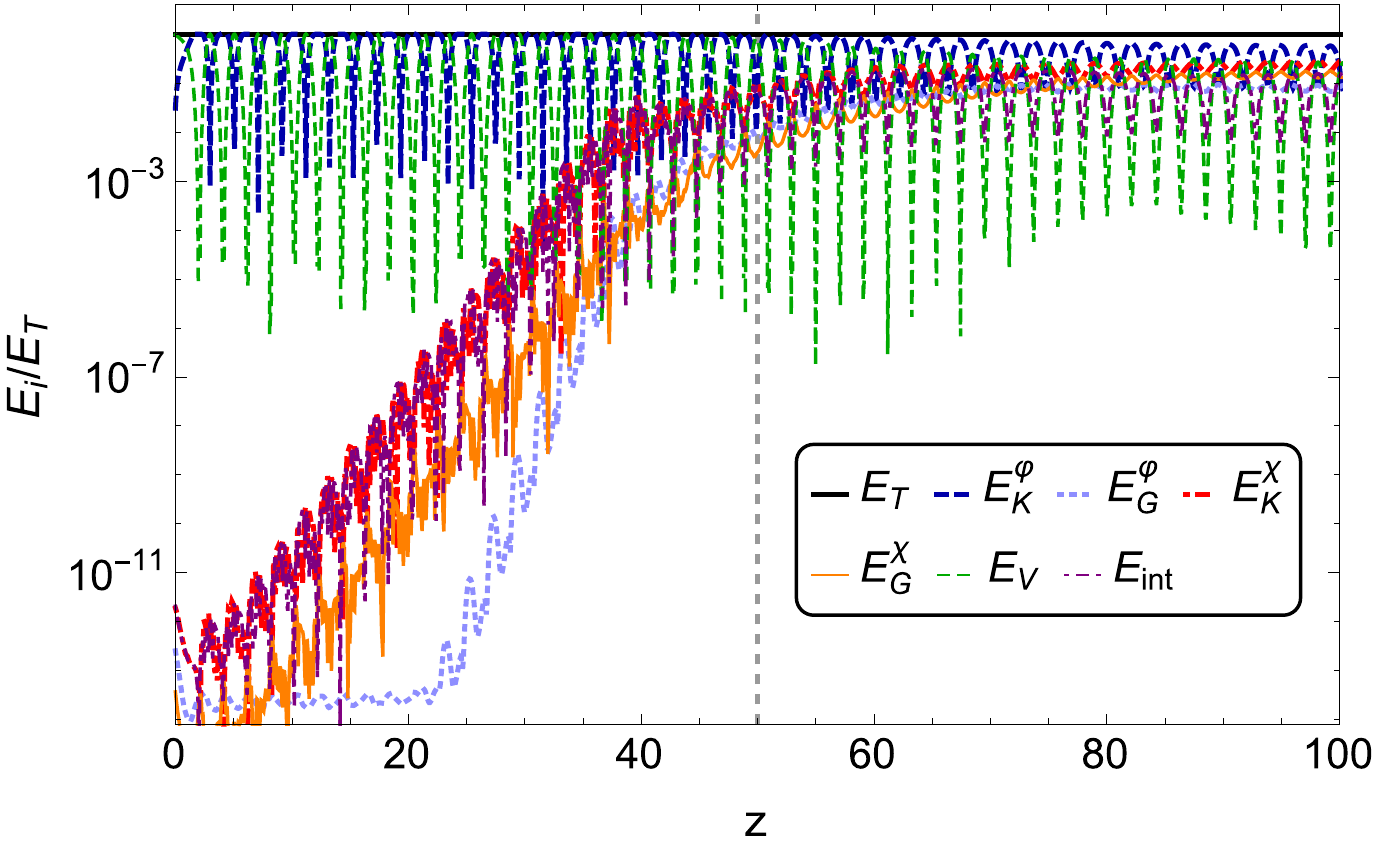}                   
                 \includegraphics[width=7.48cm]{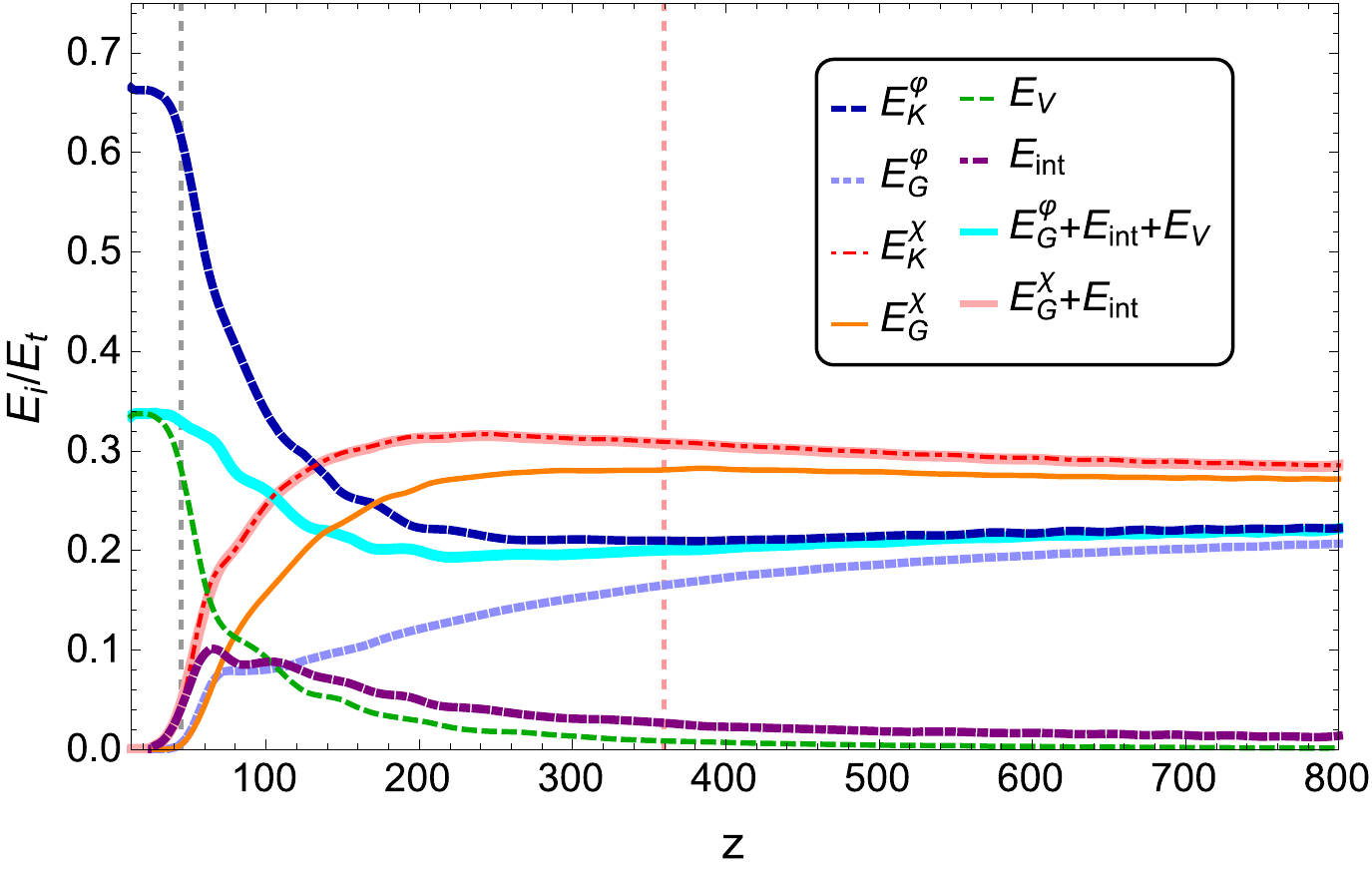}
     \end{center}
                \caption{Evolution of the different energy components of the system as a function of time, see Eq.~(\ref{eq:lphi4-energy}), for the inflationary scenario $V(\phi) \propto \phi^4$, where $q = 500$. Left: We plot $E_i / E_t$  for the initial stages of the inflaton decay, and we have indicated $z_{\rm br}$ with a vertical dashed red line. Right: We plot the same case for later times. To see better how the equipartition regime holds, we have removed the oscillations by taking the oscillation average of the different functions. We have added two new lines that indicate the sums $E_{G,\varphi} + E_{\rm int}  + E_V$ and $E_{G, \chi} + E_{\rm int}$, see Eq.~(\ref{eq:lame-equip}).}
            \label{fig:lame-energy}
 \end{figure}

In the left panel of Fig.~\ref{fig:lame-energy} we show the evolution of the volume-averaged amplitude of the different energy components of the system. There we can clearly observe how, at first, the inflaton energy dominates the energy budget of the system, alternating between kinetic and potential energies as the oscillations go on. Short after the onset of the simulation, the rest of energies start growing (including the inflaton gradient energy, which indicates the formation of inhomogeneities), becoming very soon an important part of the total energy.  At time $z_{\rm br}$, these energies have grown enough so that they start backreacting onto the inflaton condensate, inducing its decay (i.e. the decrease of the inflaton kinetic and potential energies). This can also be appreciated in Fig.~\ref{fig:lphi4-init}, where from $z \gtrsim z_{\rm br}$ the (conformal) inflaton amplitude starts decreasing significantly.

Let us note that, although the energy fractions at $z \simeq z_{\rm br}$ show some scattered dependence on $q$, in reality they are quite independent of the resonance parameter. From the numerical outcome we find
\begin{center}
Energy Fractions at $z_{\rm br}$:\vspace*{-2mm}
\end{center}
\begin{eqnarray}\label{eq:EnergiesPhi4atZi}
{E_{K,\varphi}\over E_t} \simeq (62.5 \pm 2.4) \%\,,\hspace*{0.2cm}{E_{V}\over E_t} \simeq (29.0 \pm 2.7) \%\,,\hspace*{0.2cm} {E_{K,\chi}\over E_t} \simeq (4.1 \pm 2.5) \%\,,\hspace*{0.2cm}{E_{\rm int}\over E_t} \simeq (3.6 \pm 2.2) \% \nonumber\\
\end{eqnarray}
with the errors $\pm\,\Delta E_{x}/E_t$, simply reflecting the scattering of energies with $q$. We see from this that at $z = z_{\rm br}$, most of the energy remains yet in the inflaton. However, we also learn that only when $\sim 1\%$ of the total energy is already transferred into the daughter field(s), does backreaction really becomes noticeable, making the inflaton amplitude to initiate its decay. The other energy components $E_{G,\varphi},E_{G,\chi}$ remain always at sub-percentage levels during $0 < z \lesssim z_{\rm br}$, independently of $q$.

At times $z \gtrsim z_{\rm br}$, the energy components evolve substantially from the given values in Eq.~(\ref{eq:EnergiesPhi4atZi}). The energies evolve towards an 'equiparted' distribution among components, until the system eventually reaches a stationary regime, where the energy components do not change appreciably. This is observed in the bottom panel of Fig.~\ref{fig:lame-energy}, where we have removed the oscillations by taking the oscillation average of the different energies. We observe different equipartition identities for the $\varphi$ and $\chi$ fields respectively,
\be E_{K,\varphi} \simeq E_{G,\varphi} + E_{\rm int} + E_V \ , \hspace{0.4cm}  E_{K,\chi} \simeq E_{G, \chi} + E_{\rm int} \ .  \label{eq:lame-equip} \ee
As it can be appreciated in Fig.~\ref{fig:lame-energy}, the second identity holds almost exactly for all times, while the first one only holds for late times (though it is not a bad approximation at earlier times). 

From the analysis of the energies we see that a new time scale, much longer than $z_{\rm br}$, can be naturally identified with the decay time of the mother field. This scale can be defined by how long it takes the system to relax from $z \gtrsim z_{\rm br}$ into the stationary regime. We shall call the moment when the stationary regime is onset as $z_{\rm dec}$. It is this time, and not $z_{\rm br}$, that signals the true end of the inflaton decay, because it is at $z \gtrsim z_{\rm dec}$ that there is no (appreciable) transfer of energy anymore from the inflaton into the daughter field(s). Although the exact definition of $z_{\rm dec}$ is more arbitrary than $z_{\rm br}$, we find appropriate to provide an operative definition based on the level of accuracy of equipartition. In particular, at the moment when the inflaton equipartition energy holds at a better level than $2 \%$, i.e. $(E_{K,\varphi} - E_{G, \varphi} - E_{\rm int } - E_V)/(E_{K,\varphi} + E_{G, \varphi} + E_{\rm int } + E_V) \gtrsim 0.02$, the inflaton kinetic and gradient energies are stabilized and do not evolve appreciably further, see Fig.~\ref{fig:lame-energy}. The stabilization of the inflaton energy components when equipartition is set to a $2 \%$ level is in fact independent of $q$. This is very relevant, as this makes $z_{\rm dec}$ defined in this way, a good indicator of the decay time of the mother field.

\begin{figure}
      \begin{center}
                  \includegraphics[width=11cm]{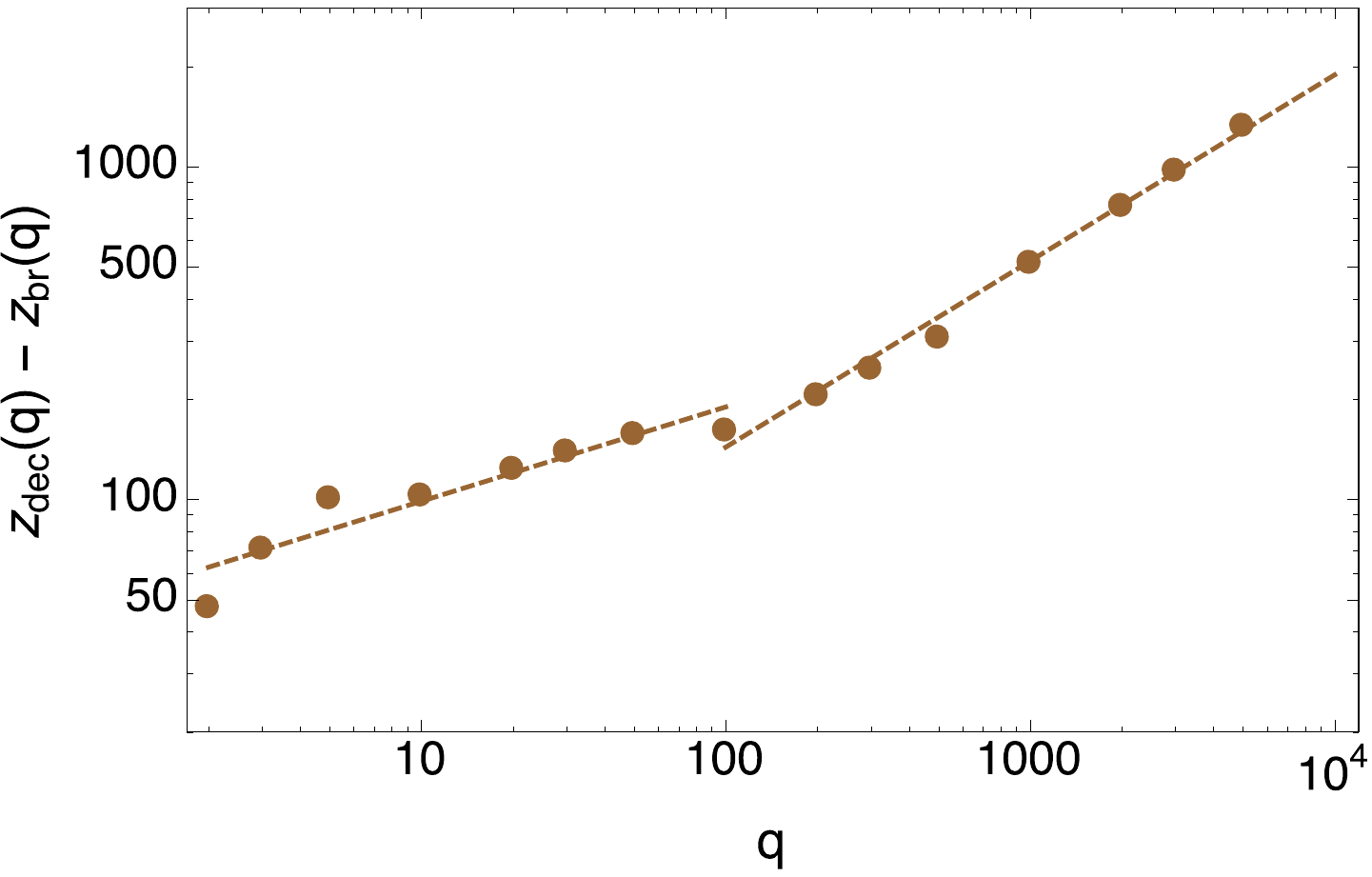}
     \end{center}
                \caption{Points show the different $z_{\rm dec}$ obtained for different lattice simulations with different values of $q$, for preheating with quartic potential. The dashed line indicates the best fit (\ref{eq:zd-decay}).} 
            \label{fig:lphi4-zdecay}
 \end{figure}

The relevant property of $z_{\rm dec}$ is that it grows with the resonance parameter $q$, following a simple power-law fit. We show in Fig.~\ref{fig:lphi4-zdecay} the value of $z_{\rm dec} - z_{\rm br}$ as a function of $q$, as extracted from our lattice simulations with different $q$'s. We obtain the following fit
\bea z_{\rm dec} (q) - z_{\rm br} (q) =
  \left\lbrace
  \begin{array}{l}
     51 q^{0.28} \hspace{0.3cm}\text{ if } q < 100 \vspace*{2mm}\\
     11 q^{0.56} \hspace{0.3cm}\text{ if } q \geq 100 \\
  \end{array}
  \right. \label{eq:zd-decay}
\eea
which we also show in Fig.~\ref{fig:lphi4-zdecay}. Note that for $q \lesssim 100$, the scales $z_{\rm br}$ and $z_{\rm dec}$ are not particularly separated, with $|z_{\rm dec}-z_{\rm br}| \lesssim z_{\rm br}$. This explains why these point must be fitted with a different power law. Note that the inflaton decay takes longer the greater the resonance parameter (i.e.~the larger the mother-daughter coupling), which is in principle counter-intuitive. Following the standard logic of the linear calculation, the larger the $q$ the shorter the decay time should be. However, once we have learned that $z_{\rm eff}$ ought not identified with the decay time, but with the onset of back-reaction $z_{\rm br}$, then the linear logic does not prevail anymore. The reason as to why the truly decay time $z_{\rm dec}$ follows the opposite trend, increasing with $q$, lies on the fact that for $z > z_{\rm br}$ the system has become non-linear. Although $a~priori$ one would tend to think that the stronger the coupling the faster the stationary regime should be achieved, our lattice simulations -- fully capturing the non-linear dynamics -- clearly prove the opposite. This was in fact, also noticed already in~\cite{Figueroa:2015hda}.

As mentioned, we can only obtain our fits for resonance parameters up to $q \sim 10^{4}$ due to the limitations of the lattice approach. However there is nothing specially different in the physics of parametric resonance for $q \gg 10^4$. Therefore, there is no impediment, in principle, to extrapolate the scaling law Eq.~(\ref{eq:zd-decay}) to higher $q$'s.

Let us note that the energy fractions at $z \gtrsim z_{\rm dec}$ do not change appreciably any more in our simulations. Some small change should be expected nonetheless, as the system approaches equilibrium. However this is not captured in our simulations. The energy from the end of the inflaton decay onwards are actually rather independent of $q$, given by the fractions
\begin{center}
Energy Fractions at $z \gtrsim z_{\rm dec}$:\vspace*{-2mm}
\begin{eqnarray}\label{eq:EnergiesPhi4atZe}
\begin{array}{c}
{E_{K,\chi}\over E_t} \simeq (29.5 \pm 3.3) \%\,,\hspace*{0.2cm}{E_{K,\varphi}\over E_t} \simeq (22.6 \pm 3.4) \%\,,\hspace*{0.2cm} {E_{G,\chi}\over E_t} \simeq (26.2 \pm 3.4) \%\,,\hspace*{0.2cm}\vspace*{3mm}\\
{E_{G,\varphi}\over E_t} \simeq (17.7 \pm 3.0) \%\,,\hspace*{0.2cm} {E_{\rm int}\over E_t} \simeq (3.2 \pm 0.7) \%\,,\hspace*{0.2cm}{E_{V}\over E_t} \simeq (0.8 \pm 0.2) \% 
\end{array}
\end{eqnarray}
\end{center}
again with the errors $\pm\,\Delta E_{j}/E_t$ reflecting some (rather random) scattering of the energies with $q$. We see from this that at $z \gtrsim z_{\rm dec}$, the energy is almost 'democratically' split between the mother and the daughter field(s), though with some more energy stored in the latter, with $E_{\chi}/E_t \simeq (E_{G, \chi} + E_{K, \chi})/E_t \sim (54.7 \pm 4.7) \%$, $E_{\varphi}/E_t \equiv (E_{K,\varphi}+E_{G,\varphi}+E_V)/E_t \simeq (41.1 \pm 4.5) \%$, and $E_{\rm int}/E_t \simeq (4.3 \pm 0.5)\%$. At these moments it is also verified the approximate equipartion $E_{K,\varphi}/E_t \simeq (E_{G,\varphi} + E_{\rm int} + E_V)/E_t \sim 21\%-23\%$ and $E_{K,\chi}/E_t \simeq (E_{G, \chi} + E_{\rm int})/E_t \sim 29\%-30\%$. 

\subsection{Lattice Simulations of preheating with quadratic potential}\label{sec:m2phi2}

Let us now consider preheating after chaotic inflation with an inflaton quadratic potential
\be V (\phi) = \frac{1}{2} m^2 \phi^2 \ . \ee
In this case, we define the onset of the oscillatory regime when the condition $H_* = m$ holds, which we take as the initial time of our lattice simulations. From a numerical calculation using the homogeneous Klein-Gordon and Friedman equations, $\ddot\phi + 3(\dot a/a)\dot\phi + {dV\over d\phi} = 0$, $3m_p^2(\dot a/a)^2 = \lbrace V_{\rm inf} (\phi)+(\dot\phi)^2/2 \rbrace$, we find $\phi_* \sim 2.32 m_p$ and $\dot{\phi}_* \sim -0.78 m m_p$. Let us define again a set of 'natural' variables as   
\be \varphi = \frac{1}{\phi_*}a^{3/2} \phi \ , \hspace{0.5cm} \chi = \frac{1}{\phi_*} a^{3/2} X \ , \hspace{0.5cm} 
z \equiv m t \ , \hspace{0.5cm} \vec{z} \equiv m \vec{x}\,, \label{eq:m2phi2-variables}\ee
where $x^{\mu} \equiv (t, \vec{x})$ are the old cosmic time and comoving coordinates. As before, we indicate differentiation with respect cosmic/natural time with a dot/prima respectively, $\dot{} \equiv d / dt$ and $' \equiv d / dz$. Spatial derivatives should be understood as taken with respect natural variables, and corresponding momenta will be referred as $\kappa \equiv k/m$. The fields' EOM in these variables are 
\be \varphi '' - \left( \frac{3}{4} \frac{a'^2}{a^2} + \frac{3}{2} \frac{a''}{a} \right) \varphi - \frac{1}{a^2} \nabla^2 \varphi + \left( 1 +  \frac{4}{a^{3}} q_*\chi^2 \right)  \varphi = 0 \ , \label{eq:m2phi2-eom}\ee
\be \chi'' - \left( \frac{3}{4} \frac{a'^2}{a^2} + \frac{3}{2} \frac{a''}{a} \right) \chi - \frac{1}{a^2} \nabla^2 \chi + \frac{4}{a^3} q_* \varphi^2 \chi = 0 \ . \label{eq:m2phi2-eom2}\ee
where the resonance parameter is defined this time as 
\begin{eqnarray}
q_* = \frac{g^2 \phi_*^2}{4 m^2} \ . \label{eq:m2phi2-resp}
\end{eqnarray}
We take $m = 6 \times 10^{-6} m_p$, as this is fixed by the observed amplitude of CMB anisotropies~\cite{Tsujikawa:2013ila}. 

Let us focus first on the case of a non-expanding universe, so we set $a = 1$ and $a' = a'' = 0$ in the equations above. As before, during some time, the $\chi$ particles are very sub-dominant with respect to the inflaton condensate, and hence the effect of their backreaction onto the inflaton can be neglected. During this regime, the mode equation of the daughter fields $\chi_k$, corresponds to the so called {\it Mathieu} equation~\cite{Kofman:1997yn}, which similarly to the set of {\it Lam\'e} equation, is characterized by a well-known structure of resonance bands. More specifically, for some regions in the $(q_*,\kappa)$ plane (with $\kappa = k / H_*$), there is a solution of the type $\chi_{\kappa} \sim e^{\mu_{\kappa} z}$  with $\mathfrak{Re}[ \mu_{\kappa}] >0$. One can distinguish two different regimes in the preheating process, depending on the particular value of $q_*$. If $q_* < 1$, the narrow resonance regime holds. In this case, the size of the resonance bands is so small that they cannot be well captured in the lattice. On the other hand, if $q_* \gg 1$, the system is in a broad resonance regime, and the bands are large enough so that lattice simulations can be applied in this case. 

When the expansion of the universe is introduced, the scale factor affects the EOM of $\varphi$ in a non-trivial way: even if the system starts in broad resonance with $q_* > 1$, as the Universe expands, the system rapidly redshifts towards neighboring bands of lower resonance parameter. This is due to the term $q_* a^{-3}$ in Eq.~(\ref{eq:m2phi2-eom2}), which makes the effective resonance parameter $q \propto 1/a^3$ to decrease as time goes by. The system does not remain therefore in a single resonance band, but redshifts due to the expansion of the universe. As a consequence, even if the system starts in a broad resonance regime, it can only be maintained as such for some finite time, until it ends up in a narrow resonance regime. For a detailed analysis of the behavior of the mode functions obeying the Mathieu equation both in Minkowski and in an expanding Universe, we recommend to read the seminal work~\cite{Kofman:1997yn}. In our present work we will just focus mostly, from now on, on the outcome from lattice simulations.

Let us note that, in principle, the coupling $g$ can be arbitrarily small, so that we could be in the regime of narrow resonance from the very beginning of the oscillations. As we cannot simulate in the lattice narrow resonance, we certainly want to avoid such cases. Furthermore, even if we start in broad resonance with $q_* \gg 1$, we need $q_*$ to be sufficiently large, so that $q= q_*/a^3$ does not turn smaller than unity before the backreaction effects from the daughter field(s) are noticed. Taking into account that the scale factor behaves as $a(z) \sim z^{2/3}$ in this scenario, a transition of broad-to-narrow resonance takes place whenever $4 q_* a^{-3} \approx 4 q_* z^{-2} = 1$, i.e.~in a time $z_{r} \approx 2 \sqrt{q_*}$ from the start of the simulation. Therefore, we want this time to be larger than the back-reaction time $z_{\rm br}$. In practice, we cannot simulate cases for $q_* < 5 \times 10^3$, because for these $z_r \lesssim z_{\rm br}$, and hence we would enter into narrow resonance before backreaction matters. We have simulated cases in the interval $q_* \in [7.5 \times 10^3 , 2.5 \times 10^6 ]$. Let us notice that the upper bound on the coupling $g$ to prevent radiative corrections, $g < 10^{-3}$, corresponds to $q_* \approx 3.7 \times 10^{4}$. Of course, in supersymmetric theories, radiative corrections from bosons and fermions tend to cancel each other. In such theories the coupling constant $g$ can be in principle much greater than $10^{-3}$. As in this work we want to be as generic as possible, we will allow ourselves to consider higher couplings. However we will only reach up to $g \leq 2\cdot(2.5 \cdot 10^6)^{1/2}(m/\phi_*) \simeq 6.875\cdot 10^{-3}$, as this corresponds to the largest resonance parameter $q_*$ we are capable of simulating. See Appendix~\ref{appen:Lattice} for an extended discussion about this.

\begin{figure}
      \begin{center} \includegraphics[width=10cm]{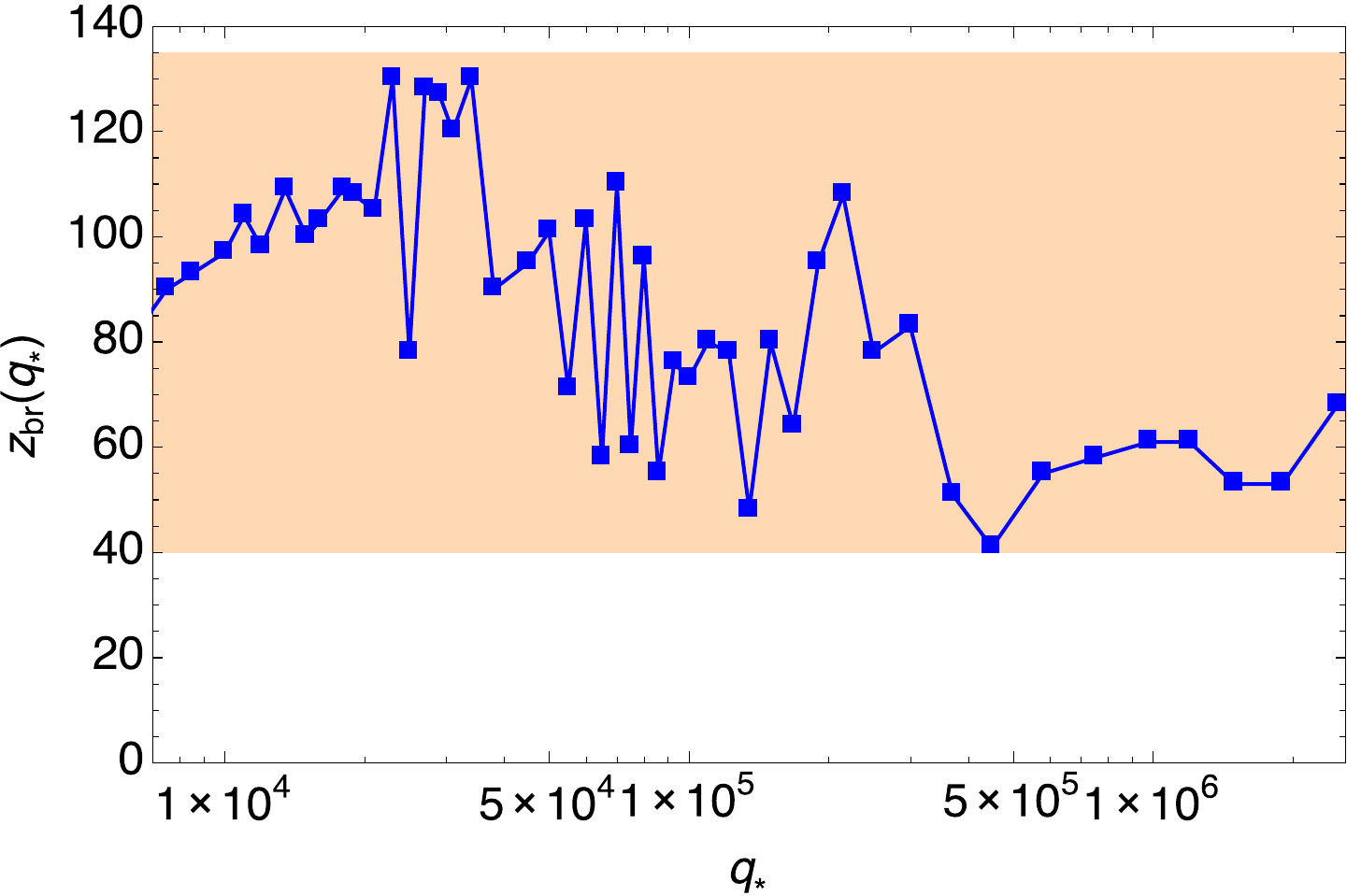}\end{center} 
      \caption{We plot the different times $z_{\rm br}$ obtained from the lattice simulations of the $m^2 \phi^2$ inflationary model with different resonance parameters. We have joined the points with a straight line, and the orange band corresponds to the values of Eq.~(\ref{eq:m2phi2-zbtime}). } 
            \label{fig:m2phi2-zbTime}
 \end{figure}

\subsubsection{Onset of non-linearities, energy evolution and decay time}

In Fig.~\ref{fig:m2phi2-zbTime} we show the backreaction time $z_{\rm br}$, obtained from our lattice simulations. We define $z_{\rm br}$ again as the moment when the inflaton conformal amplitude $\varphi$ starts decreasing abruptly, due to the back-reaction of the excited $\chi$ fields. We show $z_{\rm br}$ as a function of the resonance parameter $q_*$. For all simulations, we see that
\be z_{\rm br} \in [40, 135] \ . \label{eq:m2phi2-zbtime}\ee
We do not observe a clear pattern for $z_{\rm br}$ as a function of $q_*$, as we saw in the $\lambda \varphi^4$ case. This is however expected. The reason is that, in the present case, we cannot differentiate whether a mode is placed in the middle of a resonance band or not. Now each mode experiences a rapid scanning of bands due to the expansion of the Universe. Actually, as described in~\cite{Kofman:1997yn}, the resonance in this system is stochastic, precisely due to the scanning over the resonance bands. In~\cite{Kofman:1997yn}, it was well appreciated that when solving the Mathieu equation for different modes $\lbrace \kappa_i \rbrace$, for the same initial resonance parameter $q_*$, the Floquet index $\mu_k$ oscillates constantly around zero as we go surveying the various modes\footnote{The Floquet index alternates between positive and negative values inside a certain envelope curve. The specific form of this envelope is however irrelevant for us now, so we just refer to the interested reader to check Fig.~10 and Eq.~(81) of~\cite{Kofman:1997yn}.}. While for a given mode $\kappa_1$ the Floquet index can be positive $\mu_{\kappa_1} > 0$, at a neighboring mode $\kappa_2$ it may become negative, $\mu_{\kappa_2} < 0$, even if $|\kappa_2-\kappa_1| \ll \kappa_{1,2}$. The occurrence of positive and negative $\mu_\kappa$'s is of course not symmetric, but in a proportion 3:1, so that overall there is always a net effect of particle creation~\cite{Kofman:1997yn}. The excitation of a given mode $\kappa$ goes receiving alternating positive and negative 'kicks' in a proportion 3:1, so that in some moments $X_k$ grows, and in others it decreases, but in the overall there is always a net growth. The 'wiggly' pattern of $z_{\rm br}$ as a function of $q_*$ is, therefore, just a reflection of the stochastic nature of the resonance in this system. To our knowledge, the pattern depicted in Fig.~\ref{fig:m2phi2-zbTime}, has never been shown before. Due to the stochastic nature of the resonance, one cannot predict exactly $z_{\rm br}$ for a specific initial resonance parameter $q_*$. 

Looking at Fig.~\ref{fig:m2phi2-zbTime}, we appreciate that the onset of the backreaction, and hence the start of the inflaton decay, happens always in a time $z_{\rm br} \sim {\rm few}\times\mathcal{O}(10)$. Similarly to the analytical calculation presented in Sect.~\ref{sec:AnalyticalParamRes} for the quartic case, one can also derive an estimation, based on the linear regime, of the time it takes for an efficient transfer of energy into the daughter field(s), for the quadratic case\footnote{Given the stochastic nature of the resonance, this calculation is perhaps less transparent, but it is expected to capture well, in principle, the order of magnitude value.}. As such computation was presented in~\cite{Kofman:1997yn}, we just quoted their result (adapted to our notation) in our Eq.~(\ref{eq:EffEnergyTransferTimeScaleApproxPhi2}). Taking $\bar\mu \simeq 0.15$ as a reasonable averaged value of the stochastic Floquet index $\mu_\kappa$, then $z_{\rm eff} \simeq 8.3(15.1-1.1\log_{10}q_*)$. For $q_* \sim 10^4-10^6$, then $68 \lesssim z_{\rm eff} \lesssim 86$. As in the quartic case, we see that $z_{\rm eff}$ is a good estimation of the back-reaction time $z_{\rm br}$ (ignoring of course the stochastic pattern seen in Fig.~\ref{fig:m2phi2-zbTime}). It is not, however, a good estimation of the decay time $z_{\rm dec}$ of the inflaton, which we estimate next. 
 
We can understand better the post-inflationary dynamics at $z \gg z_{\rm br}$ if we analyze again how the different energy contributions evolve as a function of time. The total energy can be written as a sum of its components as
\be \rho = \frac{m^2 \phi_*^2}{a^3} E_t = \frac{m^2 \phi_*^2}{a^3} \left( E_{K,\varphi} + E_{K,\chi} +  E_{G,\varphi} + E_{G,\chi} + E_{\rm int} + E_{\rm V} \right) \label{eq:energy-m2phi2} \ee
with
\be E_{K,f} = \frac{1}{2} \left(  f' - \frac{3}{2} \frac{a'}{a} f \right)^2 \ , \hspace{0.5cm} E_{G,f} = \frac{1}{2 a^2} | \nabla f |^2 \ , \hspace{0.5cm} E_{\rm int} = \frac{2 q_*}{a^3} \chi^2 \varphi^2 \ , \hspace{0.5cm} E_V = \frac{1}{2} \varphi^2 \ , \label{eq:energy-m2phi2B} \ee
where $E_{K,f}$ and $E_{G,f}$ are the kinetic and gradient energy of the fields $\phi, X$ ($f = \varphi,\chi$ labeling their conformal amplitude), and $E_{\rm int}$ and $E_{V}$ are the interaction and potential energies. 

\begin{figure}
      \includegraphics[width=7.5cm]{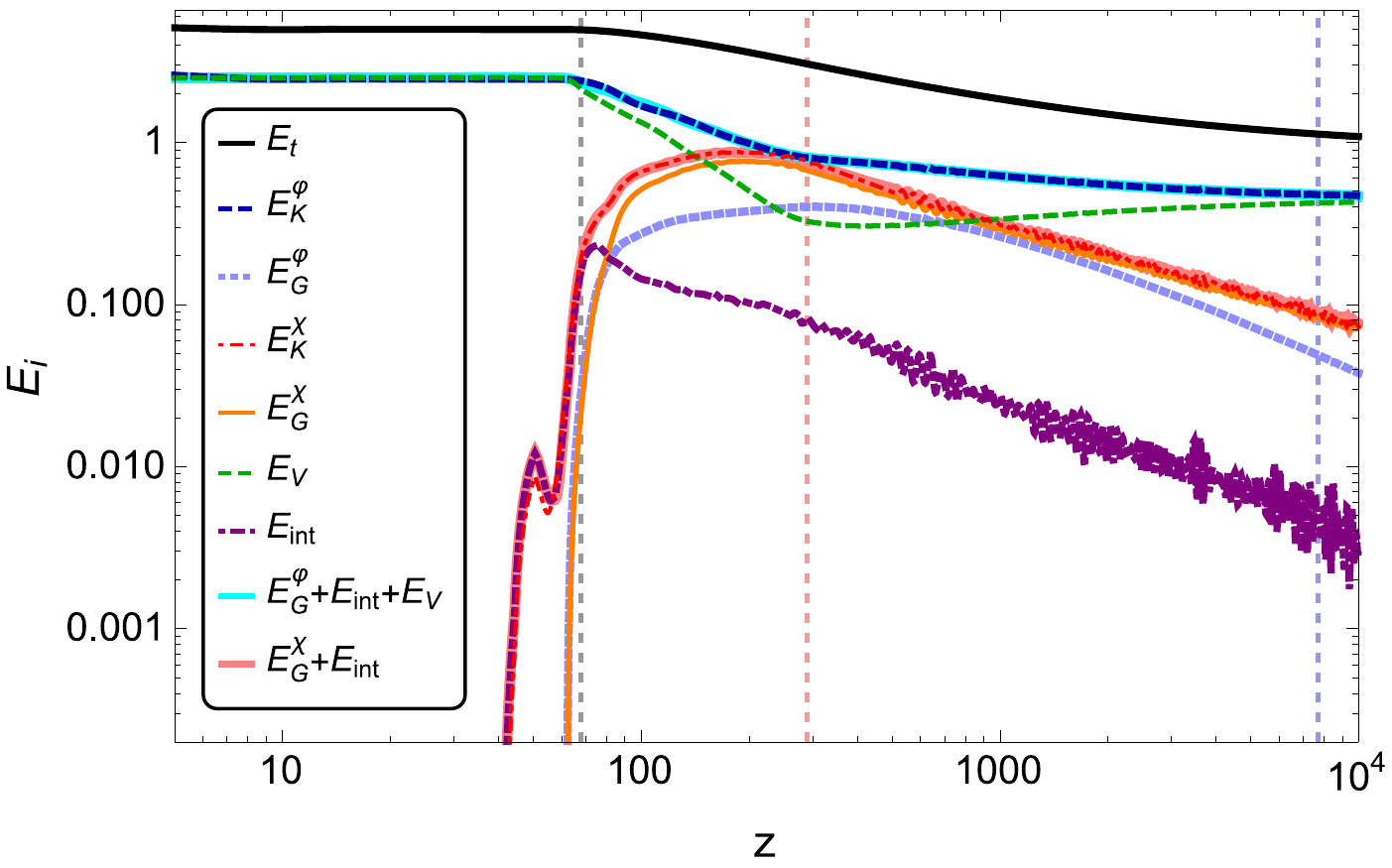}
      \includegraphics[width=7.5cm]{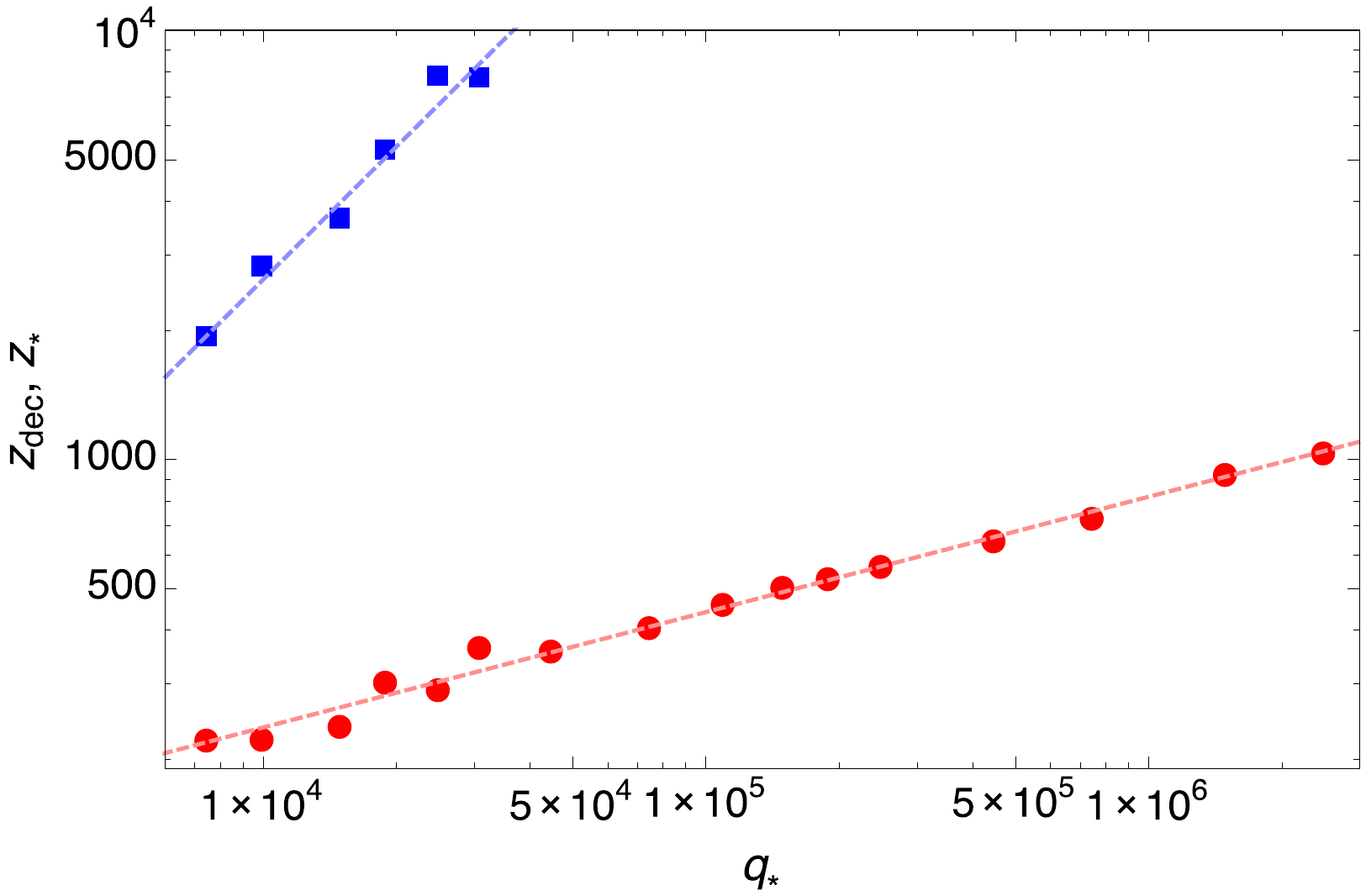}
      \caption{Left: We show for the quadratic preheating case and $q_*=25000$, the evolution of the different energy components of the system as a function of time, see Eq.~(\ref{eq:mathieu-equip}). We normalize them to the total energy at initial times, $E_i$. The gray, red, and blue vertical dashed lines indicate the times $z_{\rm br}$, $z_{\rm dec}$ and $z_{0.80}$. Right: We show the times $z_{\rm dec}$ (red circles) and $z_{0.80}$ (blue squares) as a function of $q_*$ obtained from lattice simulations.} 
            \label{fig:m2phi2-energy}
 \end{figure}

In Fig.~\ref{fig:m2phi2-energy} we show the evolution of the energy contributions as a function of time for a particular resonance parameter. We take, as before, the oscillation average of the different functions. One of the most interesting properties of this system is that the equipartition identities 
\be E_{K,\varphi} \simeq E_{G,\varphi} + E_{\rm int} + E_V \ , \hspace{0.4cm}  E_{K,\chi} \simeq E_{G, \chi} + E_{\rm int} \ ,  \label{eq:mathieu-equip} \ee
hold for all times. This can be observed in Fig.~\ref{fig:m2phi2-energy}. 

Let us begin by noting that, despite the spiky patter of $z_{\rm br}$ exhibited in Fig.~\ref{fig:m2phi2-zbTime}, the dominant energy fractions at $z \simeq z_{\rm br}$ show however, much less scattering with $q_*$ than in the case of $\phi^4$. The energy fractions are mostly independent of the resonance parameter, and are given by
\begin{center}
Energy Fractions at $z_{\rm br}$:\vspace*{-2mm}
\end{center}
\begin{eqnarray}\label{eq:EnergiesM2Phi2atZi}
{E_{K,\varphi}\over E_t} \simeq (49.4 \pm 0.1) \%\,,\hspace*{0.2cm}{E_{V}\over E_t} \simeq (48.7 \pm 0.6) \%\,,\hspace*{0.2cm} {E_{K,\chi}\over E_t} \simeq (0.9 \pm 0.3) \%\,,\hspace*{0.2cm}{E_{\rm int}\over E_t} \simeq (0.8 \pm 0.3) \% \ .  \nonumber\\
\end{eqnarray}
The errors $\pm\,\Delta E_{x}/E_t$ simply reflect the (random) scattering of energies with $q_*$. We see again that at $z = z_{\rm br}$, almost all of the energy remains yet in the inflaton. When the tresshold of $\sim 0.5\%$ of energy transferred is surpassed, backreaction then becomes noticeable, and the inflaton amplitude starts decaying. The other energy components, $E_{G,\varphi}/E_t$, $E_{G,\chi}/E_t$, remain always at less than $\sim 0.1\%$ levels during $0 < z \lesssim z_{\rm br}$, independently of $q_*$. 

We can define again a time scale $z_{\rm dec}$ characterizing the moment when the system enters into a stationary regime. As equipartition holds all the time, we cannot determine now a specific moment when equipartition is verified to better than a certain degree (as we did in the inflationary $\lambda \phi^4$ case). However, we can define $z_{\rm dec}$ at the onset of the stationary regime, understanding the latter now as the regime when the inflaton kinetic and potential energies do not evolve appreciably anymore within one inflaton oscillation period. In practice, we define $z_{\rm dec}$ at the moment when these energies do not change more than $\sim 0.5\%$ within one oscillation. This threshold is not as arbitrary as it seems: at the earlier times $z_{\rm br} \lesssim  z \lesssim z_{\rm dec}$, the relative change of the dominant energies not only is bigger than $\sim 0.5\%$, but also changes in time. However, at times $z \gtrsim z_{\rm dec}$, with $z_{\rm dec}$ defined as just said, the relative change simply remains always below the $\sim 0.5\%$ threshold. Let us note, however, that this does not mean that these energies do not evolve in time at $z \gtrsim z_{\rm dec}$. Actually they evolve smoothly, but the relative change (within an oscillation time scale) is simply very small. Extracting $z_{\rm dec}$ that way from our lattice simulations, we find the data to be very well fitted (see right panel of Fig.~\ref{fig:m2phi2-energy}) by,
\be z_{\rm dec} (q_*) \approx 19.9\,q_*^{0.27} \ . \label{eq:m2phi2-zd} \ee
Once again, we see that the larger the resonance parameter $q_*$, the longer it takes the flow of energy from the inflaton to the daughter fields to cease. At this time, the dominant energy components are actually rather independent of the resonance parameter for $q_* \gtrsim 5\cdot 10^4$. Their relative fractions are given by
\begin{center}
Dominant Energy Fractions at $z \gtrsim z_{\rm dec}$ ($q_* \gtrsim 5\cdot 10^4$):\vspace*{-0mm}
\begin{eqnarray}\label{eq:EnergiesM2phi2atZe}
\begin{array}{c}
{E_{K,\chi}\over E_t} \simeq (25.2 \pm 2.2) \%\,,\hspace*{0.2cm}{E_{K,\varphi}\over E_t} \simeq (26.0 \pm 2.3) \%\,,\hspace*{0.2cm} {E_{G,\chi}\over E_t} \simeq (22.9 \pm 2.5) \%\,,\hspace*{0.2cm}
\end{array}
\end{eqnarray}
\end{center}
again with the errors $\pm \Delta E_{j}/E_t$ reflecting some scattering of the energies with $q_*$. The interaction energy $E_{\rm int} /E_t$ is a very sub-dominant component which remains also almost constant after $z \gtrsim z_{\rm dec}$. The inflaton gradient energy $E_{G,\varphi} /E_t$ and the potential energy density $E_V /E_t$, also sub-dominant components, show however some trend of energy exchange: as $q_*$ increases, $E_{G,\varphi} /E_t $ grows and $E_V /E_t$ decreases. 
We provide the following estimations based on fits obtained within the range $q_* \in [7500, 2.5\cdot 10^{6}]$,
\begin{center}
Sub-dominant Energy Fractions at $z \approx z_{\rm dec}$ ($q_* \gtrsim 7\cdot 10^3$):\vspace*{-0mm}
\begin{eqnarray}
{E_{G,\varphi}\over E_t} \simeq {19\over(1+30000/q_*)^{1/2}} \% \,,\hspace*{0.2cm}{E_{V}\over E_t} \simeq {27\over(q_*/2000 -1)^{1/3}}\%\,, \hspace*{0.2cm} {E_{\rm int}\over E_t} \simeq (2.3 \pm 0.5) \% \ . 
\end{eqnarray}
\end{center}
For $q_* \gtrsim 5\cdot 10^5$, we observe that the potential energy becomes marginal, with $E_V/E_t \lesssim 5\%$, while the inflaton gradient energy seems to saturate to a fraction $E_{G,\varphi}/E_t \simeq 19\%-20\%$, which still remains subdominant as compared to $E_{K,\chi}, E_{G,\chi}, E_{K,\varphi}$. In other words, at $z \approx z_{\rm dec}$, the energy is 'democratically' split between the mother and the daughter fields, with final fractions given as $E_{\chi}/E_t \sim E_{\varphi}/E_t \sim 50\%$, where we have defined $E_{\chi} \equiv (E_{K,\chi}+E_{G,\chi}+{1\over2}E_{\rm int})$ and $E_{\varphi} \equiv (E_{K,\varphi}+E_{G,\varphi}+E_V+{1\over2}E_{\rm int})$.

Finally, let us note that at times $z > z_{\rm dec}$, the energy fractions $E_{K,\varphi}/E_t$ and $E_V/E_t$ still evolve, slowly, but monotonically growing. At this stage, the total energy density is not scaling anymore as $1/a^3$, so the total contribution $E_t = E_{K,\varphi} + E_{K,\chi} +  E_{G,\varphi} + E_{G,\chi} + E_{\rm int} + E_{\rm V}$ [see Eq.~(\ref{eq:energy-m2phi2})] decreases further in time after $z \gtrsim z_{\rm dec}$. This is clearly seen in the left panel of Fig.~\ref{fig:m2phi2-energy}. Actually, at very late times $z \gg z_{\rm dec}$, the inflaton dominant energies seem to evolve very slowly towards some value close to (but presumably smaller than) $E_{K,\varphi} /E_t \simeq 50 \%$, $E_V /E_t \simeq 50\%$. Correspondingly, the rest of energy fractions decrease gradually to very small values. Our simulations however do not capture the very long times required to probe the final asymptotic values of the inflaton energy components. It is very likely that neither $E_{K,\varphi} /E_t$ or $E_V /E_t$ really reach $50\%$, but a somewhat smaller value. To quantify this, we have introduced a new time scale $z_{X}$, indicating the time it takes for the inflaton energy components (kinetic and potential energies) to represent a given $X \%$ of the total energy of the system. Within our simulation capabilities, the latest time we have been able to reach is $z_{0.80}$, when $(E_{K,\varphi} + E_V) /E_t \simeq 80 \%$ (i.e.~when $E_{K,\varphi} /E_t$ and $E_V /E_t$ reach individually $\sim 40\%$, as there is equipartition). Even though $80\%$ does not represent the final asymptotic value of the inflaton energy, it clearly signals a moment where the total energy density is well dominated by the inflaton. We observe in our simulations that the rate of growth of the inflaton energy components (between some time after $z_{\rm dec}$ and $z_{0.80}$) follows a well defined power-law in time. Extrapolating such growth to later times, we can in principle predict the moment $z_{0.99}$. In Eqs.~(\ref{eq:z_X}) we provide fits to $z_{0.80}$ and to $z_{0.99}$. Whereas $z_{0.80}$ is measured directly from the numerical simulations, $z_{0.99}$ should be taken only as indicative, as it is only an extrapolation based on the growth of the inflaton energy components at $z \leq z_{0.80}$. In reality, we do not know if eventually the inflaton will dominate up to $\sim 99\%$, or whether it will saturate (most likely) to a somewhat smaller fraction. The time scales are
\begin{eqnarray}\label{eq:z_X}
z_{0.80} \simeq 0.26\,q_*~ ({\rm measured})\,~~~~~~ \rightarrow \,~~~~~~~  z_{0.99} \sim 30 \,q_*~ ({\rm extrapolated})
\end{eqnarray}
The values of $z_{0.80}$ follow a well defined power law, see right panel of Fig.~\ref{fig:m2phi2-energy}. The fit is obtained only for the cases $q_* \lesssim 40000$, since for bigger resonance parameters we cannot reach $z_{0.80}$ in our simulations (as the larger the $q_*$ the longer it takes the simulation). Assuming the fit of $z_{0.80}$ in Eq.~(\ref{eq:z_X}) is valid for every resonance parameter, we then expect $z_{0.80} \sim 10^5$ for $q_* \sim 10^5$, or $z_{0.80} \sim 10^6$ for $q_* \sim 10^6$. This explains, {\it a posteriori}, why we could not reach these time scales in the simulations for large resonance parameters.

In conclusion, even though the system manages to transfer like $\sim 50\%$ of the inflaton energy into the daughter field(s) at $z_{\rm dec}$, unless some new ingredient is added into the scenario (e.g.~new coupling to new particle species), the system tends to go back, slowly but systematically, to a complete inflaton energy domination in the long term $z \gg z_{\rm dec}$. Contrary to the $\phi^4$ case, the energy density in the daughter field(s) is eventually red-shifted away. 
  
\subsection{Lattice simulations of the decay of spectator fields}\label{sec:specfields}

We move now into the study of scenarios where the oscillating field $\phi$ does not dominate the energy budget of the Universe. This is the case of any scalar field with a monomial potential that was a spectator field during inflation. We will assume again that $\phi$ is coupled to some extra species, in particular to another scalar field $X$, with coupling $g^2\phi^2 X^2$. A
paradigmatic example of a spectator-field in cosmology is the curvaton~\cite{Enqvist:Curvaton,Lyth:Curvaton,Takahashi:Curvaton}, which is typically assumed to have a quadratic potential. Another example of relevance is the SM Higgs in a Higgs-inflaton weak coupling regime~\cite{Enqvist:2013kaa,Enqvist:2014tta,Figueroa:2014aya,Figueroa:2015hda,Enqvist:2015sua,Figueroa:2016dsc}, which has a quartic potential. As the analysis of the latter has already been presented in~\cite{Figueroa:2015hda}, we do not repeat it in this section. We will restrict our numerical study to a spectator-field with a quadratic potential,
\be V (\phi) = \frac{1}{2} m^2 \phi^2 \,. \ee
In Sect.~\ref{sec:Summary} we will add nonetheless, the fitted formulas corresponding to a spectator field with quartic potential, based on the results obtained in \cite{Figueroa:2015hda}.

 \begin{figure}
      \begin{center}
                  \includegraphics[width=11cm]{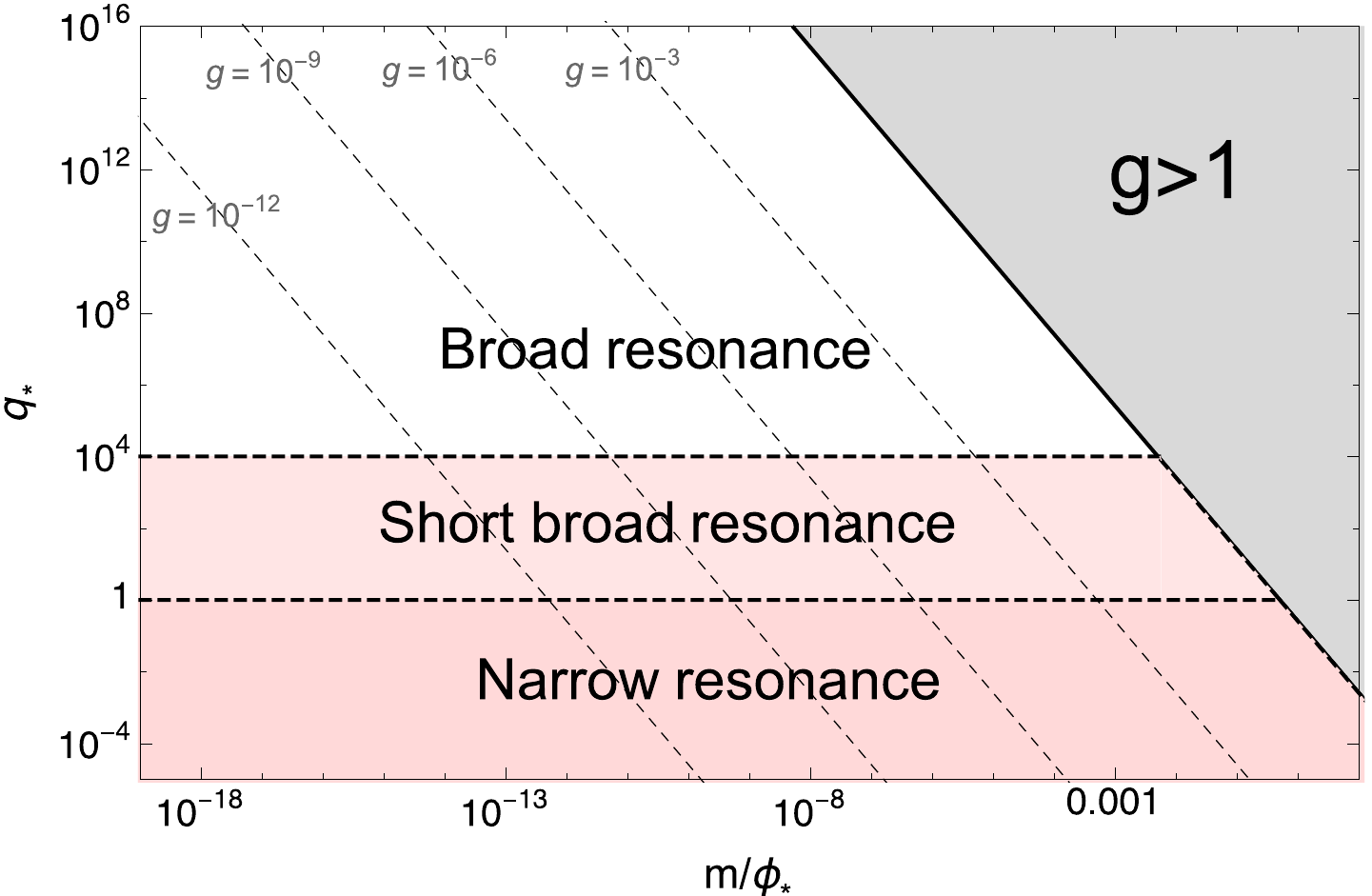}
     \end{center}
                \caption{We show different regions in the $(q_*,m/\phi_*)$ parameter space of a spectator field with $\propto \phi^2$ potential, according to their different dynamics. Note that the coupling is $g=2 (m/\phi_*) \sqrt{q_*}$ from (\ref{eq:spec-qres}). Explanation of the meaning of these regions is given in the bulk text.} 
            \label{fig:spec-parameters}
 \end{figure}

The case of a spectator-field with potential $V \propto \phi^2$ can be analyzed in a very similar way to the quadratic preheating case studied in Section~\ref{sec:m2phi2}. If we redefine the spacetime and field variables as in Eq.~(\ref{eq:m2phi2-variables}), the field EOM are identical to Eqs.~(\ref{eq:m2phi2-eom})-(\ref{eq:m2phi2-eom2}), with the resonance parameter defined as (we rewrite Eq.~(\ref{eq:m2phi2-resp}) for convenience)
\be q_* = \frac{g^2}{4} \left({\phi_*\over m}\right)^2 \ . \label{eq:spec-qres} \ee
As before, we choose the initial time of our simulations at the onset of the oscillatory regime of the spectator field, which we set to the moment when the Hubble rate equals the frequency of oscillations, $H(t=t_*) \equiv H_* = m$. We define, from now on, all quantities at this time with the subindex $*$.

There are two essential differences with respect to the analogue inflationary case. In the latter, we obtain the time-evolution of the scale factor by solving the Friedmann equations self-consistently with the fields' EOM. However, in our present scenario neither of the fields $\phi$ or $X$ dominate the energy content of the Universe. The evolution of the background expansion rate is determined by the inflationary sector, which we do not model explicitly. We will simply fix the expansion rate as a power law characterized by an equation of state $w$, i.e. 
\be a(t) = a_* \left( 1 + \frac{1}{p} H_* (t - t_*) \right)^p = \left( 1 + \frac{1}{p} z \right)^p \ , \hspace{1cm} p = {2\over3(1+w)} \ . \label{eq:exprate} \ee
We will consider $w = 1/3$ for a RD background, $w = 0$ for a MD background, and $w = 1$ for a KD background. In practice, for the quadratic potential scenario we will focus mostly in the RD case, as this represents the most relevant cosmological case of viable curvaton~\cite{Enqvist:Curvaton,Lyth:Curvaton,Takahashi:Curvaton}. For completeness, we will present some results of the quadratic spectator field in a MD background, even if this case seems not to have any cosmological relevance. For the quartic potential scenario analyzed in~\cite{Figueroa:2015hda}, we will quote in Section~\ref{sec:Summary} the results for all MD, RD and KD cases, showing also a parametrization of the results with respect a generic $w$.

  \begin{figure}
      \begin{center}
                  \includegraphics[width=11cm]{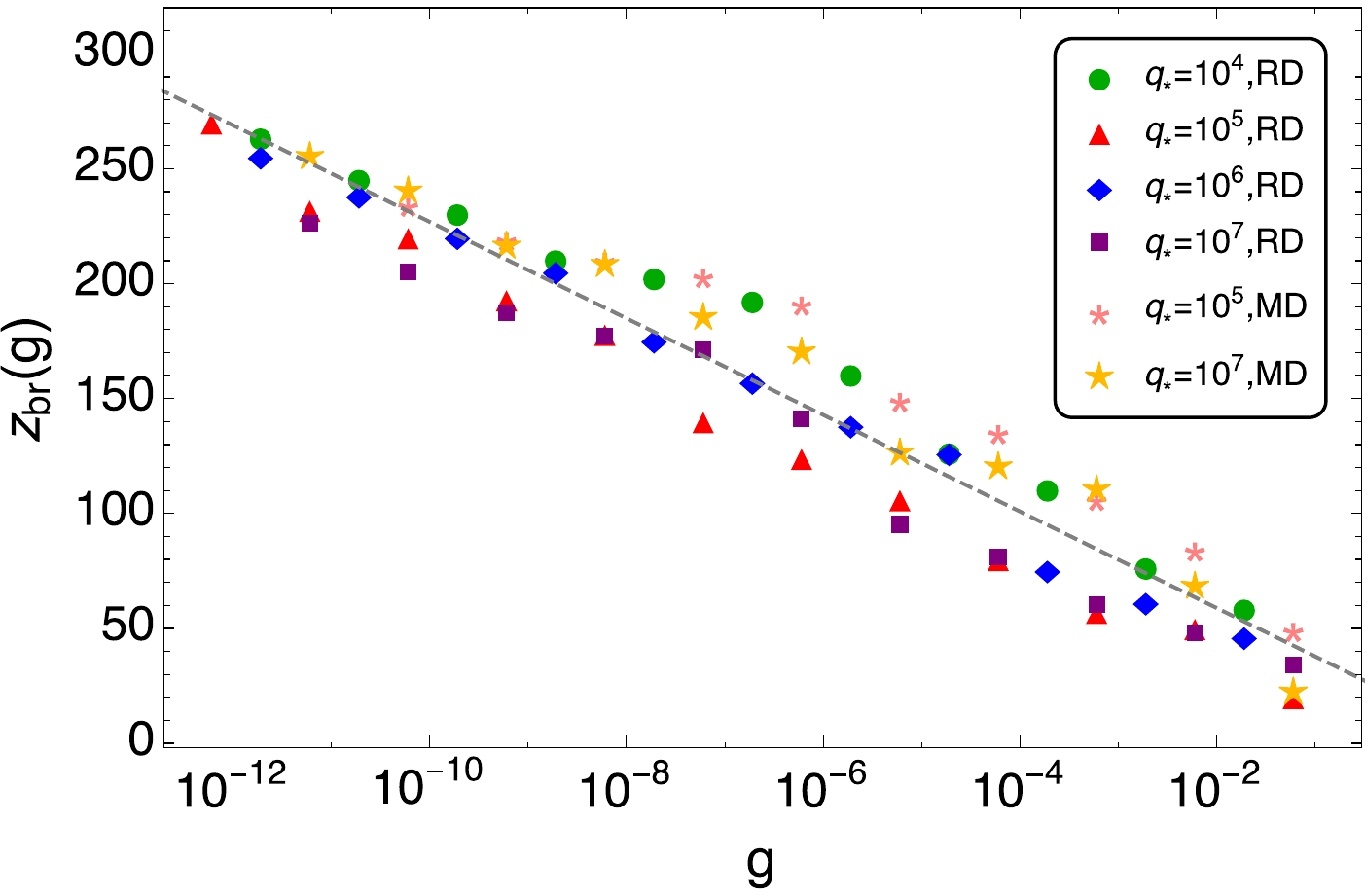}
             
     \end{center}
                \caption{We show $z_{\rm br}$ as a function of coupling $g$ obtained from lattice simulations, for an oscillating spectator-field with quadratic potential. Each symbol corresponds to a specific resonance parameter $q_*$ and expansion rate (RD or MD). We see that independently of the particular case, all values coincide approximately in a single straight line, which we fit in Eq.~(\ref{eq:spec-zitime}) and show with a dashed line.} 
            \label{fig:spec-zizetime}
 \end{figure}

The second difference with respect the quadratic inflaton is that now there are more free parameters, which makes the parametrization of the system in principle more complex. In the inflationary case the mass $m$ and the amplitude $\phi_*$ were constrained by the CMB observations, whereas now these are free parameters. Fortunately, if we look at the EOM Eqs.~(\ref{eq:m2phi2-eom})-(\ref{eq:m2phi2-eom2}), we notice that the dynamics only depend on the combination $g^2(\phi_*/m)^2$ through $q_*$. At the same time, one can check that the spectrum of the initial modes mimicking quantum fluctuations, when written in natural units, only depends on the ratio $\phi_*/m$, see Appendix~\ref{appen:Lattice}. Therefore, the system only depends ultimately on two independent parameters, $\phi_*/m$ and $g^2$ (or alternatively $\phi_*/m$ and $q_*$). Whereas in the inflationary case $\phi_*/m$ was fixed, now this ratio represents an extra free parameter. Finally, the velocity of the field at the onset of the oscillatory regime is determined from the slow-roll condition, which still holds approximately when $H_* = m$. We take therefore as initial velocity the approximation $\dot{\phi}_* \simeq - m^2 \phi /(3 H_*) = - m \phi_* /3$, which in natural units reads $\varphi^{'}_{*} = 7/6$.

Fig.~\ref{fig:spec-parameters} is a diagram of the $(q_*,m/\phi_*)$ parameter space, where the coupling strength can be read as $g = 2 (m/\phi_*) \sqrt{q_*}$. We have excluded the region $g>1$, depicted in gray in the figure, as this corresponds to non-perturbative coupling strengths. There are different regions in the parameter space $(q_*,m/\phi_*)$, according to the different dynamics of the system discussed in Section \ref{sec:m2phi2}. The narrow resonance region correspond to values $q_* < 1$, which lattice simulations cannot capture well. For $1 < q_* \lesssim 10^4$, the inflaton is in broad resonance regime initially, but due to the expansion of the Universe it enters into narrow resonance before enough energy have been transferred into the daughter fields to affect the mother field through backreaction. Hence, we denote this region as 'short broad resonance'. A broad resonance regime sustained for a sufficiently long time, corresponds to $q_* \gtrsim 10^4$ values. We will only study in the lattice this regime, sampling $q_*$ from $\sim 10^4$ to $\sim 10^7$.

\subsubsection{Onset of non-linearities, energy evolution and decay time}

We will parametrize the system as a function of $g^2$ and $m /\phi_*$, in light of the previous discussion. In Figure~\ref{fig:spec-zizetime} we show the backreaction time $z_{\rm br}$, as a function of the coupling $g$, for different combinations of $q_*$ and post-inflationary expansion rates. We define again $z_{\rm br}$ as the moment when the conformal amplitude of the mother field starts decaying compared to its previously constant value, i.e. when it really feels the back-reaction of its decay products. We see that the dependence of $z_{\rm br}$ on $g$ is mostly insensitive (within some scatter) to the choice of $q_*$ and expansion rate. We find the following fit to the data
\be z_{\rm br} (g) \approx 16.9 - 20.9 \log_{10} g \ . \label{eq:spec-zitime}\ee
As detailed in Section \ref{sec:AnalyticalParamRes}, the logarithmic dependence appears as a consequence of the initial linear behavior of the mode functions, which obey the Mathieu equation until their backreaction into $\phi$ is noticed. The scattering of the points with respect the fit reflects some mild dependence on $q_*$ and the expansion rate, but also the stochastic nature of the resonance, recall our discussions in Section~\ref{sec:m2phi2}. The reason as to why we see now the logarithmic dependence in this Figure, but not in the inflationary case (recall Fig.~\ref{fig:m2phi2-zbTime}), is that now we have the freedom to vary $g$ across many orders of magnitude, by means of increasing $m/\phi_*$, but not $q_*$.

   \begin{figure}
      \begin{center}
                  \includegraphics[width=11cm]{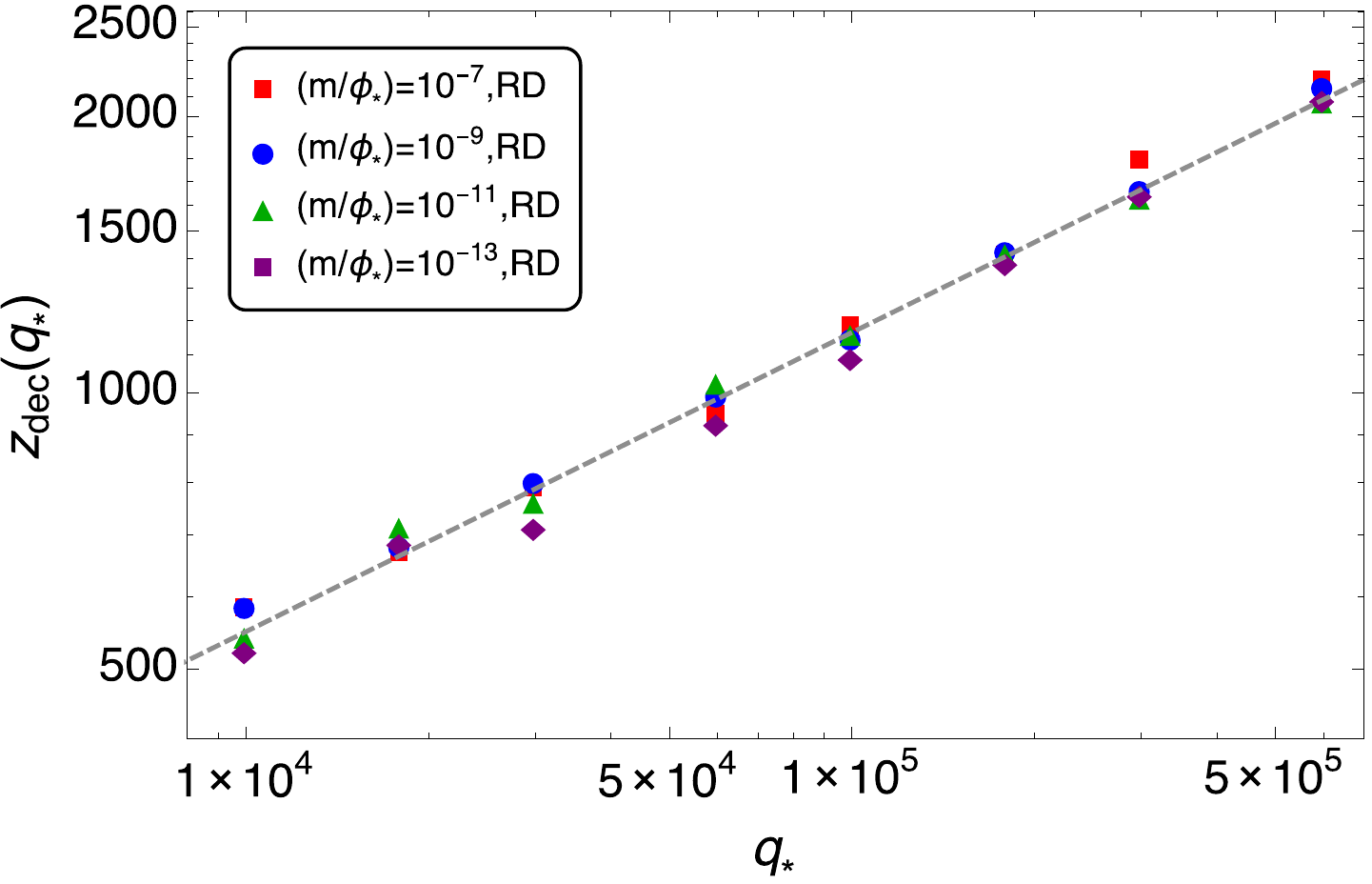}
     \end{center}
                \caption{We plot $z_{\rm dec}$ and $z_*$ as a function of $q_*$, for an oscillating spectator-field with quadratic potential and a RD Universe. For $z_{\rm dec}$ we consider different values of $m/\phi_*$, while for $z_*$ we take $m/\phi_* = 10^{-7}$. The lower dashed line corresponds to the fit of $z_{\rm dec}$ [Eq.~(\ref{eq:spec-zetime})], while the upper dotted-dashed line indicates the fit of $z_*$. } 
            \label{fig:spec-zizetime2}
 \end{figure}
 
In Fig.~\ref{fig:spec-zizetime2} we plot $z_{\rm dec}$, defined in an identical way to the inflationary $m^2 \phi^2$ case. In this case, we only provide fits for the RD case. We see that independently of the numerical value of $(m / \phi_*)$, all points can be fitted very well to
\be z_{\rm dec} (q_*) \approx 27.3 q_*^{0.33} \label{eq:spec-zetime} \ . \ee

The energy of this system can be written in terms of its different contributions in the same way as in the quadratic preheating case [Eqs.~(\ref{eq:energy-m2phi2}) and (\ref{eq:energy-m2phi2B})]. Their time-evolution is also very similar to the one seen in Fig.~\ref{fig:m2phi2-energy} for chaotic inflation, so we just specify the different energy contributions at both $z_{\rm br}$ and $z_{\rm dec}$. We find that the numbers are quite independent from $\phi_* /m$ and $q_*$. At $z_{\rm br}$, we have
\begin{center}
Energy Fractions at $z_{\rm br}$ ($q_* \gtrsim 10^4$) :\vspace*{-2mm}
\begin{eqnarray}\label{eq:EnergiesSpecatZi}
{E_{K,\varphi}\over E_t} \simeq (49.8 \pm 0.5) \%\,,\hspace*{0.2cm}{E_{V}\over E_t} \simeq (48.7 \pm 1.0) \%\,, \\ {E_{K,\chi}\over E_t} \simeq (0.7 \pm 0.7) \%\,,\hspace*{0.2cm}{E_{\rm int}\over E_t} \simeq (0.7 \pm 0.7) \% \ , 
\end{eqnarray}
\end{center}
with the other energies contributing less than $0.1\%$. The error bars $\Delta E_i / E_t$ account for the dispersion due to different choices of $q_*$ and $\phi_* / m$. As in the quadratic preheating case, at $z_{\rm br}$ most of the energy is stored in the mother field (in the kinetic and potential energies), while only $\sim 1 \%$ is stored in the daughter field. This percentage is enough to induce the onset of the mother field decay due to backreaction effects.

On the other hand, at $z_{\rm dec}$, the energies are distributed in the following manner,
\begin{center}
Dominant Energy Fractions at $z_{\rm dec}$ ($q_* \gtrsim 10^4$):\vspace*{-2mm}
\begin{eqnarray}\label{eq:EnergiesSpecatZe}
{E_{K,\varphi}\over E_t} \simeq (24.3 \pm 0.9) \%\,,\hspace*{0.2cm}{E_{G,\varphi}\over E_t} \simeq (20.0 \pm 0.8) \%\,, \\ {E_{K,\chi}\over E_t} \simeq (26.4 \pm 1.0) \%\,,\hspace*{0.2cm}{E_{G,\chi}\over E_t} \simeq (24.8 \pm 1.2) \% \ ,  
\end{eqnarray}
\end{center}
which are also approximately independent on $q_*$ and $\phi_* / m$. The other two energies are subdominant and have a certain dependence in $q_*$, which we have fitted as
\begin{center}
Sub-dominant Energy Fractions at $z_{\rm dec}$ ($q_* \gtrsim 10^4$):\vspace*{-2mm}
\begin{eqnarray}
{E_{V}\over E_t} \simeq {80 \over(1 + q_*)^{0.3}}\%\,, \hspace*{0.2cm} {E_{\rm int}\over E_t} \simeq {13 \over(1 + q_*)^{0.2}}\% \ .
\end{eqnarray}
\end{center}
Note that, unlike the quadratic preheating case, for the spectator-field both the potential and interaction energy contributions have a decreasing behavior with $q_*$.

At $z \gtrsim z_{\rm dec}$ the system enters into a stationary regime, where the energies $E_{K,\varphi}$ and $E_V$ evolve very slowly in time. However, similarly as to the analogous preheating scenario, each of the energy fractions $E_{K,\varphi} /E_t$ and $E_V / E_t$, still grow slowly but monotonically, towards some value of the order of, but (presumably) somewhat smaller than, $\sim 50\%$. This asymptotic regime is however attained at very large times, much larger than in the quadratic inflaton case for the same $q_*$'s. Due to this, we have only been able to capture partially this regime in our lattice simulations with spectator fields. We first define $z_X$ analogously as in the preheating case, as the moment when the mother field energy components represent a fraction $X\%$ of the total energy of the mother-daughter fields system. We can only reach up to $z_{0.40}$ in our numerical simulations of spectator fields (let us recall that in the case of preheating we reached $z_{0.80}$). However the trend of growth of $E_{K,\varphi} /E_t$ and $E_V / E_t$ between $z_{\rm dec}$ and $z_{0.40}$ follows again a well defined power-law, which is expected to hold at later times. Thus, extrapolating the behavior of the energy fractions at later times, we can predict again $z_{0.99}$. The fits we obtain are
\begin{eqnarray}\label{eq:z_Xcurvaton}
z_{0.4} \simeq 0.18\,q_* \ ({\rm measured})\,~~~~~~ \rightarrow \,~~~~~~~  z_{0.99} \sim 8 \cdot 10^{-6} \,q_*^{3} \ ({\rm extrapolated})
\end{eqnarray}
In reality, as in the preheating case, we do not know to which final value $E_{K,\varphi} /E_t$ and $E_V / E_t$ settle eventually down, and hence the extrapolated $z_{0.99}$ must be considered only as indicative of the time scale of the final asymptotic state.

\section{Collection of Fitted formulas}\label{sec:Summary}

In this section we just collect together the fitted formulas from all the scenarios considered. In the case of a spectator-field with a quartic potential we just quote the results from~\cite{Figueroa:2015hda}. The interested reader can, in this manner, access rapid and easily to the key results from this paper (complemented with those from~\cite{Figueroa:2015hda}). 

For self-consistent reading of this Section, let us first summarize the dynamics of parametric resonance, and define the variables to which we provide fits. In parametric resonance with $q > 1$, as soon as the mother field $\phi$ starts oscillating, there is a fast transfer of energy into the daughter species. This occurs independently of whether the mother field dominates or not the energy budget of the universe. During few oscillations, the energy of the daughter fields $X$ remain orders of magnitude smaller than the energy of the mother field. Hence the $\phi$ field oscillates initially almost unaffected by the presence of its decay products. This corresponds to a linear regime, where the mode functions of the daughter field grow exponentially fast in some range of momenta. Due to this exponential excitation, there is always a time for any given resonance parameter, when the energy transferred becomes large enough so that the backreaction from the daughter species into the mother field cannot be further ignored. We refer to this moment as $z_{\rm br}$. From that moment onwards, the (conformal) amplitude of the mother field starts decreasing in a noticeable manner, see Fig.~\ref{fig:lphi4-init} for an example of this. The time scale $z_{\rm br}$ defines therefore the onset of the mother field decay, and not the time scale of the decay itself, as the linear calculation suggests. From then on, at $z \gtrsim z_{\rm br}$, the system becomes non-linear, and evolves towards a stationary state. The latter is characterized by the different energy fractions of the fields (kinetic, gradient and potential energies) evolving very slowly,
while at the same time an equipartition distribution of energies is set. This regime is attained at a time $z_{\rm dec}$. We consider this moment as the truly decay time scale of the mother field: while during the non-linear regime $z_{\rm br} \leq z \leq z_{\rm dec}$ energy is significantly exchanged between the mother and the daughter fields, at $z \geq z_{\rm dec}$ the energy exchange ceases and the energy fractions evolve in a stationary regime. In the case of quadratic potentials, the system tends very slowly to restore, at long times $z \gg z_{\rm dec}$, the mother field energy dominance. Hence, we also provide the time scales $z_{0.80}$ and $z_{0.99}$ corresponding to moments when the mother field represents $\sim 80\%$ and $\sim 99\%$ of the energy budget of the mother-daughter system.

In the following we summarize our fits for $z_{\rm br}$, $z_{\rm dec}$ (and $z_{0.8}$, $z_{0.99}$ when applicable) as a function of the resonance parameter, for all the scenarios we have analyzed. We take the coupling between the mother and the daughter field to be of the form $g^2 \phi^2 \chi^2$. In the case of preheating, where the mother field -- the inflaton -- dominates the energy budget of the universe, we also provide due to its interest, the stationary energy fractions.

\begin{itemize}
\item Preheating with inflationary potential $V_{\rm inf} (\phi) = \frac{1}{4} \lambda \phi^4$:
\bea z_{\rm br} (q) \in  [40, 250 ] &\ ;& \hspace{0.2cm} \text{ See Fig.~\ref{fig:lame-zbtime} } \\ \nonumber\\
 z_{\rm dec} (q) - z_{\rm br} (q) \simeq
  \left\lbrace
  \begin{array}{l}
     51 q^{0.28} \hspace{0.3cm}\text{ if } q < 100 \vspace*{2mm}\\
     11 q^{0.56} \hspace{0.3cm}\text{ if } q \geq 100 \\
  \end{array}
  \right. &\ ;& \hspace{0.2cm} \text{ See Fig.~\ref{fig:lphi4-zdecay} }\label{eq:ze1}
\eea
where $q \equiv \frac{g^2}{\lambda}$.\vspace*{-2mm}
\begin{center}
Energy Fractions at $z \gtrsim z_{\rm dec}$:
\begin{eqnarray}\label{eq:EnergiesPhi4atZeV2}
\begin{array}{c}
{E_{K,\chi}\over E_t} \simeq (29.5 \pm 3.3) \%\,,\hspace*{0.2cm}{E_{K,\varphi}\over E_t} \simeq (22.6 \pm 3.4) \%\,,\hspace*{0.2cm} {E_{G,\chi}\over E_t} \simeq (26.2 \pm 3.4) \%\,,\hspace*{0.2cm}\vspace*{3mm}\\
{E_{G,\varphi}\over E_t} \simeq (17.7 \pm 3.0) \%\,,\hspace*{0.2cm} {E_{\rm int}\over E_t} \simeq (3.2 \pm 0.7) \%\,,\hspace*{0.2cm}{E_{V}\over E_t} \simeq (0.8 \pm 0.2) \% 
\end{array}
\end{eqnarray}
\end{center}
\vspace*{0.5cm}

\item Preheating with inflationary potential $V_{\rm inf} (\phi) = \frac{1}{2} m^2 \phi^2$:
\bea z_{\rm br} (q) \in  [40, 135 ] &\ ;& \hspace{0.2cm} \text{ See Fig.~\ref{fig:m2phi2-zbTime} } \\
z_{\rm dec} (q) \simeq 19.9 q_*^{0.27} &\ ;& \hspace{0.2cm} \text{ See Fig.~\ref{fig:m2phi2-energy} }
\\
z_{0.8} \simeq 0.26\,q_* &\ ;& \hspace{0.2cm} \text{ See Fig.~\ref{fig:m2phi2-energy} }\\
z_{0.99} \sim 30 \,q_* &\ ;& \hspace{0.2cm} \text{ (extrapolated) }
\label{eq:ze2}
\eea
where $q_* \equiv \frac{g^2 \phi_*^2}{4 m^2}$, with $\phi_*$ the initial value of the inflaton field.\vspace*{0.0cm}\\
\begin{center}
Dominant Energy Fractions at $z \simeq z_{\rm dec}$ ($q_* \gtrsim 5\cdot 10^4$):\vspace*{-0mm}
\begin{eqnarray}\label{eq:EnergiesM2phi2atZeV2}
\begin{array}{c}
{E_{K,\chi}\over E_t} \simeq (25.2 \pm 2.2) \%\,,\hspace*{0.2cm}{E_{K,\varphi}\over E_t} \simeq (26.0 \pm 2.3) \%\,,\hspace*{0.2cm} {E_{G,\chi}\over E_t} \simeq (22.9 \pm 2.5) \%\,,\hspace*{0.2cm}
\end{array}
\end{eqnarray}
\vspace*{0.1cm}\\
Sub-dominant Energy Fractions at $z \simeq z_{\rm dec}$ ($q_* \gtrsim 7.5\cdot 10^3$):\vspace*{-0mm}
\begin{eqnarray}
\begin{array}{c}
{E_{G,\varphi}\over E_t} \simeq {19\over(1+ 30000/q_*)^{1/2}} \% \,,\hspace*{0.2cm}{E_{V}\over E_t} \simeq {27\over(q_*/2000 -1)^{1/3}}\%\,, \hspace*{0.2cm} {E_{\rm int}\over E_t} \simeq (2.3 \pm 0.5) \%
\end{array} 
\end{eqnarray}
\end{center}

\item Spectator-field with potential $V (\phi) = \frac{1}{2} m^2 \phi^2$ and RD expansion rate:
\bea z_{\rm br} (g) \simeq  16.9 - 20.9 \log_{10} g &\ ;& \hspace{0.2cm} \text{ See Fig.~\ref{fig:spec-zizetime} } \\
z_{\rm dec} (q) \simeq 27.3 q_*^{0.33} &\ ;& \hspace{0.2cm} \text{ See Fig.~\ref{fig:spec-zizetime2} }\\
z_{0.40} \simeq 0.18\,q_* &\ ;& \hspace{0.2cm} \text{ (measured) }\\
z_{0.99} \sim 8 \cdot 10^{-6} \,q_*^{3} &\ ;& \hspace{0.2cm} \text{ (extrapolated) }
\label{eq:ze3}
\eea
where $q_* \equiv \frac{g^2 \phi_*^2}{4 m^2}$ with $\phi_*$ the initial value of the mother field.\vspace*{0.5cm}

\item Spectator-field with potential $V (\phi) = \frac{1}{4} \lambda \phi^4$ (Standard Model Higgs, see \cite{Figueroa:2015hda}\footnote{In the notation of that reference, we use $z_i$ instead of $z_{\rm br}$, and $z_e$ instead of $z_{\rm dec}$.}):
\bea z_{\rm br} (q) &\approx &
  \left\lbrace
  \begin{array}{l}
     16 \beta^{\frac{-(1 + 3 \omega)}{3 (1 + \omega)}}\hspace{2.65cm} \text{ if } q \in \text{Resonance Band} \vspace*{2mm}\\
     ( 86.9- 9.2 \log{q} )  \beta^{\frac{-(1 + 3 \omega)}{3 (1 + \omega)}} \hspace{0.31cm}\text{ if } q \notin \text{Resonance Band}  \\
  \end{array}
  \right.  \\
  z_{\rm dec} (q) & \approx & 50.7 \beta^{\frac{-(1 + 3 \omega)}{3 (1 + \omega)}} q^{0.44} \ ,\label{eq:ze4}
\eea
where $\beta \equiv \frac{\sqrt{\lambda} \phi_*}{H_*}$, $q \equiv \frac{g^2}{\lambda}$, and $\omega$ is the equation of state ($\omega=0, 1/3, 1$, for MD, RD, and KD respectively).

\end{itemize}

Note that the time scales reported here depend on our choice of å mother-daughter interaction $g^2 \phi^2 X^2$, representing this the only interaction the daughter field experiences. The time scales may change, for instance, in the presence of self-interactions of the $X$ field \cite{Podolsky:2005bw}.

\section{Discussion}\label{sec:discussion}

Thoughtful analysis of parametric resonance, including analytical calculations of the the Floquet index and analysis of the Floquet theorem, can be found e.g.~in~\cite{Kofman:1997yn,Greene:1997fu,Amin:2014eta}. In this work we rather concentrate in the study of parametric resonance using classical real time field theory lattice simulations. We have simulated an oscillating mother field $\phi$ coupled to a daughter field $X$, which is excited due to an interaction term $g^2\phi^2 X^2$. We have considered two main scenarios. First, when the mother field is the inflaton field, oscillating around the minimum of its potential after inflation. We have considered the case of chaotic inflation with $V \propto \phi^2$ and $V \propto \phi^4$ potentials. In a second type of scenarios, the oscillating field was just a spectator-field during inflation, playing no dynamical role on the expansion of the Universe. We have considered also $V \propto \phi^2$ and $V \propto \phi^4$ potentials, but analyzed only numerically the former, as the latter was already analyzed in~\cite{Figueroa:2015hda}.

Our results show very clearly that the computation in the linear regime of the moment of efficient transfer of energy $z_{\rm eff}$, see Eqs.~(\ref{eq:EffEnergyTransferTimeScaleApproxPhi4}), (\ref{eq:EffEnergyTransferTimeScaleApproxPhi2}), does not represent a good estimation of the decay time scale $z_{\rm dec}$ of the mother field. Instead, $z_{\rm eff}$ indicates well (up to $\mathcal{O}(1)$ factors) the onset of the mother field decay at $z_{\rm br}$, when the back-reaction of the daughter field becomes noticeable. Despite the exponential transfer of energy into the daughter fields during the time $z < z_{\rm br}$, the daughter field fluctuations follow a linear equation, whilst the mother field amplitude remains almost unperturbed. At $z \gtrsim z_{\rm br}$, the presence of the excited daughter fields makes the amplitude and energy of the mother field to abruptly decrease. At $z \gtrsim z_{\rm br}$ the dynamics become non-linear, and there is a noticeable transfer of energy between the mother and the daughter fields. Eventually, at $z \gtrsim z_{\rm dec}$ the amplitude of the fields settle down to stationary values, with the energy equiparted among the different components. As for $z \geq z_{\rm dec}$ the dominant energy components do not evolve any more noticeably, we identify the onset of that stationary stage as the truly time scale of the decay of the mother field. In the case of a quadratic potential, at $z \gtrsim z_{\rm dec}$, in reality only the mother field kinetic and potential (conformal) terms remain almost constant, as the (conformal) energy components of the daughter fields decay slowly at long times. 

The linear calculation of $z_{\rm eff} \sim z_{\rm br}$ indicates that the stronger the coupling between mother and daughter field, the faster the system becomes non-linear. The dependence is however only logarithmic, so in practice $z_{\rm br}$ only changes by a factor $\mathcal{O}(1)$ when varying the strength of the coupling in more than 10 orders of magnitude, see e.g.~Figure~\ref{fig:spec-zizetime}. As the system becomes however non-linear after $z \gtrsim z_{\rm br}$, our numerical results show the rather counter-intuitive result, opposite to the intuition gained from the analytic estimations: the stronger the mother-daughter coupling, the longer the time decay $z_{\rm dec}$ scale is, with a typical power-law behavior with respect the resonance parameter, $z_{\rm dec} \propto q^r$, with $r \sim 1/4, 1/3$ or $1/2$, depending on the case, see Eqs.~(\ref{eq:ze1}), (\ref{eq:ze2}), (\ref{eq:ze3}), (\ref{eq:ze4}). 

Let us note that we have defined and obtained the decay time scale $z_{\rm dec}$ at the onset of the stationary regime, but we have not analyzed the evolution of the equation of state or the departure from thermal equilibrium. For a study of the subsequent evolution of the system at $z \gtrsim z_{\rm dec}$ towards thermalization, see~\cite{Micha:2002ey,Micha:2004bv,Podolsky:2005bw,Lozanov:2016hid}. We have found nonetheless a remarkable result: in the case of quadratic potentials, the energy components of the daughter field tends to decay at the very late times $z \gg z_{\rm dec}$, so that slowly but monotonically the mother field tends to dominate the energy budget of the mother-daughter system.

Let us remark that in this work we have considered the decay products to be scalar fields. However, parametric resonance can also take place for all bosonic species, including gauge fields (either Abelian and non-Abelian). There are many scenarios where the decay products are gauge fields, see e.g.~\cite{GarciaBellido:1999sv,Rajantie:2000nj,GarciaBellido:2003wd,Bezrukov:2008ut,GarciaBellido:2008ab,Bezrukov:2014ipa,Dufaux:2010cf,Deskins:2013lfx,Adshead:2015pva,Figueroa:2015hda,Enqvist:2015sua,Lozanov:2016pac,Figueroa:2016dsc}, although not in all of them the driving particle production mechanism is parametric resonance. As we demonstrated in~\cite{Figueroa:2015hda}, the dynamics of parametric resonance into Abelian gauge fields (at least for a mother field with quartic potential), is only slightly modified in the linear regime, i.e.~$z_{\rm br}$ is only marginally changed. The late time non-linear dynamics remain however basically unchanged. Therefore, in principle, our fitted formulas can be applied equally to the case of parametric resonance of gauge bosons. 
In the case of non-abelian gauge fields, the non-linearities in the gauge boson EOM due to the non-abelian structure, may block parametric resonance before reaching $z_{\rm br}$, if the resonance parameter is sufficiently large, see e.g.~\cite{Enqvist:2015sua}.

It is well known that violent out-of-equilibrium phenomena like particle production via parametric resonance, can produce scalar metric perturbations~\cite{Bassett:1998wg,Bassett:1999mt,Bassett:1999ta,Finelli:2000ya,Chambers:2007se,Bond:2009xx} and a significant amount of gravitational waves (GW)~\cite{Khlebnikov:1997di,Easther:2006gt,Easther:2006vd,GarciaBellido:2007dg,GarciaBellido:2007af,Dufaux:2007pt,Dufaux:2008dn,Figueroa:2011ye,Bethke:2013aba,Bethke:2013vca,Figueroa:2016ojl,Antusch:2016con}. A natural extension of our present fitting analysis is to parametrize the production of GW from parametric resonance in the early Universe. Although GW production in preheating after chaotic inflation models has been widely considered in the literature, there is still lacking a systematic parametrization of the GW spectrum today as a function of the different couplings\footnote{A parameter-fitting analysis exists however for the GW production from Hybrid preheating, see~\cite{Dufaux:2008dn}.}. We plan to do this in a forthcoming publication. 

There are some scenarios of preheating where the daughter fields are scalar fields, but the mechanism responsible for the particle production is not parametric resonance, e.g.~hybrid preheating~\cite{Felder:2000hj,Felder:2001kt,Copeland:2002ku,GarciaBellido:2002aj,GarciaBellido:2007dg,GarciaBellido:2007af,Dufaux:2008dn}. Our fitted formulas, unfortunately, cannot be applied to these scenarios. The case of trilinear or non-renormalizable interactions between the mother and the daughter field(s)~\cite{Dufaux:2006ee,Croon:2015naa,Antusch:2015vna,Enqvist:2016mqj} are neither captured by our analysis. The case of oscillations of a multi-component field is neither captured well by our analysis\footnote{In the case of super-symmetric flat directions, it may well happen that the flat directions are never really excited in first place~\cite{Enqvist:2011pt}, and therefore it makes no sense to speak about oscillations after inflation.}, see e.g.~\cite{Tkachev:1998dc,Olive:2006uw,Gumrukcuoglu:2008fk,DeCross:2015uza,Ballesteros:2016euj}. We speculate nonetheless, that the non-linear dynamics after the initial excitation in all these scenarios, is probably very similar to the one after parametric resonance. However, only proper lattice simulations can prove this.

As a final thought, it is interesting to note that, in the case of an inflaton with quartic potential, our results may challenge somehow the application of the standard perturbative calculation of the reheating temperature $T_{\rm RH} \sim 0.1\sqrt{\Gamma m_p}$, where $\Gamma$ is the inflaton decay width. It is often argued that, as preheating does not deplete completely the energy from the inflaton, 
reheating will only be completed when the inflaton decays perturbatively into other matter fields. Our simulations for a potential $V \propto \phi^4$ show however, that at the onset of the stationary regime at $z \simeq z_{\rm dec}$, the energy fractions do not evolve significantly anymore, and the inflaton energy never represents more than $50$\% of the total energy budget. Therefore, even if eventually the inflaton decays perturbatively into some species, the originally produced daughter field from parametric resonance (which also represent $50\%$ of the energy budget), may have already thermalized and reheated the universe. As we expect equipartition in the different field components at the onset of the stationary regime, if there were several daughter fields experiencing parametric resonance (and not just one like in our simulations), in principle the fraction of energy stored in the inflaton at the stationary regime, should be approximately suppressed by the total number of species (i.e.~the number of daughter fields plus one inflaton). In that case, whether the inflaton decays perturbatively later or not, should be mostly irrelevant, since by then most of the energy is stored in the parametrically excited daughter fields, which may very well have thermalized before. In the case of a quadratic potential, our results tend however to reinforce the opposite circumstance, as the system approaches at long times $z \gg z_{\rm dec}$, slowly but monotonically, towards a complete energy dominance of the mother field over the daughter field(s). This reinforces the idea that for a quadratic potential, a perturbative decay (or further interactions besides $g^2\phi^2X^2$) are necessary ingredients in order for the mother field to decay at all.

\acknowledgments
We thank Juan Garc\'ia-Bellido for very useful discussions and collaboration on related projects. This work is supported by the Research Project of the Spanish MINECO  FPA2013-47986-03-3P and the Centro de Excelencia Severo Ochoa Program SEV-2012-0249. F.T. is supported by the FPI-Severo Ochoa Ph.D. fellowship SVP-2013-067697. We acknowledge the use of the IFT Hydra cluster for the development of this work.

\appendix

\section{Spectral Analysis}\label{appen:Spectra}

We describe in this appendix the essential features of the field spectra during preheating with both quartic and quadratic potential (the case of a spectator-field with quadratic potential is similar to its inflationary analogue, so we do not refer to it specifically in this appendix). This will help us understand better the meaning of the time scales $z_{\rm br}$ and $z_{\rm dec}$ we have parametrized above, and how to relate them with the structure of resonance bands of the linearized \emph{Lam\'e} and \emph{Mathieu} equations. More specifically, we will look at the spectra of the energy density, which can be written for the mother and daughter fields in the quartic case as ($\kappa = k/H_*$)

\be \rho_{\kappa, \varphi} = \frac{\lambda \phi_*^4}{2} (|\varphi_{\kappa}^{'}|^2 + \omega_{\kappa, \varphi}^2 | \varphi_{\kappa} |^2 ) \ , \hspace{0.3cm} \rho_{\kappa, \chi} = \frac{\lambda \phi_*^4}{2} (|\chi_{\kappa}^{'}|^2 + \omega_{\kappa, \chi}^2 | \chi_{\kappa} |^2 ) \ , \label{eq:lphi4-spectra}\ee
with $\omega_{\kappa, \varphi} = \sqrt{ \kappa^2 + 3 \varphi^2 + q \chi^2 - \frac{a''}{a} }$ and $\omega_{\kappa, \chi} = \sqrt{\kappa^2 + q \varphi^2 - \frac{a''}{a} }$, and in the quadratic case as
\be \rho_{\kappa, \varphi} = \frac{m^2 \phi_*^2}{2} a \left( \left| \varphi_{\kappa}^{'} -  \frac{a'}{2a} \varphi_{\kappa} \right|^2 + \omega_{\kappa, \varphi}^2 |\varphi_{\kappa}|^2 \right) \ , \hspace{0.3cm} \rho_{\kappa, \chi} = \frac{m^2 \phi_*^2}{2} a \left(  \left| \chi_{\kappa}^{'} -  \frac{a'}{2a} \chi_{\kappa} \right|^2 + \omega_{\kappa, \chi}^2 | \chi_{\kappa}|^2 \right) \ , \label{eq:m2phi2-spectra} \ee
with $\omega_{\kappa, \varphi} = \sqrt{\frac{\kappa^2}{a^2} + 1 + \frac{4}{a^3} q_* \chi^2 - \frac{a''}{a} + \left(  \frac{a'}{a} \right)^2}$ and $\omega_{\kappa, \chi} = \sqrt{\frac{\kappa^2}{a^2} + \frac{4}{a^3} q_* \varphi^2 - \frac{a''}{a} + \left(  \frac{a'}{a} \right)^2}$.

\begin{figure}
      \begin{center}
                  \includegraphics[width=7.5cm]{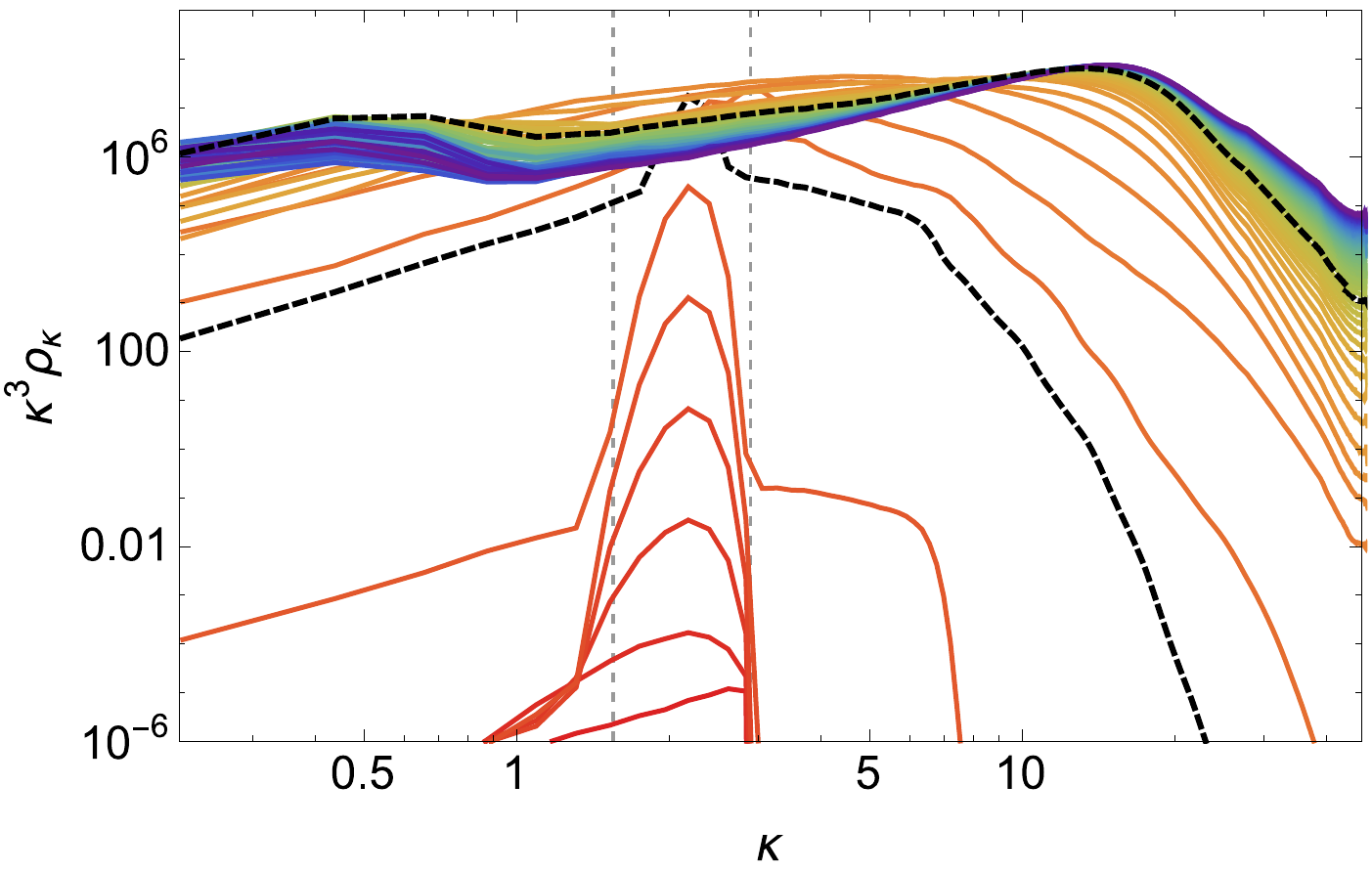}
					 \hspace{0.2cm}
                  \includegraphics[width=7.5cm]{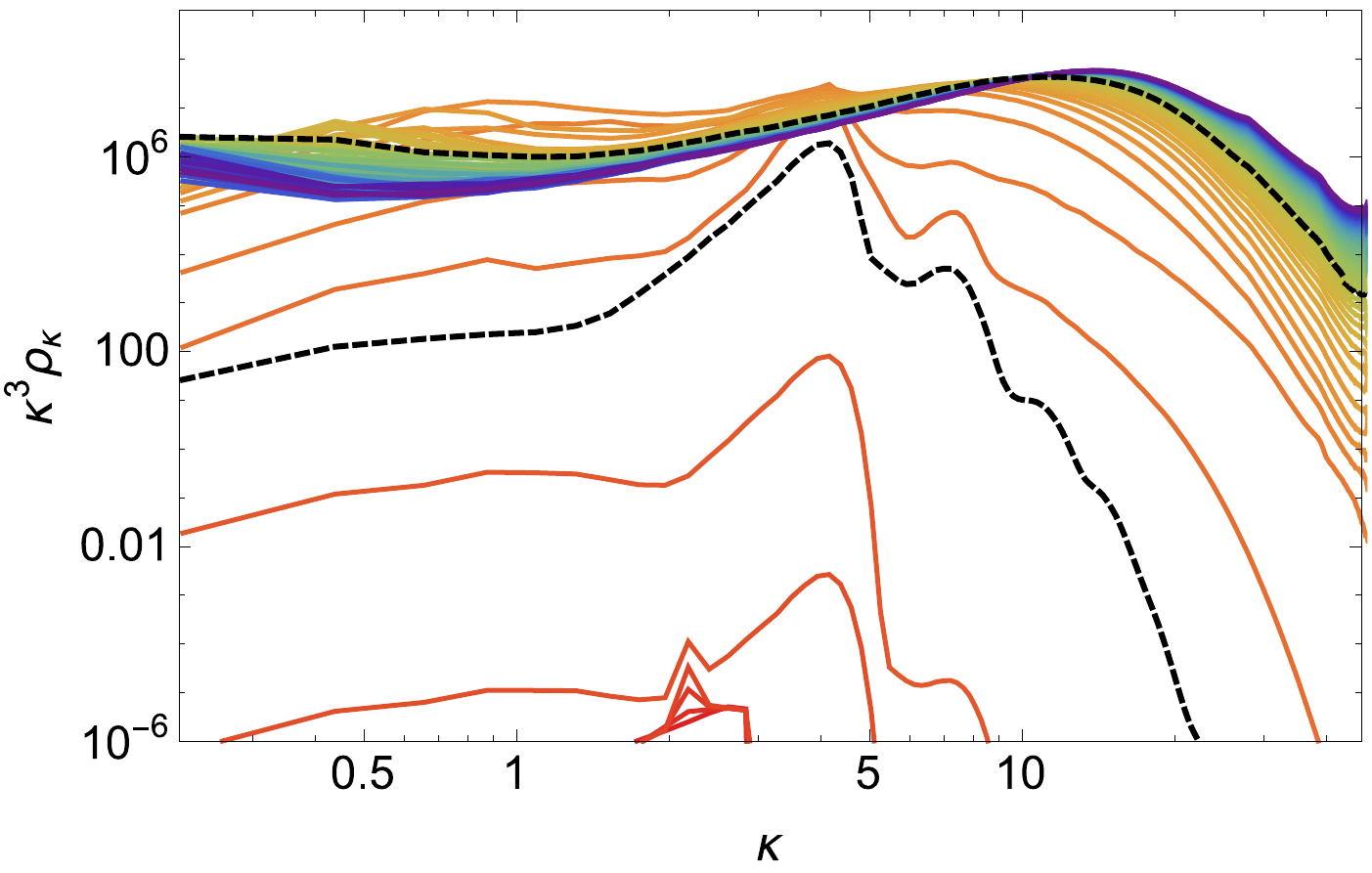}
                    \includegraphics[width=7.5cm]{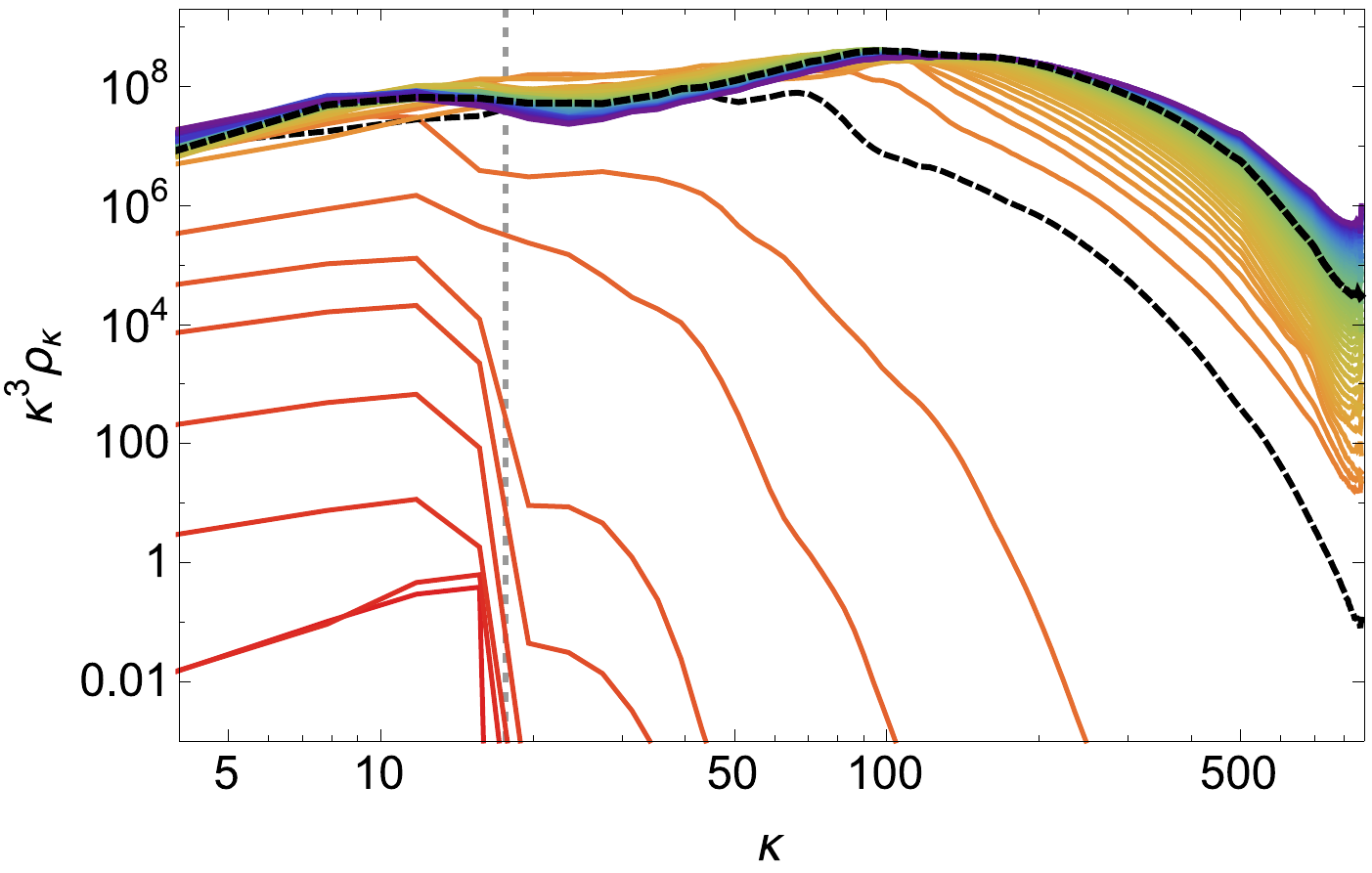} \hspace{0.2cm}
                  \includegraphics[width=7.5cm]{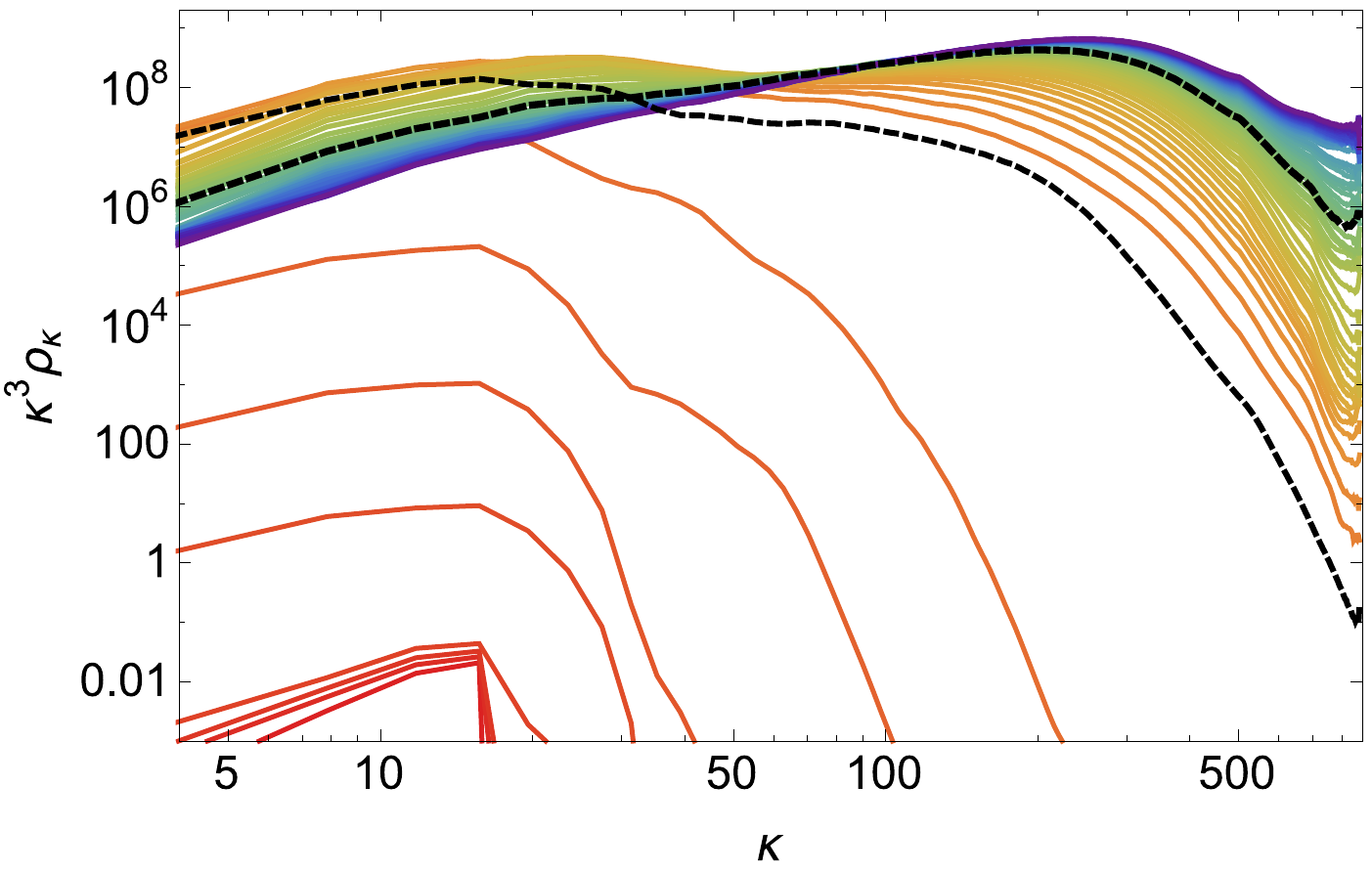}
     \end{center}
                \caption{
We show the energy density spectra as a function of $\kappa = k / H_*$ for preheating with quartic potential [upper panels, Eq.~(\ref{eq:lphi4-spectra})] and quadratic potentials [lower panels, Eq.~(\ref{eq:m2phi2-spectra})]. In each case, the left and right panels show the spectra of the daughter and mother fields respectively. Each line corresponds to the spectra at a given time, going from red (early times) to purple (late times). The time step between lines is $\Delta z \approx 10$, so $z \approx 0, 10, 20, \cdots 600$. We have drawn with a dashed black line the spectra corresponding to $z_{\rm br}$ and $z_{\rm dec}$. For the quartic case we show $q = 43$, with $z_{\rm br} \sim 60$ and $z_{\rm dec} \sim 220$. For the quadratic case we show $q_* = 25000$, with $z_{\rm br} \approx 80$ and $z_{\rm dec} \approx 300$. For the daughter fields we also show with gray dashed lines the position of the resonance bands: $\kappa_- < \kappa < \kappa_+$ for the quartic case, and $0 < \kappa < \kappa_{M}$ for the quadratic case. }    \label{fig:spectra-table}
 \end{figure}
 
Let us recall first that the time-evolultion of the excited fields during parametric resonance can be broadly divided in two regimes. The first one is the linear regime, in which the non-linear terms of the field equations of motion are subdominant, and hence the excited range of momenta coincide with the position of the resonance bands derived analytically from the \emph{Lam\'e} and \emph{Mathieu} equations. The second one is the non-linear regime, in which these terms become relevant for the dynamics. The transition from the first to the second regime takes place at time $z_{\rm br}$ approximately.

In Fig.~\ref{fig:spectra-table} we show the time-evolution of $\rho_k$ for the mother and daughter fields, for both preheating with quartic potential (upper panels) and quadratic potential (lower panels). The black dashed lines in each of the four panels indicate the spectra at times $z_{\rm br}$ and $z_{\rm dec}$. The vertical gray dashed lines indicate the position of the resonance bands. For quartic preheating we have chosen the parameter $q = 43$, which possesses a band of the type $k_- < k < k_+$, with $k_-$ and $k_+$ being two particular numbers obtained from the stability/instability chart of the Lam\'e equation (see Fig.~\ref{fig:lame-bands}). For $q \gg 1$, $k_+ \approx k_L$ [Eq.~(\ref{eq:momentumX})]. For quadratic preheating we have chosen $q_* = 25000$, which has a resonance band of the type $0 < k < k_M$, with the definition of $k_M$ given in Eq.~(\ref{eq:mathieu-kcut}).

In the figure, it can be clearly observed that during the linear regime $z \lesssim z_{\rm br}$, the spectra of the daughter fields gets excited precisely at the momenta corresponding to its resonance band. Due to this, the mother fields are dragged by the daughter fields, and hence their spectra also grow. During this regime, a initial structure of peaks appear in the four spectra. However, when we enter into the non-linear regime $z \approx z_{\rm br}$, the spectra starts growing outside the resonance bands. Due to rescattering effects, there is a population of modes of higher and higher momenta as times goes on, which makes the whole spectra to move to the UV. This process makes vanish the peaks formed during the linear-regime. Finally, when we arrive to the time $z \approx z_{\rm dec}$ approximately, the spectra have stabilized, having developed another peak at greater momenta with a hunchback shape.

We can therefore identify the initial linear dynamics as an IR effect, and the subsequent non-linear dynamics as an UV effect. One can check that, as we increase $q$, the separation between the IR and UV scales also grows as $\propto q^{\alpha}$, with $\alpha$ a numerical coefficient dependent on the particular parametric system we study. The parametrization of the position and amplitude of the different peaks of the spectra is relevant, for example, for the study of Gravitational Wave production during this process, but this goes beyond the scope of this publication.

\section{Lattice formulation and initial conditions}\label{appen:Lattice}

In this appendix we provide information about the lattice formulation of our work, as well as about how initial conditions are set in the lattice.

\subsection{Lattice formulation: General considerations}

For the development of this work, we have solved a discretized version of the EOM of the different fields in lattice expanding cubes in $3 + 1$ dimensions. Let us denote the number of points per length dimension in the lattice by $N$, the length of the cube by $L$, the time step of the numerical solver by ${\rm dt}$, and the lattice spacing by ${\rm dx} \equiv L /N$. The discrete momenta defined in this lattice are
\be p_n = n p_{\rm min} \equiv n \frac{2 \pi}{L}, \hspace{0.3cm} n = 1, 2, \cdots , \frac{\sqrt{3} N}{2} , \ee
where the mininum and maximum momenta captured by the lattice are respectively $p_{\rm min} \equiv (2 \pi) /L$ and $p_{\rm max} \equiv (\sqrt{3} N / 2) p_{\rm min}$. We will use $p$ for lattice (discrete) momenta, and $k$ for physical (continuous) momenta from now on.

One must choose the set of parameters $(N,L)$ so that all the relevant momenta for the dynamics of the system are well captured. In the parametric cases we have studied, there are two basically two regimes: the initial linear dynamics, in which the excited range of momenta of the fields coincide with their corresponding resonance bands; and the later non-linear evolution, in which the spectra of the different fields move to the UV, populating modes of higher and higher momenta. This has been described in Appendix~\ref{appen:Spectra}. Therefore, we must ensure that $p_{\rm min} \lesssim \mathcal{O} (0.1) k_L$  for the $\lambda \phi^4$ model [Eq.~(\ref{eq:momentumX})] or $p_{\rm min} \lesssim \mathcal{O} (0.1) k_M$ for the $m^2 \phi^2$ model [Eq.~(\ref{eq:mathieu-kcut})], but also allow $p_{\rm max}$ to be great enough to capture well the subsequent non-linear regime.

This in fact poses two important limitations when simulating parametric resonance in
the lattice for low and great resonance parameters:

\begin{itemize}

\item On the one hand, for lower resonance parameters, the size of the resonance bands is too small, so we cannot introduce an appropriate number of nodes inside the corresponding bands, and hence we cannot simulate well the linear dynamics. This can be clearly see in Fig.~\ref{fig:lame-bands} for the quartic case, where resonance band become extremely narrow for $q< 1$ [definition in Eq.~(\ref{eq:lphi4-qres})]. This is also the case for the quadratic case when the effective resonance parameter becomes less than one, $q = q_*/a^3 (t) < 1$ [definition in Eq.~(\ref{eq:m2phi2-resp})].

\item On the other hand, as we increase the resonance parameter, simulations require higher and higher running time due to several reasons. First, the mother-field decay time grows with a power-law in $q$ [see Eqs.~(\ref{eq:zd-decay}), (\ref{eq:m2phi2-zd}), and (\ref{eq:spec-zetime})], which makes necessary to increase the running time if we want to observe well the inflaton decay. Second, as we described in Appendix~\ref{appen:Spectra}, the larger the $q$, the broader the separation between infrared
(IR) and ultraviolet (UV) scales in momentum space, making necessary the use of an
increasing number of lattice points in the box. Finally, as we increase the resonance
parameter, the rescattering process during the non-linear regime populate modes of
higher and higher momenta, which makes necessary to increase the UV cutoff of the
lattice, and hence reducing ${\rm dx}$. As we need to ensure the stability condition of the differential equation iterative solver ${\rm dt}/{\rm dx} < 1/\sqrt{3}$, this implies reducing ${\rm dt}$, which hence also increases the necessary running time.

\end{itemize}

\begin{figure}
      \begin{center}
                  \includegraphics[width=7.5cm]{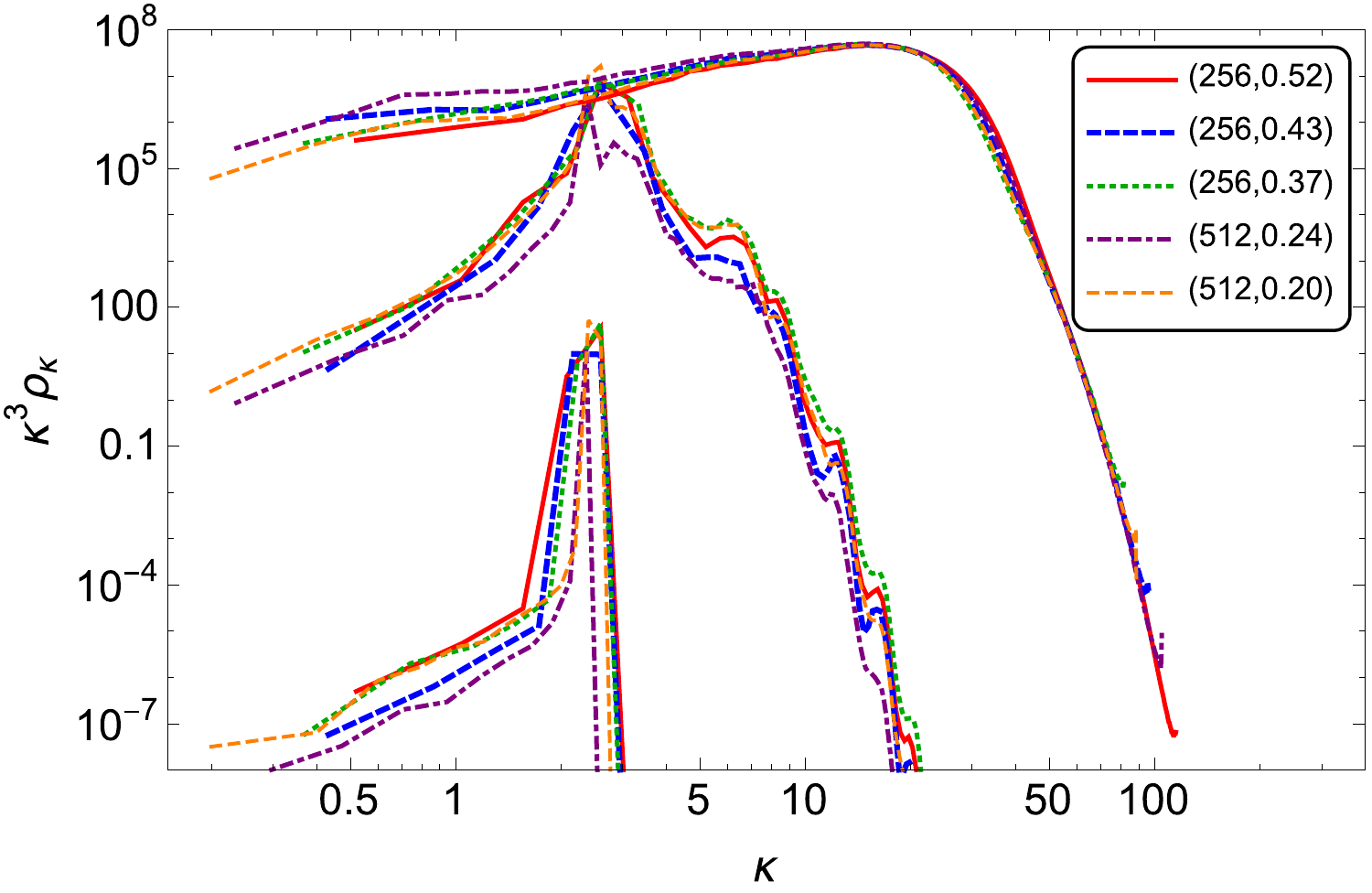}
					 \hspace{0.2cm}
                  \includegraphics[width=7.5cm]{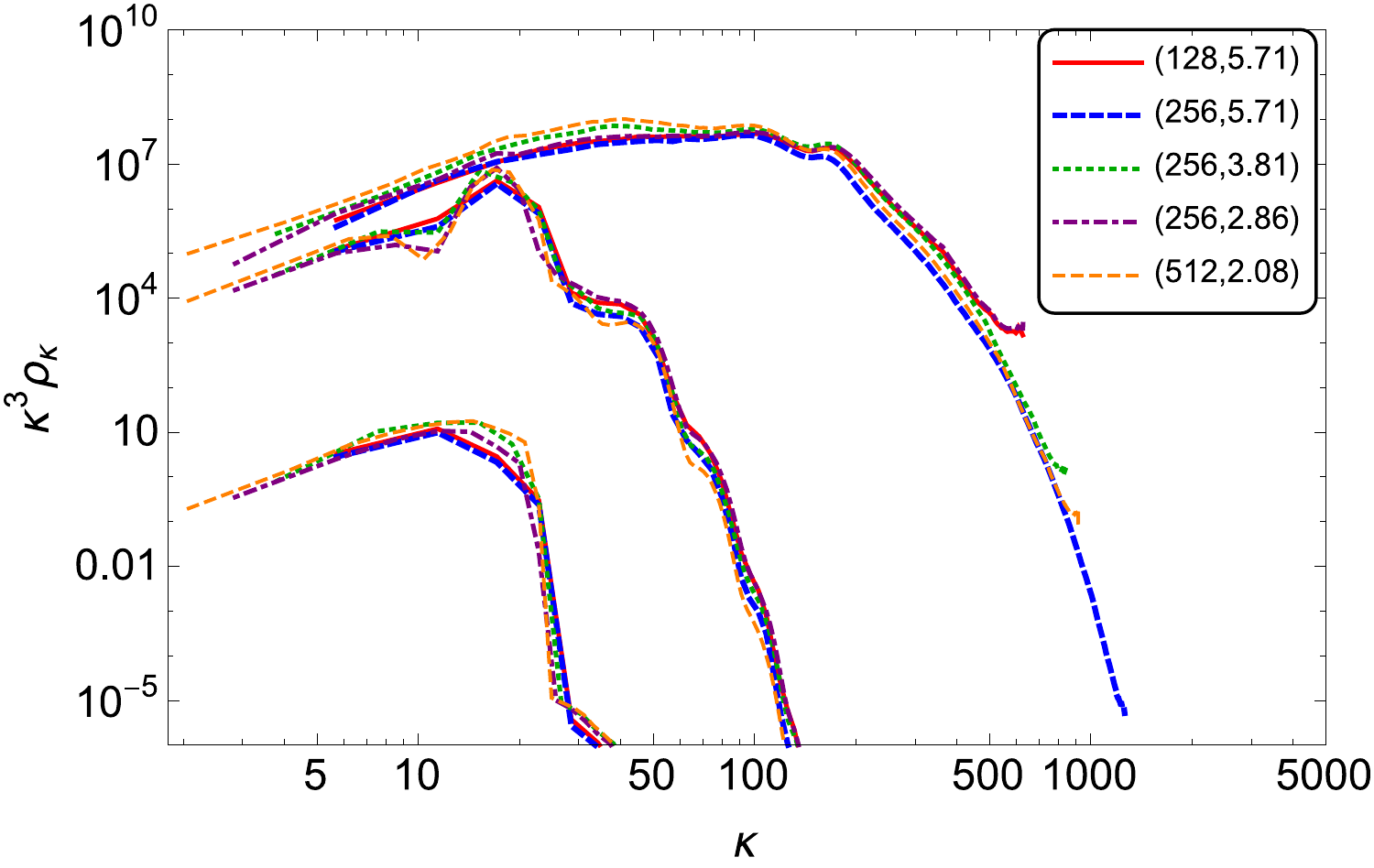}
     \end{center}
                \caption{We compare the daughter-field spectra for different sets of $(N,p_{\rm min} /H_*)$, where $N$ is the number of lattice points per length of the cube, and $p_{\rm min}$ is the minimum momentum covered by the lattice. Left panel
shows the spectra for $\lambda \phi^4$ preheating and times $z = 50, 101, 211$. Right panel shows the spectra for $m^2 \phi^2$ preheating and times $z = 25, 81, 116$.}    \label{fig:lattice-depen}
 \end{figure}

For the obtention of this paper results, we have used lattices with $N^3 = (256)^3$ points, changing $L$ in each simulation so that the lattice covers both the IR and the UV dynamics. Some simulations have been also done with $N^3 = (128)^3$ and $N^3 = (512)^3$ to check the independence of the results on the particular lattice characteristics, see Fig.~\ref{fig:lattice-depen}. With these parameters, and our available computational resources, we have been able to simulate $0.4 < q < 5000$ for quartic preheating, $6000 < q_* < 2.5 \times 10^6$ for quadratic preheating, and $10^4 < q_* < 10^7$ for a spectator field with quadratic potential. Increasing the power of the applied computational resources can push up the upper limit, but despite that, these limitations cannot be avoided if one increases the resonance parameter indefinitely. Fortunately, our results have shown that the most relevant time scales $z_{\rm br}$ and $z_{\rm dec}$ can be nicely fitted and easily extrapolated oustide the range of cases directly simulated.

\subsection{Discretized equations}

Let $ \hat{n}$ be a point in the lattice. We define the discrete derivatives as
\be (\Delta_{\mu}^+ \varphi) (\hat{n} ) = \frac{1}{ {\rm dx}_{\mu}} ( \varphi ( \hat{n} +{\rm dx}_{\mu} ) - \varphi (\hat{n}) ) \ , \hspace{0.3cm} (\Delta_{\mu}^- \varphi) (\hat{n} ) = \frac{1}{ {\rm dx}_{\mu}} ( \varphi ( \hat{n}) - \varphi (\hat{n} - {\rm dx}_{\mu}  ) ) \ , \ee
with $\mu=0,1,2,3$, ${\rm dx_0} \equiv {\rm dt}$ the time step, and ${\rm dx_i} \equiv {\rm dx}$ ($i=1,2,3$).

For preheating with quartic potential, we wrote the field EOM in the continuum in
Eq.~(\ref{eq:eom-lame}). In the lattice, we solve the following discrete version of them:
\bea 
\Delta_0^- \Delta_0^+ \varphi - \frac{(\Delta_0^- \Delta_0^+ a)}{a} \varphi - \sum_i \Delta_i^- \Delta_i^+ \varphi + [\varphi^2 + q \chi^2] \varphi &=& 0 \ , \\
\Delta_0^- \Delta_0^+ \chi - \frac{(\Delta_0^- \Delta_0^+ a)}{a}  \chi - \sum_i \Delta_i^- \Delta_i^+ \chi + q \varphi^2 \chi &=& 0 \ .
\eea

For preheating with quadratic potential, we wrote the field EOM in the continuum in
Eqs.~(\ref{eq:m2phi2-eom})-(\ref{eq:m2phi2-eom2}). In the lattice, we solve the following discrete version of them:
\bea \Delta_0^- \Delta_0^+ \varphi + \left[ - \frac{3(\Delta_0^+ a)^2}{4 a^2} -  \frac{3 (\Delta_0^- \Delta_0^+ a)}{2 a} \right] \varphi - \frac{1}{a^2} \sum_i \Delta_i^- \Delta_i^+ \varphi + \left( 1 + \frac{4}{a^3} q_* \chi^2 \right) \varphi &=& 0 \ , \label{eq:discr-m2phi2a}\\
\Delta_0^- \Delta_0^+ \chi + \left[ - \frac{3 (\Delta_0^+ a)^2}{4 a^2} -  \frac{3 (\Delta_0^- \Delta_0^+ a)}{2 a} \right] \varphi - \frac{1}{a^2} \sum_i \Delta_i^- \Delta_i^+ \chi + \frac{4}{a^3} q_* \varphi^2 \chi &=& 0 \ .\label{eq:discr-m2phi2b}
\eea
These two couples of equations are solved self-consistently with a discrete version of the Friedmann equations, or more specifically, a particular combination of them. The algorithm to solve the discrete equations in time, as well as the output functions (means, energy, spectra...), are identical to the ones used in the Latticeeasy code \cite{Latticeeasy-paper}, so we refer to its documentation for more details.

Finally, for the case in which the mother field is a spectator field with quadratic potential, the equations of motion are the same as in Eq.~(\ref{eq:discr-m2phi2a}), but with the scale factor evolving as the fixed power-law equation (\ref{eq:exprate}), instead of being solved self-consistently as in the preheating cases. 

\subsection{Initial conditions}

We now discuss how we set the initial conditions in the lattice, and check if the lattice results presented in this work depend on its intrinsic random behaviour. For simplicity, we will use conformal variables in this section, instead of natural variables.

We set the initial time of the lattice simulations at $t_*$, which is given by the onset of the oscillatory regime of the mother field. For the quartic model, this time is defined when the condition $H(t_*) = \sqrt{\lambda} \phi (t_*)$ holds, while for the quadratic model, this condition is $H(t_*) = m$. At this time, we impose to both the daughter and mother fields the initial homogeneous modes $X (t_*) =0$, and $\phi (t_*) \equiv \phi_*$. On top of these, we put to both fields a spectra of initial modes mimicking quantum fluctuations . Let us call $f$ to either of the fields (i.e. $f=\phi, X$), and call $f(k)$ to its Fourier transform in momentum space. The spectra we impose is

\bea
f(k) =
  \left\lbrace
  \begin{array}{l}
     \frac{|f_k|}{\sqrt{2}} (e^{i \theta_1} + e^{i \theta_2} ) \hspace{0.2cm}\text{ if } k < k_c , \vspace*{2mm}\\
     0 \hspace{2.5cm} \text{ if } k \geq k_c , \\  \end{array} \right. \ , \hspace{0.4cm}     
     \dot{f}(k) =
  \left\lbrace
  \begin{array}{l}
     \frac{|f_k|}{\sqrt{2}} i \omega_k (e^{i \theta_1} - e^{i \theta_2} ) \hspace{0.2cm}\text{ if } k < k_c , \vspace*{2mm}\\
     0 \hspace{3.1cm} \text{ if } k \geq k_c , \\  \end{array} \right. \label{eq:initfluct}      
      \eea
where the two independent solutions for $f (k)$ and $\dot{f} (k)$ account for left-moving and right-moving waves. Here, $\omega_k = \sqrt{(k/a_*)^2 + m_f^2}$ is the frequency of the mode $k$, with $m_f$ the initial effective mass of the field $f$, and $a_*$ the initial scale factor. For the quartic model in Section \ref{sec:lphi4}, we have $m_{\phi}^{2}=3 \lambda \phi_*^2$ and $m_{\chi}^2 = g^2 \phi_*^2$, while for the quadratic model of Sections \ref{sec:m2phi2} and \ref{sec:specfields}, we have $m_{\phi}^2 = m^2$ and $m_{\chi}^2 = g^2 \phi_*^2$. 

In (\ref{eq:initfluct}), we have included a cutoff $k_c$ in the initial fluctuations, so that only modes with $k < k_c$ are excited initially. Here, $|f_k|$ is a quantity that changes from point to point of the lattice in momentum space, following the probability distribution function
\be P(|f_k|) d |f_k| = \frac{2 |f_k|}{\langle |f_k|^2 \rangle} e^{- \frac{| f_k|^2}{\langle |f_k|^2 \rangle}} d |f_k| \ , \hspace{0.5cm} \langle |f_k|^2 \rangle = \frac{1}{2 a_*^3 \omega_k} \ .\ee
On the other hand, we also let the phases $\theta_{1,2}$ vary randomly throughout the lattice  in the interval $\theta_{1,2} \in [0,2\pi)$. In practice, the randomness of both $|f_k|$ and $\theta_{1,2}$ is implemented in the code with a pseudo-random number generator, so that different seeds generate different realizations for these quantities.

\begin{figure}
      \begin{center}
                  \includegraphics[width=7.5cm]{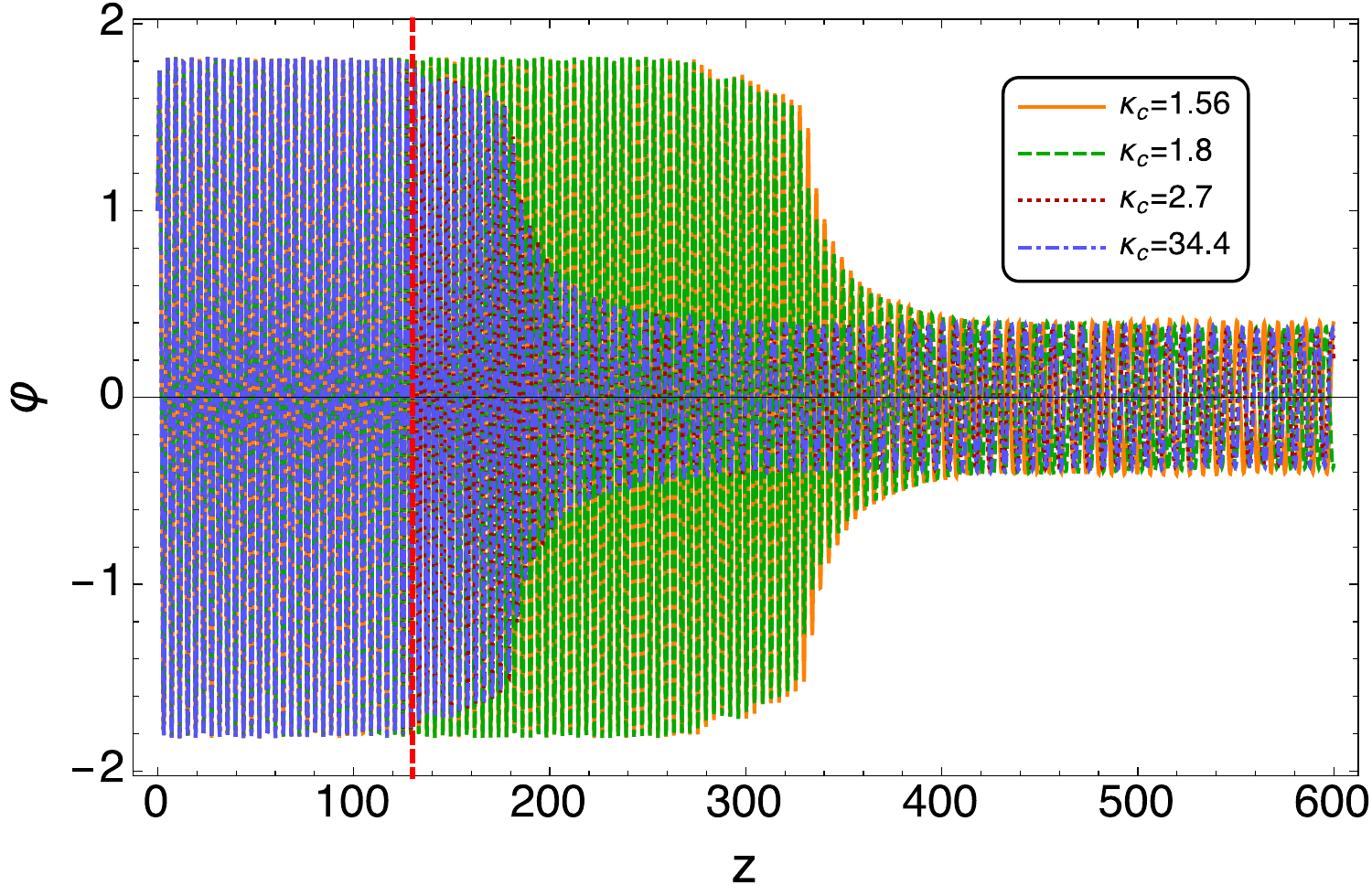}
					 \hspace{0.2cm}
                  \includegraphics[width=7.5cm]{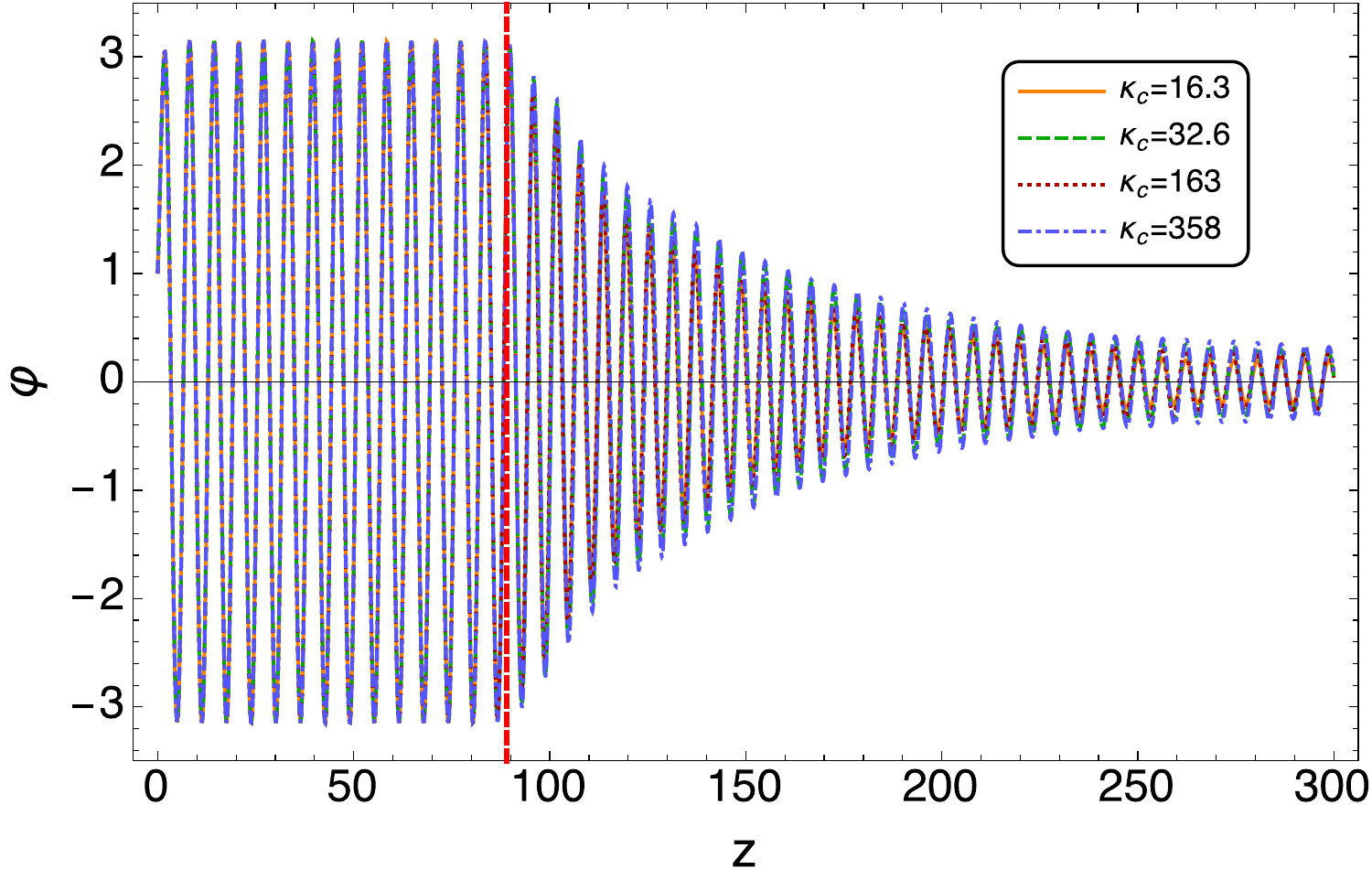}
                    \includegraphics[width=7.5cm]{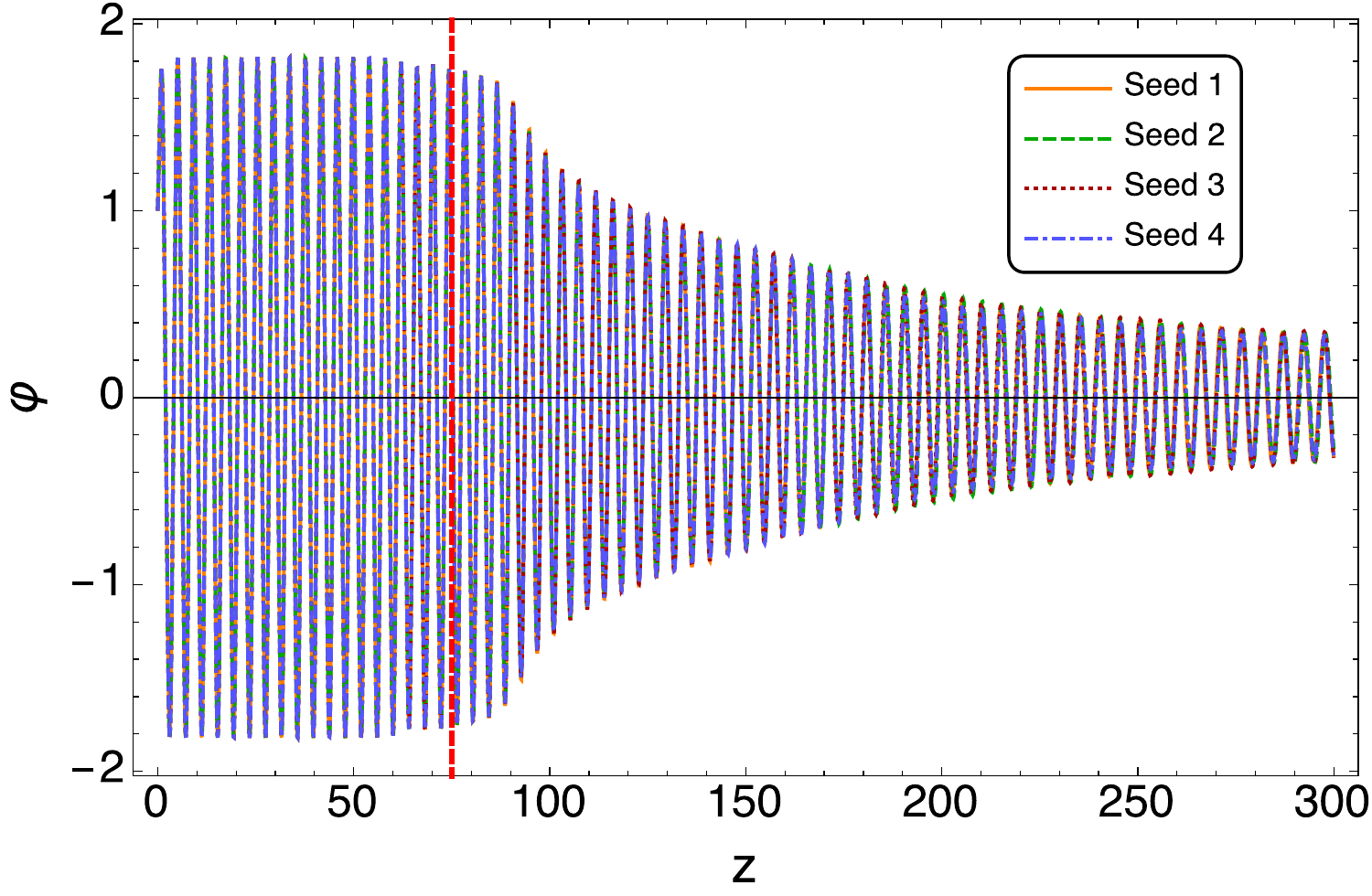} \hspace{0.2cm}
                  \includegraphics[width=7.5cm]{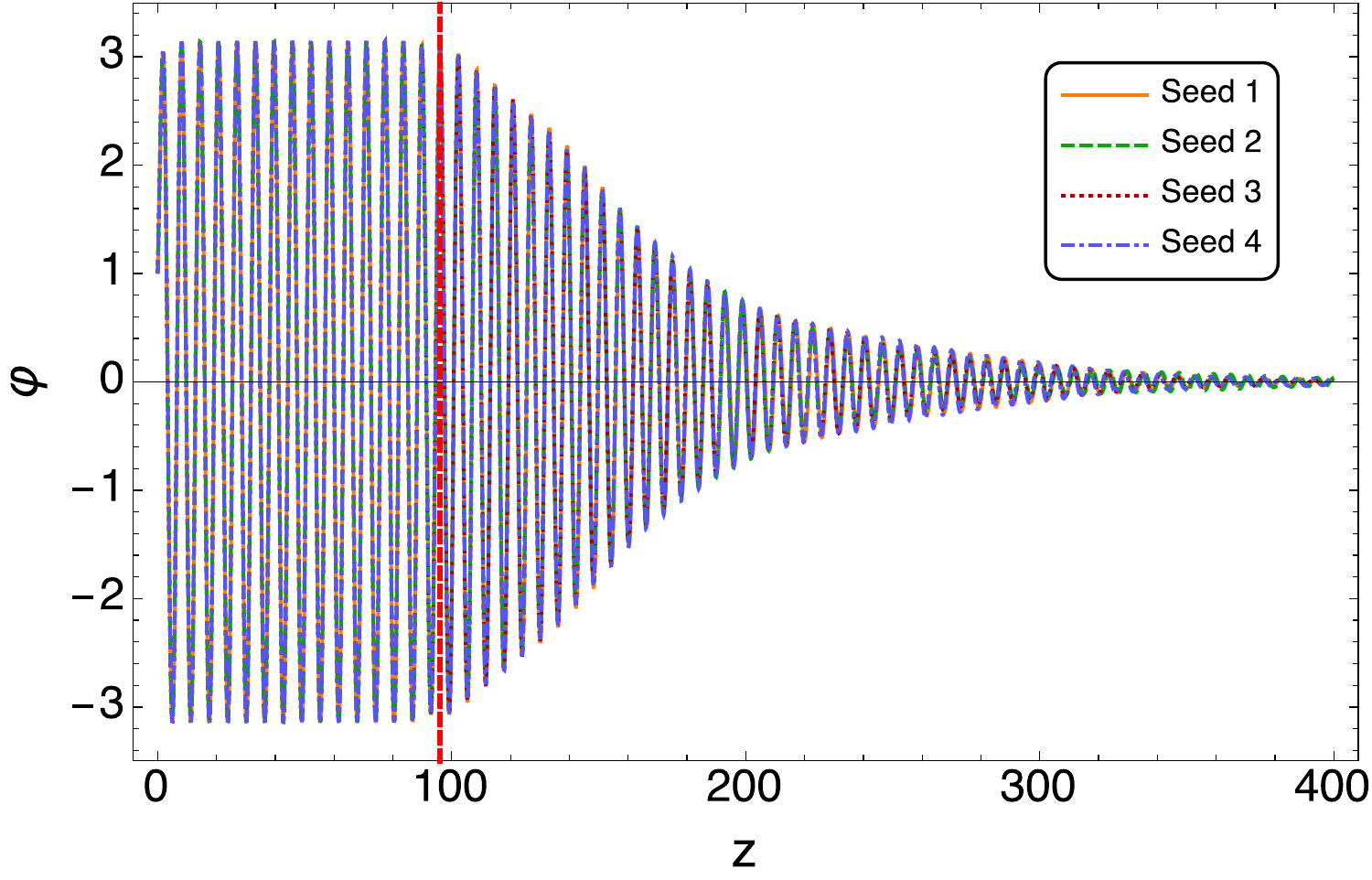}
     \end{center}
                \caption{We show the volume-average amplitude of the inflaton field, and how it changes when varying the cutoff $\kappa_c \equiv k /H_*$ (top panels), and the seed that generates the different realizations of $|f_k|$ and $\theta_{1,2}$ (bottom panels). The top-left panel shows the inflaton amplitude for $\lambda \phi^4$ preheating and $q=11$, for different values of $\kappa_c$. The top-right panel shows the same for $m^2 \phi^2$ preheating, with the choice $q_*=30000$. The bottom-left panel shows, for $q = 86$, the average of the inflaton for $\lambda phi^4$ preheating and different realizations (different seeds) of $\theta_{1,2}$ and $|f_k|$. The bottom-right panel shows the same for $m^2 \phi^2$ preheating for $q_*=80000$. In the four panels, the vertical red dashed line indicates the estimated time $z_{\rm br}$.}    \label{fig:seeds-kcutoff}
 \end{figure}

Therefore, there are two sources of uncertainty with respect the initial conditions. The first one is how to fix the position $k_c$ of the cutoff. A proper choice allows to filtrate only those modes that are being exponentially excited, and hence can be treated as classical. This means that for parametric resonance with quadratic potential, one should in principle fix $k_c \approx k_M$ [Eq.~(\ref{eq:mathieu-kcut})], while for the quartic case the choice should be $k_c \approx k_+$, with $k_+$ the maximum momentum of the main resonance band of the \emph{Lam\'e} equation [Eq.~(\ref{eq:modeEQ})]. The second one comes from the different choices of seed, which generate different realizations for $|f_k|$ and $\theta_{1,2}$ throughout the lattice. However, if we want to trust the results from our lattice simulations, these should not depend very much on the particular choice of initial conditions. It is important, therefore, to check this issue thorougly.

We show in Fig.~\ref{fig:seeds-kcutoff} the time-evolution of the volume-average amplitude of the inflaton field for different choices of $k_c$ and seed. The top two panels show the inflaton average for different choices of $k_c$, for both quartic preheating (top-left panel), and quadratic preheating (top-right panel). The two bottom panels show the same, but varying in this case the seed in the pseudo-random number generator for quartic preheating (bottom-left panel), and quadratic preheating (bottom-right panel). In all panels, we indicate the estimated time $z_{\rm br}$ with a vertical red dashed line.

Let us focus first in the top-left panel. We have chosen the particular case $q=11$, which has a main resonance band of the type $\kappa_- < \kappa < \kappa_+$ ($\kappa \equiv k /H_*$), with $\kappa_-=2.36$ and $\kappa_+=2.78$. One can observe that for $\kappa_c = 1.56$ and $\kappa_c =1.8$, the onset of the inflaton decay takes much longer than for the other cases, giving the estimated time $z_{\rm br} \approx 300$. This happens because in these two cases, we have $\kappa_c < \kappa_{-,+}$, and hence we are not exciting the main resonance band initially. This falsifies the dynamics. On the other hand, for the simulations with $\kappa_c = 2.7$ and $\kappa_c = 34.4$, we get the same behaviour for the inflaton, coinciding both with the estimated (shorter) time $z_{\rm br} \approx 130$. This shows that in the quartic preheating case, the dynamics of the system is very independent on the position of $\kappa_c$, as long as $\kappa_c > \kappa_+$\footnote{It is important to mention that fixing the cutoff of the initial fluctuations as $\kappa_c = \kappa_L = q^{1/2} / \sqrt{ 2 \pi^2}$ [Eq.~(\ref{eq:momentumX})] is wrong. For certain choices of $q$, we have $\kappa_L < \kappa_{-,+}$, and hence we would not capture well the initial resonance band. This is clearly seen in the top-left panel of Fig.~\ref{fig:seeds-kcutoff}, where for $q=11$ we have $\kappa_L \approx 1.56 < \kappa_{-} \approx 2.36$.}.

If we now focus on the quadratic preheating case of the top-right panel, we observe that the inflaton dynamics are also independent on the choice of $\kappa_c$, giving all simulations the estimated value $z_{\rm br} \approx 89$. Note that we have depicted here the case $q_*=30000$, which corresponds to an estimated cutoff of $\kappa_M \approx 32.6$.

Finally, as we can observe in the two bottom panels, the inflaton dynamics also remain unchanged when varying the seed of the pseudo-random number generator, and hence for different realizations of $|f_k |$ and $\theta_{1,2}$. Therefore, we conclude that the source of error in the estimation of the time scales coming from the uncertainty in the initial conditions is negligible.

\bibliography{FitParamResBiblio}
\bibliographystyle{h-physrev4}

\end{document}